\documentclass[a4paper,dvipsnames]{article}

\usepackage{geometry}
\newcommand{\Version}{1}

\RequirePackage{ifthen}
\ifthenelse{\Version=0}{\usepackage[notcite,notref]{showkeys}}{}

\newcommand{\OnlyPreprint}[1]{%
	\ifthenelse{\Version=0}{%
		\medskip
		\todo[color=green!50]{\ding{247}\ding{247}\ding{247}\ding{247}\ding{247} ONLY in preprint}
		\hspace*{0pt}\\
		{#1}
		\hspace*{0pt}\\
		\todo[nolist,color=green!50]{\ding{249}\ding{249}\ding{249}\ding{249}\ding{249}}
		\medskip
	}%
	{\ifthenelse{\Version=2}{{#1}}{}}%
}%

\newcommand{\OnlySubmission}[1]{%
	\ifthenelse{\Version=0}{%
		\medskip
		\todo[color=brown!80]{\ding{247}\ding{247}\ding{247}\ding{247}\ding{247} ONLY in submission}
		\hspace*{0pt}\\
		{#1}
		\hspace*{0pt}\\
		\todo[nolist,color=brown!80]{\ding{249}\ding{249}\ding{249}\ding{249}\ding{249}}
		\medskip
	}%
	{\ifthenelse{\Version=1}{{#1}}{}}%
}%

\usepackage[british]{babel}
\usepackage{amsmath}
\usepackage{amssymb}
\usepackage{amsthm}
\usepackage[UKenglish]{isodate}
\usepackage{bm}
\usepackage{dsfont}
\usepackage{enumitem}
\usepackage{float}
\usepackage[OT2,OT1]{fontenc}
\usepackage{ifthen}
\usepackage[cal=cm,scr=boondoxo]{mathalfa}
\usepackage{pifont}
\usepackage{rotating}
\usepackage{textcomp}
\usepackage{theoremref}
\usepackage{tikz}
	\usetikzlibrary{arrows}
	\usetikzlibrary{patterns}
\usepackage[textwidth=3cm,colorinlistoftodos]{todonotes}
\usepackage{xcolor}
\usepackage{stackrel}

\newcommand{\mc}[1]{{\mathcal{#1}}}			
\newcommand{\ms}[1]{{\mathscr{#1}}}			
\newcommand{\bb}[1]{{\mathbb{#1}}}			
\newcommand{\qu}{\overline}				
\newcommand{\mr}{\mathring}				
\newcommand{\wt}{\widetilde}			
\newcommand{\wh}{\widehat}			

\DeclareMathOperator{\RE}{Re}\renewcommand{\Re}{\RE}	
\DeclareMathOperator{\IM}{Im}\renewcommand{\Im}{\IM}	
\DeclareMathOperator{\tr}{tr}				
\newcommand{\Poi}[1]{\mc P[{#1}]}			
\newcommand{\Smallo}{{\rm o}}				
\newcommand{\BigO}{{\rm O}}				
\DeclareMathOperator{\B}{B}				

\newcommand{\DS}{\colon\mkern3mu}			
\newcommand{\DQ}{\mkern6mu}				
\newcommand{\DD}{\mkern4mu{\mathrm d}}			
\newcommand{\RD}{\mathrm{d}}				
\newcommand{\DP}{{.\kern7pt}}				
\newcommand{\DF}{\colon}				
\newcommand{\DE}{\mathrel{\mathop:}=}			
\newcommand{\DI}{\mathrel{\mathop:}\Leftrightarrow}	

\DeclareMathOperator{\ran}{ran}

\newcommand{\smmatrix}[4]{\Bigl(			
\begin{smallmatrix}
\hspace*{-0.2ex} #1 \hspace*{0.2ex} & \hspace*{0.2ex} #2 \hspace*{-0.2ex}
\\[0.5ex]
\hspace*{-0.2ex} #3 \hspace*{0.2ex} & \hspace*{0.2ex} #4 \hspace*{-0.2ex}
\end{smallmatrix}
\Bigr)}

\newcommand{\String}[2]{\ms S[{#1},{\ms #2}]}		

\newlength{\maxlabwidth}

\newcounter{Enum}					
\newenvironment{Enumerate}{\begin{enumerate}[label={\rm({\roman*})}]}{\end{enumerate}}
\newcommand{\Enumref}[1]{{\setcounter{Enum}{#1}{\rm(\roman{Enum})}}}

\newcommand{\descriptionlabelsave}{}			
\newenvironment{Itemize}{%
	\renewcommand{\descriptionlabelsave}{\descriptionlabel}\renewcommand{\descriptionlabel}{$\triangleright$}%
	\begin{description}[leftmargin=15pt,itemindent=-5.2pt]}{%
	\end{description}\renewcommand{\descriptionlabel}{\descriptionlabelsave}}

\newcounter{StepsCount}					
\newenvironment{Steps}{%
	\begin{list}{\ding{\value{StepsCount}}}{\usecounter{StepsCount} \leftmargin=0pt \labelwidth=12pt \itemindent=\labelwidth%
	\itemsep=5pt\listparindent=\parindent} \setcounter{StepsCount}{171}}{\end{list}}

\numberwithin{equation}{section}
\swapnumbers
\theoremstyle{plain}
	\newtheorem{lemma}{Lemma}[section]
	\newtheorem{proposition}[lemma]{Proposition}
	\newtheorem{theorem}[lemma]{Theorem}
	\newtheorem{corollary}[lemma]{Corollary}
	\newtheorem{ntheoreM}[lemma]{}
\theoremstyle{definition}
	\newtheorem{definitioN}[lemma]{Definition}
	\newtheorem{ndefinitioN}[lemma]{}
	\newtheorem{examplE}[lemma]{Example}
\theoremstyle{remark}
	\newtheorem{remarK}[lemma]{Remark}
	\newtheorem{nremarK}[lemma]{}
\newcommand{\thlab}[1]{\thlabel{#1}\label{#1 }}
\renewcommand{\qedsymbol}{\raisebox{-2pt}{\large\ding{113}}}
\newcommand{\defendsymbol}{$\lozenge$}
\newcommand{\qedsymbolsave}{\qedsymbol}

\newenvironment{definition}{\begin{definitioN}}{
	\renewcommand{\qedsymbolsave}{\qedsymbol}\renewcommand{\qedsymbol}{\defendsymbol}
	\popQED{\qed}\renewcommand{\qedsymbol}{\qedsymbolsave}\end{definitioN}}

\newenvironment{remark}{\begin{remarK}}{
	\renewcommand{\qedsymbolsave}{\qedsymbol}\renewcommand{\qedsymbol}{\defendsymbol}
	\popQED{\qed}\renewcommand{\qedsymbol}{\qedsymbolsave}\end{remarK}}
\newenvironment{example}{\begin{examplE}}{
	\renewcommand{\qedsymbolsave}{\qedsymbol}\renewcommand{\qedsymbol}{\defendsymbol}
	\popQED{\qed}\renewcommand{\qedsymbol}{\qedsymbolsave}\end{examplE}}

\newcommand\cyr{%
\renewcommand\rmdefault{wncyr}%
\renewcommand\sfdefault{wncyss}%
\renewcommand\encodingdefault{OT2}%
\normalfont
\selectfont}
\DeclareTextFontCommand{\textcyr}{\cyr}

\newcommand{\Qlabel}{\framebox{\vbox to 6pt{\vfil\hbox to 6pt{\hfil{\bf\alph*}\hfil}\vfil}}}

\newcommand{\dblarrow}[1]{\overset{\text{\tiny\raisebox{-0.2ex}{$\leftrightarrow$}}}{#1}}
\newcommand{\hatcalM}{\rule{0ex}{1ex}\hspace{0.7ex}\widehat{\rule[1.5ex]{1.5ex}{0ex}}\hspace{-2.4ex}{\mathcal M}}
\newcommand{\tildM}{\rule{0ex}{1ex}\hspace{0.6ex}\widetilde{\rule[1.5ex]{1.5ex}{0ex}}\hspace{-2.2ex}M}
\newcommand{\tildmsm}{\rule{0ex}{1ex}\hspace{0.5ex}\widetilde{\rule[0.9ex]{0.9ex}{0ex}}\hspace{-1.6ex}\mathscr m}
\DeclareMathOperator{\conv}{conv}
\newcommand{\ranf}[1]{\mathcal R_{#1}}
\newcommand{\qD}{q_{\textsf{\textup{\text{D}}}}}
\newcommand{\qN}{q_{\textsf{\textup{\text{N}}}}}
\newcommand{\qS}{q_{\textsf{\textup{\text{S}}}}}
\newcommand{\muS}{\mu_{\textsf{\textup{\text{S}}}}}
\newcommand{\alphaS}{\alpha_{\textsf{\textup{\text{S}}}}}
\newcommand{\wtqD}{\widetilde{q}_{\textsf{\textup{\text{D}}}}}
\newcommand{\AD}{A_{\textsf{\textup{\text{D}}}}}
\newcommand{\AN}{A_{\textsf{\textup{\text{N}}}}}

\begin{document}

\ifthenelse{\Version=0}{%
	\pagenumbering{roman}
	\fbox{
	\parbox{100mm}{
	\hspace*{0pt}\\[1mm]
	\centerline{{\Large\ding{45}}\quad\,{\large\sc Draft}\quad{\Large\ding{45}}}
	\hspace*{0pt}\\[-2mm]
	\textcircledP\ \ Preliminary version Mon 6 Feb 2023 12:24
	\\[2mm]
	\cleanlookdateon
	\hspace*{5mm} Compilation date: \today
	\\[2mm]
	\ding{233}\quad In final version set variable \textsf{Version} to 1.
	\\[-1mm]
	}
	}
	\IfFileExists{bibtex-database-woracek.bib}
	{}
	{
	\hspace*{0pt}\\[4mm]\centerline{\large\sf !!! File bibtex-database-woracek.bib does not exist !!!}
	\\[-4mm]
	}
	\tableofcontents
	\listoftodos
	\newpage
	\pagenumbering{arabic}
	\setcounter{page}{1}
}
{}

\begin{flushleft}
	{\Large\textbf{Estimates for the Weyl coefficient of a \\[2mm] two-dimensional canonical system}}
\end{flushleft}
	\vspace*{5mm}
	\textsc{
	Matthias Langer
	\,\ $\ast$\,\
	Raphael Pruckner
	\,\ $\ast$\,\
	Harald Woracek
	}
	\\[6mm]
	{\small
	\textbf{Abstract.}
		For a two-dimensional canonical system $y'(t)=zJH(t)y(t)$ on some interval $(a,b)$
		whose Hamiltonian $H$ is a.e.\ positive semi-definite and which is regular
		at $a$ and in the limit point case at $b$, denote by $q_H$ its Weyl coefficient.
		De~Branges' inverse spectral theorem states that the assignment
		$H\mapsto q_H$ is a bijection between Hamiltonians (suitably normalised)
		and Nevanlinna functions.

		We give upper and lower bounds for $|q_H(z)|$ and $\Im q_H(z)$
		when $z$ tends to $i\infty$ non-tangentially.
		These bounds depend on the Hamiltonian $H$ near the left endpoint $a$
		and determine $|q_H(z)|$ up to universal multiplicative constants.
		We obtain that the growth of $|q_H(z)|$ is independent of the off-diagonal entries
		of $H$ and depends monotonically on the diagonal entries in a natural way.
		The imaginary part is, in general, not fully determined by our bounds
		(in forthcoming work we shall prove that for ``most'' Hamiltonians also $\Im q_H(z)$
		is fully determined).

		We translate the asymptotic behaviour of $q_H$ to the behaviour of
		the spectral measure $\mu_H$ of $H$ by means of Abelian--Tauberian results
		and obtain conditions for membership of growth classes defined by
		weighted integrability condition (Kac classes)
		or by boundedness of tails at $\pm\infty$ w.r.t.\ a weight function.
		Moreover, we apply our results to Krein strings and Sturm--Liouville equations.
	\\[3mm]
	}

\begin{center}
	{\textbf{Table of contents}}
\end{center}
\vspace*{-1ex}
{\footnotesize
\begin{itemize}
\item[\ref{Y56}.] Introduction\hfill p.~\pageref{Y56}
\item[\ref{Y40}.] Weyl discs and matrix algebra\hfill p.~\pageref{Y40}
\item[\ref{Y41}.] Proof of Weyl coefficient estimates\hfill p.~\pageref{Y41}
\item[\ref{Y43}.] Asymptotic behaviour of the spectral measure\hfill p.~\pageref{Y43}
\item[\ref{Y34}.] Further discussion of the main theorem\hfill p.~\pageref{Y34}
\item[\ref{Y81}.] Relation to previous work\hfill p.~\pageref{Y81}
\item[\ref{Y50}.] Regularly varying functions\hfill p.~\pageref{Y50}
\item[\ref{Y108}.] The generalised inverse of a non-decreasing function\hfill p.~\pageref{Y108}
\end{itemize}
}


%
%
%
\section{Introduction}
\label{Y56}

We study two-dimensional canonical systems
\begin{equation}\label{Y99}
	y'(t)=zJH(t)y(t),\qquad t\in[a,b),
\end{equation}
where $-\infty<a<b\le\infty$, $J\DE\smmatrix 0{-1}10$, $z\in\bb C$,
and where the \emph{Hamiltonian} $H$ is assumed to satisfy
\begin{Itemize}
\item
	$H\in L^1_{\textrm{loc}}\big([a,b),\bb R^{2\times 2}\big)$
	and $\{t\in[a,b)\DS H(t)=0\}$ has measure $0$;
\item
	$H(t)\ge 0$, $t\in[a,b)$ a.e.\ (in the sense of positive semi-definiteness);
\item
	$H$ is in the limit point case at $b$, i.e.\
	\begin{equation}\label{Y138}
		\int_a^b\tr H(t)\DD t=\infty.
	\end{equation}
\end{Itemize}
Differential equations of this form appear frequently in theory and applications.
They can be shown to be a unifying framework for classical equations
like Schr\"odinger equations, Krein strings, Dirac systems, and others;
see, e.g.\ \cite{remling:2002,kaltenbaeck.winkler.woracek:bimmel,sakhnovich:2002,romanov:1408.6022v1}.
For the relevance (and origins) of canonical systems in natural sciences we refer to
\cite{arnold:1989,faddeev.takhtajan:2007,flanders:1989,ramm:2005,schechter:1981}.

In the spectral theory of equation \eqref{Y99} the notion of the \emph{Weyl coefficient}
$q_H$ associated with a Hamiltonian $H$ plays a crucial role.
The construction of $q_H$ goes back to H.~Weyl \cite{weyl:1910}
and is based on a nested discs argument; see \eqref{Y139} below for the definition.
The Weyl coefficient is a Nevanlinna function, i.e.\ it is analytic in the
open upper and lower half-planes $\bb C^+$ and $\bb C^-$,
it is symmetric in the sense that $q_H(\qu z)=\qu{q_H(z)}$,
and $\Im q_H(z)\ge0$ for $z\in\bb C^+$.
Being a Nevanlinna function, $q_H$ admits the Herglotz integral representation
\begin{equation}\label{Y156}
	q_H(z) = \alpha_H + \beta_Hz
	+ \int_{\bb R}\Bigl(\frac{1}{t-z}-\frac{t}{1+t^2}\Bigr)\DD\mu_H(t),
	\qquad z\in\bb C\setminus\bb R,
\end{equation}
where $\alpha_H\in\bb R$, $\beta_H\ge0$ and $\mu_H$ is a Borel measure on $\bb R$
that satisfies $\int_{\bb R}\frac{\DD\mu_H(t)}{1+t^2}<\infty$.
The differential operator corresponding to \eqref{Y99} is unitarily equivalent
to the multiplication operator with the independent variable in the space $L^2(\mu_H)$,
and the unitary equivalence is established by a natural integral operator
(if $\beta_H>0$, then the differential operator is multi-valued and a point mass
at infinity has to be added).
For this reason $\mu_H$ is called the \emph{spectral measure} of $H$.
The famous inverse spectral theorem of L.~de~Branges \cite{debranges:1968}
states that for every Nevanlinna function $q$ there exists an essentially unique
Hamiltonian $H$ such that $q=q_H$.
Up-to-date references for the spectral theory of canonical systems are
\cite{hassi.snoo.winkler:2000,remling:2018,romanov:1408.6022v1}.

Having the\:---\:essentially one-to-one\:---\:correspondence between Hamiltonians
on the one side and their Weyl coefficients or spectral measures on the other side,
it is a natural task to relate properties of the one to properties of the other.
Both directions in the de~Branges correspondence involve limiting processes.
This makes the correspondence difficult to handle, but also intriguing to investigate.

There are indeed several properties which can\:---\:fully or partially\:---\:be translated.
Some of them instantiate the following principle.
\begin{quote}
	\textit{The behaviour of the distribution function of the spectral measure
	towards $\pm\infty$ corresponds to the behaviour of the Hamiltonian
	locally at the left endpoint $a$.}
\end{quote}
Usually it is not difficult to find Abelian--Tauberian theorems which allow to
translate the behaviour of tails of $\mu_H$ at $\pm\infty$ to the behaviour of
its Herglotz integral $q_H$ at $i\infty$ (also called the high-energy behaviour of $q_H$).
Thus, when seeking theorems which instantiate the quoted principle,
the essence is to translate properties of $q_H$ locally at $i\infty$ to
properties of $H$ locally at $a$.
In the present paper we contribute to this family of results.
Namely, we prove upper and lower estimates for the modulus and the imaginary part
of $q_H$ locally at $i\infty$ in terms of integrals of the Hamiltonian
in a neighbourhood of the left endpoint $a$.

For Sturm--Liouville equations such estimates go back to, at least,
\cite{hille:1963}, \cite{atkinson:1988} and \cite{bennewitz:1989};
for Krein strings estimates for the principal Titchmarsh--Weyl coefficient
were proved in \cite{kasahara:1975} and \cite{kasahara.watanabe:2010};
for Jacobi operators in \cite{jitomirskaya.last:1999};
for canonical systems some results were obtained in \cite{winkler:2000a}.
Estimates of the distribution function of the spectral measure
are studied in, e.g.\ \cite{marchenko:1952} for Sturm--Liouville equations
and \cite{kac:1982} for strings.
A detailed discussion of related work is given in Section~\ref{Y81}.

The following theorem is our main result.
We establish an explicit quantitative relation between $q_H$ and $H$.

\begin{theorem}\thlab{Y98}
	Let $H$ be a Hamiltonian defined on some interval $[a,b)$ as at the
	beginning of the introduction, and write
	\begin{equation}\label{Y08}
		H(t) = \begin{pmatrix} h_1(t) & h_3(t) \\ h_3(t) & h_2(t) \end{pmatrix},\quad
		M(t) = \begin{pmatrix} m_1(t) & m_3(t) \\ m_3(t) & m_2(t) \end{pmatrix}
		\DE\int_a^t H(s)\DD s.
	\end{equation}
	Assume that neither $h_1=0$ a.e.\ nor $h_2=0$ a.e.
	Fix a parameter $\eta\in\bigl(0,1-\frac{1}{\sqrt{2}\,}\bigr)$ and
	set $\sigma\DE\frac{1}{(1-\eta)^2}-1\in(0,1)$.
	For $r>0$, let $\mr t(r)\in(a,b)$ be the unique number that satisfies
	\begin{equation}\label{Y30}
		(m_1m_2)\bigl(\mr t(r)\bigr)=\frac{\eta^2}{4r^2}\,.
	\end{equation}
	Further, set
	\[
		A(r) \DE \sqrt{\frac{m_1(\mr t(r))}{m_2(\mr t(r))}}\,, \qquad
		L(r) \DE A(r)\cdot\frac{\det M(\mr t(r))}{(m_1m_2)(\mr t(r))}\,.
	\]
	Then the Weyl coefficient $q_H$ associated with the Hamiltonian $H$ satisfies,
	for each $\vartheta\in(0,\pi)$ and $r>0$,
	\begin{align}
		\biggl(\frac{1+\sigma+\frac{2}{\eta\sin\vartheta}}{1-\sigma}\biggr)^{-1}\cdot A(r)
		\le \big|q_H\big(re^{i\vartheta}\big)\big|
		&\le \frac{1+\sigma+\frac{2}{\eta\sin\vartheta}}{1-\sigma}\cdot A(r),
		\label{Y35}
		\\
		\big|\Re q_H\bigl(re^{i\vartheta}\bigr)\big|
		&\le \frac{1+\sigma+\frac{1}{\eta\sin\vartheta}}{1-\sigma}\cdot A(r),
		\label{Y92}
		\\
		\frac{\frac{\eta\sin\vartheta}2}{1+|\cos\vartheta|}\cdot\frac{1-\sigma}{1+\sigma}
		\cdot L(r)
		\le \Im q_H\big(re^{i\vartheta}\big)
		&\le \frac{\sigma+\frac 2{\eta\sin\vartheta}}{1-\sigma}\cdot A(r).
		\label{Y96}
	\end{align}
\end{theorem}

\medskip

\noindent
Let us add some comments.

\begin{remark}\thlab{Y19}
	\hfill
	\begin{Enumerate}
	\item
		Let us note that the inequalities in \eqref{Y35} can also be derived from \cite[Theorem~3.2]{hassi.remling.snoo:2000},
		which is based on a completely different proof.
		However, the most involved part of our proof is the proof of the first inequality in \eqref{Y96},
		which cannot be deduced from \cite{hassi.remling.snoo:2000}.
	\item
		By the assumption that neither $h_1$ nor $h_2$ vanishes a.e.,
		the equation \eqref{Y30} indeed has a unique solution for every $r>0$.
		To see this, set
		\begin{equation}\label{Y196}
			\mr a \DE \sup\bigl\{t\in[a,b)\DS m_1(t)m_2(t)=0\bigr\},
		\end{equation}
		which is equal to the right endpoint of a maximal interval of the form $(a,c)$
		where $h_1=0$ a.e.\ or $h_2=0$ a.e.\ if such an interval exists and
		equal to $a$ otherwise.
		Then $(m_1m_2)'=h_1m_2+h_2m_1>0$ a.e.\ on $(\mr a,b)$, and
		\[
			\lim_{t\to b}\bigl(m_1(t)+m_2(t)\bigr) = \int_a^b\tr H(s)\DD s = \infty.
		\]
		This implies that $m_1m_2$ is a strictly increasing bijection
		from $[\mr a,b)$ onto $[0,\infty)$.
		It follows then from \eqref{Y30} that $\mr t$ is a strictly decreasing
		bijection from $(0,\infty)$ onto $(\mr a,b)$.
		Since $\lim_{r\to\infty}\mr t(r)=\mr a$, the inequalities
		in \eqref{Y35}--\eqref{Y96} relate the behaviour of $q_H(re^{i\vartheta})$
		as $r\to\infty$ to the behaviour of $H(t)$ when $t\searrow\mr a$.
		This shows that the theorem is a perfect instance of the above quoted principle
		at the level of the Weyl coefficient, in particular, in the case
		when $\mr a=a$.
		Note that $\mr a>a$ if and only if one of $h_1$ and $h_2$ vanishes locally at $a$;
		this case is discussed in more detail in \S\ref{Y127}.
	\item
		The constants in \eqref{Y35}--\eqref{Y96} are symmetric about $\frac{\pi}{2}$
		and depend monotonically on $\vartheta$
		on $(0,\frac{\pi}{2}]$ and $[\frac{\pi}{2},\pi)$ respectively.
		Hence the estimates are valid in a sector around the positive
		imaginary axis.
	\item
		Alternative forms of $A(r)$ and $L(r)$ are given in Section~\ref{Y192}.
		Examples to show how \thref{Y98} can be applied are discussed in \thref{Y65}
		and Section~\ref{Y193}.
	\item
		Solving equation \eqref{Y30} may not be possible explicitly or
		computationally difficult.
		Using monotonicity properties we can show weaker estimates for $q_H$ already
		from an estimate of the solution.
		Details are given in Section~\ref{Y78}.
	\item
		Some properties of $A(r)$ and $L(r)$ and hence of $q_H(re^{i\vartheta})$
		can be seen directly from properties of the functions $m_j$ without finding $\mr t(r)$,
		e.g.\ $\lim_{r\to\infty}A(r)=0$ if and only if $\lim_{t\to\mr a}\frac{m_1(t)}{m_2(t)}=0$;
		similar characterisations hold for $\liminf_{r\to\infty}A(r)=0$
		or $A$ being bounded.
	\item
		Some results about the spectral measure $\mu_H$ (see \eqref{Y156})
		are deduced in \thref{Y86,Y74}.
		There we deal with Kac classes, i.e.\ weighted integrability of $\mu_H$,
		and boundedness of tails of $\mu_H$ relative to a weight function.
	\item
		Since the estimates in \eqref{Y35}--\eqref{Y96} are valid for all $r\in(0,\infty)$,
		one can also obtain information about the asymptotic behaviour
		of $q_H(re^{i\vartheta})$ as $r\to0$. For this one needs information
		about $m_i(t)$ as $t\to b$.
		This relates to the considerations in \cite[\S5]{winkler:2000a}.
	\item
		The two cases that are excluded in the theorem are trivial:
		if $h_1=0$ a.e., then $q_H\equiv0$; if $h_2=0$ a.e., then $q_H\equiv\infty$.
	\vspace*{-3ex}
	\end{Enumerate}
\end{remark}

\medskip

\noindent
Using \thref{Y98} we can prove an independence result and a comparison result
for the absolute value of $q_H$.
The first one is an immediate corollary, while the second one requires some arguments;
the proof is given in Section~\ref{Y29}.

\begin{corollary}\thlab{Y80}
	Let $H$ be a Hamiltonian defined on some interval $[a,b)$. Then the absolute value $|q_H(ir)|$ is, up to
	universal multiplicative constants, independent of the off-diagonal entries of $H$.
\end{corollary}

\begin{proposition}\thlab{Y09}
	Let $H$ and $\widetilde H$ be two Hamiltonians defined on some interval $[a,b)$
	such that none of the respective diagonal entries $h_1,h_2,\tilde h_1,\tilde h_2$
	vanishes a.e., let $\mr a$ and $\mr{\tilde a}$ be as in \eqref{Y196}
	corresponding to $H$ and $\widetilde H$ respectively,
	and define $\widetilde m_j$ and $\tildM$ analogously to \eqref{Y08}.

	Assume that there exist constants $c_1,c_2,\gamma_1,\gamma_2>0$ and
	a point $a'\in(\max\{\mr a,\mr{\tilde a}\},b]$
	such that, for all $t\in(\max\{\mr a,\mr{\tilde a}\},a')$,
	\begin{align}
		& \frac 1{c_1}\tr M(t) \le \tr\tildM(t) \le c_2\tr M(t),
		\label{Y21}
		\\
		& m_1(t)\le\gamma_1\widetilde m_1(t),\qquad \widetilde m_2(t)\le\gamma_2 m_2(t).
		\label{Y13}
	\end{align}
	Then there exist $C>0$ and $r_0\ge0$, such that for all $r>r_0$,
	\[
		|q_H(ir)|\le C|q_{\widetilde H}(ir)|.
	\]
	The constant $C$ depends on $c_1,c_2,\gamma_1,\gamma_2$, but not on $a',H,\widetilde H$.
	Moreover, $r_0=0$ when $a'=b$.
\end{proposition}

\noindent
Let us point out that some assumption on the absolute sizes of $M$ and $\tildM$
has to be made in order to have a chance for any kind of comparison result
because otherwise, one could rescale the independent variable without changing
the Weyl coefficient.
We use \eqref{Y21} since this is sufficiently flexible in applications;
note that it is clearly satisfied if both $H$ and $\widetilde H$ are trace-normed.
Also note that none of the conditions in \eqref{Y13} can be removed
as simple examples show.

As a corollary of \thref{Y09} we obtain a stability result.

\begin{corollary}\thlab{Y87}
	Let $H$ and $\widetilde H$ be Hamiltonians defined on some interval $[a,b)$
	such that none of the respective diagonal entries $h_1,h_2,\tilde h_1,\tilde h_2$
	vanishes a.e.
	Further, let $a'\in(\max\{\mr a,\mr{\tilde a}\},b]$ and assume
	that $m_1\asymp\widetilde m_1$ and $m_2\asymp\widetilde m_2$
	on $(\max\{\mr a,\mr{\tilde a}\},a')$.
	Then there exists $r_0\ge0$, with $r_0=0$ when $a'=b$, such
	that $|q_H(ir)|\asymp|q_{\widetilde H}(ir)|$ on $(r_0,\infty)$.
\end{corollary}

\noindent
\textbf{Notation.}
In \thref{Y87} and for the rest of the paper we use the following notation.
We write $f\lesssim g$ if there exists $c>0$ such that $f(r)\le cg(r)$ for all $r$,
and we write $f\asymp g$ if $f\lesssim g\,\wedge\, g\lesssim f$.
Moreover, we use the notation $f\ll g$ for $\frac fg\to 0$.
We deliberately do not always specify the range of values of $r$;
one can think of $r$ belonging to a certain portion of the ray $(0,\infty)$,
or to some sequence tending to $\infty$, or similar.

\subsection*{About the method of proof}

The proof of \thref{Y98} is based on the geometric idea to directly estimate Weyl discs.
It follows the approach of H.~Winkler in \cite{winkler:2000a}.

Recall Weyl's nested discs construction:
from the fundamental solution of the system \eqref{Y99} a family of
discs $\Omega_{t,z}$ is built; see \S\ref{Y160} for details.
Here the parameter $t$ ranges over $[a,b)$,
and the spectral parameter $z$ lies in the open upper half-plane.
For each fixed $z\in\bb C^+$, the discs $\Omega_{t,z}$ form a nested family,
and, due to our assumption that $\int_a^b\tr H(s)\DD s=\infty$, they shrink to a
single point when $t$ approaches $b$.
The function $q_H$ is then defined by $\bigcap_{t\in(a,b)}\Omega_{t,z}=\{q_H(z)\}$.

The radius of the Weyl disc $\Omega_{t,ir}$ built at a point in $t$ and a point $ir$
in the plane decays to $0$ not only for each fixed $r$ when $t$ increases to $b$,
but also for each fixed $t$ when $r$ increases to $\infty$.
Hence, when looking at large $r$ one can afford to use $t$ close to $a$ and
still have a relatively small Weyl disc\footnote{%
	This can be seen as a mathematical reason behind the principle that high-energy behaviour
	of $q_H$ corresponds to behaviour of $H$ locally at the left endpoint,
	which we quoted at the very beginning.
	Another, maybe even more striking, reason is established by a certain group action
	on the set of Hamiltonians, namely, by the group of rescaling operators.
	This ``rescaling trick'' goes probably back to Y.~Kasahara in \cite{kasahara:1975}.%
	}.

To produce estimates for $q_H$, we start from the fact that centres and radii of
Weyl discs $\Omega_{t,ir}$ can be expressed explicitly in terms of the fundamental solution.
Then we
\begin{Itemize}
\item
	prove estimates for the power series coefficients of the fundamental solution;
\item
	use these estimates to determine the leading terms of centres and radii of Weyl discs;
\item
	deduce estimates for $q_H$.
\end{Itemize}
The first step is the technical core of the proof.
It can be carried out only when $(t,r)$ lies in the region indicated in the following picture, and
this explains the role of equation \eqref{Y30}: it describes the border of
our method to prove coefficient estimates.
\begin{figure}[H]
\begin{center}
{\footnotesize
%
\begin{tikzpicture}[x={(20pt,0pt)},y={(0pt,20pt)},scale=0.8]
\draw[-triangle 60,thick] (2,1.5) -- (2,13.5);				
\draw[-triangle 60,thick] (1.5,2) -- (18.5,2);
\draw (1.25,2) node {$0$};
\draw (2,1.25) node {$a$};
\draw (18.75,1.25) node {$t$};
\draw (1.25,13.6) node {$r$};

\draw (4.5,4.5) circle (2);						
\draw (9,4.5) circle (1.2);
\draw (13,4.5) circle (0.8);
\draw (16,4.5) circle (0.5);
\draw (4.5,8.5) circle (1);
\draw (4.5,11) circle (0.5);
\draw (10.5,6) node {$\Omega_{t,ir}$};

\draw[domain=3:17,smooth,variable=\t]		 			
	plot ({\t},{36/\t+0.5});
\draw (7.5,13.2) node {border of the method: $4r^2(m_1m_2)(t)=\eta^2$};

\fill[pattern=dots,opacity=0.4,domain=3:17,smooth,variable=\t] 		
	(2,2) -- (2,12.5) -- plot ({\t},{36/\t+0.5}) -- (17,2)-- (2,2);

\draw[double,-stealth] (7,10.5) -- (7,11.5);				
\draw (7,12) node {\emph{shrink}};
\draw[double,-stealth] (15,6) -- (16,6);
\draw (17,6.07) node {\emph{shrink}};
\end{tikzpicture}
%
}
\end{center}
\vspace*{-7mm}
\caption{admissible region for coefficient estimates}
\label{Y37}
\end{figure}
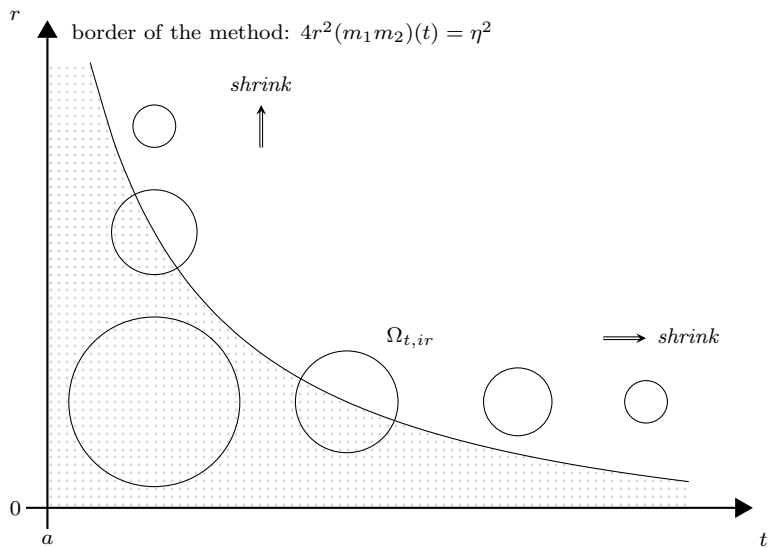
\noindent
Since $A(r)$ always correctly describes the absolute value of $q_H(ir)$,
and\:---\:as we shall see later\:---\:the lower bound $L(r)$ is in many situations correct,
it seems that this border occurs not ``just because of the method'',
but for intrinsic reasons.

\subsection*{Further perspective}

The present work is part of a series of papers where we investigate the
high-energy behaviour of the Weyl coefficient.
The other parts which are already\:---\:or will very soon be\:---\:available, are
\cite{pruckner.woracek:limp-arXiv,langer.pruckner.woracek:gapsatz-arXiv,langer.pruckner.woracek:asysupp-arXiv}.
In \cite{pruckner.woracek:limp-arXiv} limit points of $q_H(ir)$ for $r\to\infty$ are investigated.
The contents of the other two papers is discussed below.

Let us first make three observations about \thref{Y98}.
\begin{Steps}
\item
	The absolute value of $q_H$ is fully determined by \eqref{Y35} up to some
	universal constants, while \eqref{Y96} gives only estimates for the imaginary part
	of the Weyl coefficient:
	\[
		L(r) \lesssim \Im q_H(ir) \le |q_H(ir)|\asymp A(r).
	\]
\item
	The lower bound $L(r)$ becomes smaller when the relative size of the
	off-diagonal entries of $H$ becomes larger.
\item
	The bounds given in \thref{Y98} involve universal constants which depend
	on the parameter $\eta$.  Consider, for example, the constant on the right-hand side
	of \eqref{Y35}.  Its value can be minimised by making an appropriate choice of $\eta$,
	but this minimum is larger than $1$.
\end{Steps}

\noindent
Each of these observations gives rise to a natural question.
\\[1ex]
\emph{Question concerning observation} \ding{172}. \\
If $L(r)\asymp A(r)$, then clearly $\Im q_H(ir)\asymp|q_H(ir)|$.
What happens if $L(r)\ll A(r)$?

Investigating the relation between the gaps $\frac{L(r)}{A(r)}$
and $\frac{\Im q_H(ir)}{|q_H(ir)|}$ more closely, requires very different methods
from the presently developed ones, namely, a refined variant of Kasahara's rescaling trick.
We settle the above question on a qualitative level in \cite{langer.pruckner.woracek:gapsatz-arXiv},
where we prove that the lower bound $L(r)$ is sharp in the sense that $L(r)\ll A(r)$
if and only if $\Im q_H(ir)\ll|q_H(ir)|$.
Moreover, we characterise this situation explicitly in terms of $H$ and see that
for ``most'' Hamiltonians $L(r)\asymp A(r)$ holds.
This confirms the intuition familiar from complex analysis that ``usually'' $\Im q_H(ir)\asymp|q_H(ir)|$.
\\[1ex]
\emph{Question concerning observation} \ding{173}. \\
Does $\Im q_H(ir)$ show the same behaviour as $L(r)$, namely that it becomes smaller
when the relative size of the off-diagonal entries of $H$ becomes larger?

There are indications that this is the case; see, for example, the main theorem in the
forthcoming \cite{langer.pruckner.woracek:asysupp-arXiv}.
However, we have no general result confirming this, and it seems that the question
is very much related to what is stated as an open problem below.
\\[1ex]
\emph{Question concerning observation} \ding{174}. \\
When does $q_H$ have an asymptotic expansion at $i\infty$?

With the present \thref{Y98} we cannot possibly exclude that $|q_H(ir)|$ oscillates
between the bounds given by \eqref{Y35}; cf.\ \thref{Y60}.
Dealing with asymptotic expansions requires different methods.
We answer this question in \cite{langer.pruckner.woracek:asysupp-arXiv}, where we characterise
the presence of a regularly varying asymptotic
explicitly in terms of $H$. The idea will again be to apply a variant of Kasahara's trick already mentioned above.
\\[1ex]
\emph{An open problem.} Is it possible to quantitatively compare the
gaps $\frac{L(r)}{A(r)}$ and $\frac{\Im q_H(ir)}{|q_H(ir)|}$?
In particular, we do not know whether there exist a Hamiltonian
with $L(r)\asymp\Im q_H(ir)\ll|q_H(ir)|$.
It seems that none of the presently available methods is suitable to attack this problem.

\medskip

Let us give an overview of the contents of the paper.
In Section~\ref{Y40} we present some preliminary material that is used in the
proof of the main theorem, in particular a discussion about Weyl discs,
a power series expansion of the fundamental solution of \eqref{Y99}
and some matrix algebra that is useful for estimates of the coefficients
of the power series.
Section~\ref{Y41} contains the proofs of \thref{Y98} and \thref{Y09}.
In Section~\ref{Y43} our estimates for the Weyl coefficient are related
to estimates of the spectral measure.  In particular, we prove
characterisations for $\mu_H$ belonging to some Kac classes
and for the distribution function of $\mu_H$ satisfying certain growth estimates.
Section~\ref{Y34} contains further results, which complement \thref{Y98}.
In particular, we prove a monotonicity property; we discuss the situation
when $h_1$ or $h_2$ vanishes on neighbourhood of $a$;
we study a rotation transformation that improves the bounds in \eqref{Y35}--\eqref{Y96}
in certain situations; we discuss some examples; and we express $A(r)$ in terms of
the mass function of a related Krein string.
Finally, in Section~\ref{Y81} we apply our theorems to Krein strings
and Sturm--Liouville equations, and relate our results to previous work
in the literature.
Two appendices collect some facts about regularly varying functions and
generalised inverses.

\medskip

\noindent
\textbf{Acknowledgements.}
The second and third authors were supported by the project P~30715-N35 of the
Austrian Science Fund (FWF). The third author was supported by the joint project I~4600 of the Austrian
Science Fund (FWF) and the Russian foundation of basic research (RFBR).

\section{Weyl discs and matrix algebra}
\label{Y40}

In this section we collect some preliminary information.

\subsection{Centre and radius of a Weyl disc}
\label{Y160}

We recall the definition of the Weyl discs $\Omega_{t,z}$ and a
basic formula for their centres and radii.
Let $W(t,z)$ be the (transpose of) the fundamental solution of the system \eqref{Y99},
i.e.\ the unique $2\times2$-matrix-valued solution of the initial value problem
\[
	\left\{
	\begin{array}{l}
		\dfrac{\partial}{\partial t}W(t,z)J = zW(t,z)H(t),\qquad t\in[a,b),
		\\[2.5ex]
		W(a,z)=I.
	\end{array}
	\right.
\]
Equivalently, $W(t,z)$ is the solution of the integral equation
\begin{equation}\label{Y95}
	W(t,z)J-J = z\int_a^t W(s,z)H(s)\DD s,\qquad t\in[a,b).
\end{equation}
Note that the transposes of the rows of $W$ are solutions of \eqref{Y99}.
Writing $W(t,z)=\smmatrix{w_{11}(t,z)}{w_{12}(t,z)}{w_{21}(t,z)}{w_{22}(t,z)}$,
the Weyl disc $\Omega_{t,z}$ is defined as the image of the closed upper half-plane
under the fractional linear transformation
\[
	\zeta \mapsto \frac{w_{11}(t,z)\zeta+w_{12}(t,z)}{w_{21}(t,z)\zeta+w_{22}(t,z)}\,.
\]
The Weyl discs are nested: $\Omega_{t_1,z}\supseteq\Omega_{t_2,z}$
when $a\le t_1\le t_2<b$.  Since we assume that $H$ is in the limit point case at $b$,
i.e.\ \eqref{Y138} holds, the Weyl discs $\Omega_{t,z}$ shrink to a point
as $t\to b$ for $z\in\bb C\setminus\bb R$.  This point is denoted by $q_H(z)$, i.e.\
\begin{equation}\label{Y139}
	q_H(z) \DE \lim_{t\to b}\frac{w_{11}(t,z)\zeta+w_{12}(t,z)}{w_{21}(t,z)\zeta+w_{22}(t,z)}\,,
	\qquad z\in\bb C\setminus\bb R,
\end{equation}
independent of $\zeta$, and $q_H$ is called the \emph{Weyl coefficient}
associated with the Hamiltonian $H$.
The centre and the radius of $\Omega_{t,z}$ can be expressed in a neat way using the function
\begin{equation}\label{Y14}
	\nabla(t,z) \DE \int_a^t W(s,z)H(s)W(s,z)^*\DD s
	= \frac{W(t,z)JW(t,z)^*-J}{2i(\Im z)}\,;
\end{equation}
the second equality in \eqref{Y14} follows from,
e.g.\ \cite[(2.5)]{hassi.snoo.winkler:2000}.

\begin{lemma}\thlab{Y51}
	Let $z\in\bb C^+$ and let $t\in (a,b)$ be such that $h_2|_{(a,t)}\ne0$.
	Then the Weyl disc $\Omega_{t,z}$ is the closed disc with
	\begin{equation}\label{Y23}
		\text{centre:} \quad \frac{b(t,z)}{a(t,z)}+i\frac 1{2(\Im z)a(t,z)}
		\qquad\Big\vert\qquad
		\text{radius:} \quad \frac 1{2(\Im z)a(t,z)}
	\end{equation}
	where $a(t,z)$ and $b(t,z)$ respectively denote the $(2,2)$ and the $(1,2)$ entries of $\nabla(t,z)$.
\end{lemma}

\noindent
This is folklore and can be found, for instance, implicitly in \cite[\S3]{winkler:2000a}.
For the convenience of the reader, we recall the argument.

\begin{proof}[Proof of Lemma~\ref{Y51}]
	Let $\smmatrix{a_{11}}{a_{12}}{a_{21}}{a_{22}}\in\bb C^{2\times 2}$
	be an invertible matrix such that $\Im(a_{22}\qu{a_{21}})>0$,
	and consider the fractional linear transformation
	$w\mapsto\frac{a_{11}w+a_{12}}{a_{21}w+a_{22}}$\,.
	The image of the closed upper half-plane under this map is the disc with
	\begin{equation}\label{Y39}
		\text{centre:} \quad \frac{a_{12}\qu{a_{21}}-a_{11}\qu{a_{22}}\,}{a_{22}\qu{a_{21}}-a_{21}\qu{a_{22}}\,}
		\qquad\Big\vert\qquad
		\text{radius:} \quad \bigg|\frac{a_{11}a_{22}-a_{12}a_{21}\,}{a_{22}\qu{a_{21}}-a_{21}\qu{a_{22}}\,}\bigg|\,.
	\end{equation}
	Note that the condition $\Im(a_{22}\qu{a_{21}})>0$ is equivalent to the fact
	that the image of the closed upper half-plane is a disc.

	For each $t\in(a,b)$ and $z\in\bb C^+$, the Weyl disc $\Omega_{t,z}$ is the image
	of the closed upper half-plane (considered on the Riemann sphere)
	under the fractional linear transformation
	$\zeta\mapsto\frac{w_{11}(t,z)\zeta+w_{12}(t,z)}{w_{21}(t,z)\zeta+w_{22}(t,z)}$.
	From \eqref{Y14} we obtain
	\begin{flalign}
		& w_{22}(t,z)\qu{w_{21}(t,z)}-w_{21}(t,z)\qu{w_{22}(t,z)}
		= \binom 01^*\Bigl[W(t,z)JW(t,z)^*-J\Bigr]\binom 01
		\nonumber\\
		& = 2i(\Im z)\binom 01^*\nabla(t,z)\binom 01
		= 2i(\Im z)a(t,z)
		\label{Y32}
		\\
		\text{and} \hspace*{1ex} && \nonumber\\
		& w_{12}(t,z)\qu{w_{21}(t,z)}-w_{11}(t,z)\qu{w_{22}(t,z)}+1
		= \binom 10^*\Bigl[W(t,z)JW(t,z)^*-J\Bigr]\binom 01
		\nonumber\\
		& = 2i(\Im z)\binom 10^*\nabla(t,z)\binom 01
		= 2i(\Im z)b(t,z).
		\label{Y33}
	\end{flalign}
	With $v(t,z)=\binom{w_{21}(t,z)}{w_{22}(t,z)}$, which satisfies \eqref{Y99},
	we have
	\[
		a(t,z) = \int_a^t v(s,z)^*H(s)v(s,z)\DD s \ge 0.
	\]
	Assume that $a(t,z)=0$.  Then $H(s)v(s,z)=0$ a.e.\ and hence $v$ is constant.
	by \eqref{Y99}.  The initial condition $W(a,z)=I$ yields $v(t,z)\equiv\binom01$.
	This, in turn, implies that $h_2(t)=0$ a.e.\ on $(a,t)$, which is a contradiction
	to the assumption.
	Therefore $\Im(w_{22}(t,z)\qu{w_{21}(t,z)})=(\Im z)a(t,z)>0$ for $z\in\bb C^+$.
	Hence we can apply \eqref{Y39}, which, together with \eqref{Y32} and \eqref{Y33},
	yields that the centre of the disc is
	\[
		\frac{w_{12}(t,z)\qu{w_{21}(t,z)}-w_{11}(t,z)\qu{w_{22}(t,z)}\,}{
			w_{22}(t,z)\qu{w_{21}(t,z)}-w_{21}(t,z)\qu{w_{22}(t,z)}\,}
		= \frac{2i(\Im z)b(t,z)-1}{2i(\Im z)a(t,z)}
		= \frac{b(t,z)}{a(t,z)} + i\frac{1}{2(\Im z)a(t,z)}\,,
	\]
	and its radius is
	\[
		\frac{1}{\big|w_{22}(t,z)\qu{w_{21}(t,z)}-w_{21}(t,z)\qu{w_{22}(t,z)}\big|}
		= \frac{1}{2(\Im z)a(t,z)}\,;
	\]
	this proves \eqref{Y23}.
\end{proof}

\subsection{The power series expansion of $\nabla\bm{(t,z)}$}

For $t\in[a,b)$ let $W_n(t)$ be the coefficients in the
power series expansion of $W(t,z)$, i.e.\
\begin{equation}\label{Y10}
	W(t,z)=\sum_{n=0}^\infty W_n(t)z^n.
\end{equation}
This series converges uniformly on every compact subset of $[a,b)\times\bb C$.
The integral equation \eqref{Y95} for $W(t,z)$ shows that the coefficient
sequence $(W_n(t))_{n=0}^\infty$ is given by the recurrence relation
\begin{equation}\label{Y03}
	\left\{
	\begin{array}{l}
		W_0(t)=I,
		\\[1ex]
		\displaystyle W_{n+1}(t)=\int_a^tW_n(s)H(s)\DD s\cdot(-J),\quad n\ge 0.
	\end{array}
	\right.
\end{equation}
The latter implies that
\begin{alignat}{2}
	& W_n(a)=0, \qquad && n\ge 1;
	\label{Y12}\\[0.5ex]
	& W_n(t)\in\bb R^{2\times 2}, \qquad && n\ge 0,\; t\in[a,b).
	\nonumber
\end{alignat}
Plugging the power series expansion of $W(t,z)$ into the
definition of $\nabla(t,z)$ in \eqref{Y14} we obtain
\[
	\nabla(t,z) = \sum_{n,m=0}^\infty\biggl(\int_a^t W_n(s)H(s)W_m(s)^*\DD s\biggr)z^n\qu z^m.
\]
Setting
\[
	\alpha_{n,m}(t)\DE\int_a^t W_n(s)H(s)W_m(s)^*\DD s
\]
we have, for $z=re^{i\vartheta}$,
\begin{equation}\label{Y94}
	\nabla\big(t,re^{i\vartheta}\big)
	= \sum_{l=0}^\infty\biggl(\,\sum_{n=0}^l\alpha_{n,l-n}(t)e^{i\vartheta(2n-l)}\biggr)r^l.
\end{equation}
Note that the coefficients $\alpha_{n,m}$ satisfy the symmetry property
$\alpha_{m,n}=\alpha_{n,m}^*$.

Let us list a couple of properties of the function $M$ defined in \eqref{Y08}.
\begin{Itemize}
\item
	The recurrence relation \eqref{Y03} implies that $W_1J=M$, which, together with
	the symmetry of $M$ shows that
	\begin{equation}\label{Y11}
		-JW_1^* = W_1J.
	\end{equation}
\item
	The symmetry of $M$ also implies that $MJM=(\det M)J$, and consequently,
	\begin{equation}\label{Y38}
		W_1W_1=-(\det M)I, \qquad W_1JW_1^*=(\det M)J.
	\end{equation}
\item
	Since $M$ is the primitive of a pointwise positive semi-definite
	matrix function, $M$ is itself positive semi-definite
	and $M(t)\le M(s)$ whenever $t\le s$.
\end{Itemize}

\subsection{Some matrix algebra}

We frequently use certain algebraic manipulations with matrices.
In order to present arguments in a clean way,
it is practical to introduce the following operations and relations for matrices.

\begin{definition}\thlab{Y85}
	For $U=(u_{ij})_{i,j=1}^2\in\bb C^{2\times 2}$ set
	\[
		|U| \DE \bigl(|u_{ij}|\bigr)_{i,j=1}^2.
	\]
	For $U=(u_{ij})_{i,j=1}^2,\tilde U=(\tilde u_{ij})_{i,j=1}^2\in\bb R^{2\times 2}$ define
	\[
		U \preceq \tilde U \;\;\DI\;\; u_{ij}\le\tilde u_{ij}, \quad i,j\in\{1,2\}.
	\]
\end{definition}

\noindent
In the next lemma we list a couple of properties and rules for these operations and relations
(their simple proofs are omitted).

\begin{lemma}\thlab{Y84}
	\hfill
	\begin{Enumerate}
	\item ${\displaystyle
		\forall U,\tilde U\in\bb C^{2\times 2}\DP
		|U+\tilde U|\preceq|U|+|\tilde U| \;\;\wedge\;\; |U\tilde U|\preceq|U|\,|\tilde U|
		}$;
	\item ${\displaystyle
		\forall U\in\bb C^{2\times 2},z\in\bb C\DP
		|zU|=|z|\,|U|
		}$;
	\item ${\displaystyle
		\forall U,\tilde U\in\bb R^{2\times 2}\DP
		}$
		\\[0.5ex]
		\hspace*{5ex}
		${\displaystyle
		U\preceq\tilde U\Leftrightarrow
		\Bigl(\forall V\in\bb R^{2\times 2}\DP U+V\preceq\tilde U+V\Bigr)
		\Leftrightarrow \Bigl(\forall x>0\DP xU\preceq x\tilde U\Bigr)
		}$;
	\item ${\displaystyle
		\forall U,\tilde U,V,\tilde V\in\bb R^{2\times 2}\DP
		}$
		\\[1ex]
		\hspace*{5ex}
		${\displaystyle
		|U|\preceq|\tilde U|\;\wedge\;|V|\preceq|\tilde V|
		\;\;\Rightarrow\;\;
		|U+V|\preceq|\tilde U|+|\tilde V|\;\wedge\;|U\cdot V|\preceq|\tilde U|\cdot|\tilde V|
		}$;
	\item ${\displaystyle
		\forall U\in L^1\big((c,d),\bb C^{2\times 2}\big)\DP
		\bigg|\int_c^d U(s)\DD s\bigg|\preceq\int_c^d|U(s)|\DD s
		}$;
	\item ${\displaystyle
		\forall U,\tilde U\in L^1\big((c,d),\bb R^{2\times 2}\big)\DP
		}$
		\\[0.5ex]
		\hspace*{5ex}
		${\displaystyle
		U(s)\preceq\tilde U(s),\ s\in(c,d)\text{ a.e.}\;\;\Rightarrow\;\;\int_c^d U(s)\DD s\preceq\int_c^d\tilde U(s)\DD s
		}$.
	\end{Enumerate}
\end{lemma}

\subsection{Alternative forms for the bounds $\bm{A(r)}$ and $\bm{L(r)}$}
\label{Y192}

In this short subsection we consider some reformulations of $\mr t$, $A(r)$ and $L(r)$,
which are used in the proofs of \thref{Y98} and other results in later sections.
Let us set
\begin{equation}\label{Y124}
	\mr r(t)\DE\frac{\eta}{2}\bigl(m_1(t)m_2(t)\bigr)^{-\frac 12},
	\qquad t\in(\mr a,b),
\end{equation}
which is the solution for $r$ of the equation \eqref{Y30}.
It follows that $\mr t$ and $\mr r$ are inverses of each other.
In particular, $r = \frac{\eta}{2}(m_1m_2)^{-\frac 12}(\mr t(r))$, which implies the
representations
\begin{align}
	\label{Y36}
	A(r) &= \frac{2r}{\eta}\cdot m_1(\mr t(r)) = \frac{\eta}{2r}\cdot\frac{1}{m_2(\mr t(r))},
	\\[1ex]
	\label{Y120}
	L(r) &= \frac{2}{\eta}\cdot\frac{r\det M(\mr t(r))}{m_2(\mr t(r))}
	= \frac{4}{\eta^2}\cdot r^2\det M(\mr t(r))\cdot A(r).
\end{align}
Let us also note that the function $\frac{\det M}{m_2}$ is non-decreasing since
\begin{align}
	\frac{\RD}{\RD t}\biggl(\frac{\det M(t)}{m_2(t)}\biggr)
	&= \frac{1}{m_2(t)^2}\Bigl[m_2(t)^2h_1(t)+m_3(t)^2h_2(t)-2m_2(t)m_3(t)h_3(t)\Bigr]
	\nonumber
	\\
	&= \frac{1}{m_2(t)^2}\binom{m_2(t)}{-m_3(t)}^*
	\begin{pmatrix} h_1(t) & h_3(t) \\[0.7ex] h_3(t) & h_2(t) \end{pmatrix}
	\binom{m_2(t)}{-m_3(t)}
	\ge 0.
	\label{Y71}
\end{align}
This is also used in later sections.

\section{Proof of Weyl coefficient estimates}
\label{Y41}

In this section we prove \thref{Y98} and \thref{Y09}.
This is done by establishing bounds for the coefficients $\alpha_{n,m}(t)$
of the expansion \eqref{Y94} and some related quantities in terms of
\[
	M^+(t)
	\DE
	\begin{pmatrix}
		m_1(t) & \sqrt{m_1(t)m_2(t)}
		\\[1ex]
		\sqrt{m_1(t)m_2(t)} & m_2(t)
	\end{pmatrix}
	,
\]
and
\[
	m^+(t) \DE 2\cdot\sqrt{m_1(t)m_2(t)}.
\]
Observe the following properties of $M^+(t)$ and $m^+(t)$.

\begin{lemma}\thlab{Y01}
	With the notation introduced in \thref{Y85} we have
	\begin{Enumerate}
	\item
		${\displaystyle
		|M(t)|\preceq\int_a^t |H(s)|\DD s\preceq M^+(t);
		}$
	\item
		${\displaystyle
		\bigl(M^+(t)|J|\bigr)^n M^+(t) = m^+(t)^n M^+(t), \quad n\ge 0;
		}$
	\item
		${\displaystyle
		\forall s \le t\DP M^+(s) \preceq M^+(t) \;\wedge\; m^+(s)\le m^+(t).
		}$
	\end{Enumerate}
\end{lemma}

\begin{proof}
	The first inequality in \Enumref{1} follows from \thref{Y84}\,(v).
	Consider the second inequality.
	The diagonal entries of the left-hand and right-hand sides actually coincide.
	Since $H(t)\ge 0$, also $|H(t)|\ge 0$, and hence $\int_a^t |H(s)|\DD s\ge 0$. Thus
	$m_1(t)m_2(t)-\bigl(\int_a^t|h_3(s)|\DD s\bigr)^2\ge 0$.

	Item \Enumref{2} follows from a general observation.
	Consider a matrix $U=(u_{ij})_{i,j=1}^2\in\bb R^{2\times 2}$ that is
	of the form $U=\alpha\xi_\phi\xi_\phi^*$ with some $\alpha,\phi\in\bb R$
	where $\xi_\phi\DE(\cos\phi,\sin\phi)^*$.  Then
	\begin{align*}
		(U|J|)U &= \alpha\xi_\phi\xi_\phi^*|J|\cdot\alpha\xi_\phi\xi_\phi^*
		= \alpha\xi_\phi\cdot\alpha\xi_\phi^*|J|\xi_\phi\cdot\xi_\phi^*
		\\[0.5ex]
		&= \alpha\xi_\phi\cdot\alpha 2\cos\phi\sin\phi\cdot\xi_\phi^*
		= 2u_{12}\cdot\alpha\xi_\phi\xi_\phi^*
		= 2u_{12} U.
	\end{align*}
	Thus, inductively,
	\[
		(U|J|)^nU=(2u_{12})^nU,\quad n \ge 0.
	\]
	The matrix $M^+(t)$ is symmetric with determinant zero, and hence of the considered form.

	Item \Enumref{3} is clear since $m_1$ and $m_2$ are non-decreasing.
\end{proof}

\subsection{Bounds involving $\bm{A(r)}$}

In order to obtain the bounds \eqref{Y35}, \eqref{Y92}, and the estimate from above
in \eqref{Y96}, it is enough to have the following crude estimate.

\begin{lemma}\thlab{Y15}
	We have
	\begin{equation}\label{Y06}
		|W_n(t)| \preceq m^+(t)^{n-1}M^+(t)|J|,\quad n \ge 1,
	\end{equation}
	and
	\[
		|\alpha_{n,m}(t)|\preceq m^+(t)^{n+m}M^+(t),\quad n,m \ge 0.
	\]
\end{lemma}

\medskip

\noindent
This statement has been extracted from the personal communication \cite{winkler:2000b} by H.~Winkler%
\footnote{%
	The proofs of the main results of \cite{winkler:2000a} use the
	estimate \cite[Lemma~3.2]{winkler:2000a} of power series coefficients
	whose proof contains a mistake (and whose assertion is most probably wrong).
	However, the theorems stated in the paper are still correct.
	They can be proved using a weaker form of \cite[Lemma~3.2]{winkler:2000a},
	which does hold.  This is shown in \cite{winkler:2000b},
	and \thref{Y15} is nothing but a clean formulation of the relevant argument.
}.
Since this communication remained unpublished, we provide a complete proof.

\begin{proof}
	To see \eqref{Y06}, we use induction on $n$. First, observe that
	\[
		|W_1(t)| = \Big|\int_a^t H(s)\DD s\cdot(-J)\Big|
		\preceq \int_a^t|H(s)|\DD s\cdot|J| \preceq M^+(t)|J|
	\]
	by Lemma~\ref{Y84}(i), (iv) and Lemma~\ref{Y01}(i).
	Second, with Lemma~\ref{Y01}(ii) we obtain
	\begin{align*}
		|W_{n+1}(t)| &= \Big|\int_a^t W_n(s)H(s)\DD s\cdot(-J)\Big|
		\preceq \int_a^t|W_n(s)||H(s)|\DD s\cdot|J|
		\\[0.5ex]
		&\preceq \int_a^t m^+(s)^{n-1}M^+(s)|J||H(s)|\DD s\cdot|J|
		\preceq m^+(t)^{n-1}M^+(t)|J|\int_a^t |H(s)|\DD s\cdot|J|
		\\[0.5ex]
		&\preceq m^+(t)^{n-1}\cdot M^+(t)|J|M^+(t)\cdot|J|
		= m^+(t)^nM^+(t)|J|,
	\end{align*}
	which proves \eqref{Y06}.

	Now we use \eqref{Y06} to estimate $\alpha_{n,m}(t)$.
	For $n,m\ge 1$ we have
	\begin{align*}
		|\alpha_{n,m}(t)| &\preceq \int_a^t|W_n(s)|\cdot|H(s)|\cdot|W_m(s)|^*\DD s
		\\
		&\preceq \int_a^t m^+(s)^{n+m-2}M^+(s)|J|\cdot|H(s)|\cdot|J|M^+(s)\DD s
		\\
		&\preceq m^+(t)^{n+m-2}M^+(t)|J|\cdot\int_a^t |H(s)|\DD s\cdot|J|M^+(t)
		\\[0.5ex]
		&\preceq m^+(t)^{n+m-2}M^+(t)|J|\cdot M^+(t)\cdot|J|M^+(t)
		\preceq m^+(t)^{n+m}M^+(t).
	\end{align*}
	For $n\ge 1$
	\begin{align*}
		|\alpha_{n,0}(t)| &\preceq \int_a^t|W_n(s)|\cdot|H(s)|\DD s
		\preceq \int_a^t m^+(s)^{n-1}M^+(s)|J|\cdot|H(s)|\DD s
		\\
		&\preceq m^+(t)^{n-1}M^+(t)|J|\cdot\int_a^t |H(s)|\DD s
		\preceq m^+(t)^nM^+(t),
	\end{align*}
	and by symmetry also $|\alpha_{0,n}|\preceq m^+(t)^nM^+(t)$.
	Finally, we have $|\alpha_{0,0}(t)|=|M(t)|\preceq M^+(t)$.
\end{proof}

\begin{proof}[Proof of Theorem~\textup{\ref{Y98}}\, (upper bounds)]
	Let $H,\eta,\sigma,\mr t(r)$ and $\vartheta\in(0,\pi)$ be as in the formulation
	of the theorem.  Note that \eqref{Y30} is equivalent to $rm^+(\mr t(r))= \eta$.
	Using \eqref{Y94} and Lemma~\ref{Y15} we obtain
	\begin{align*}
		&\big|\nabla\bigl(\mr t(r),re^{i\vartheta}\bigr)-M(\mr t(r))\big|
		= \bigg|\sum_{l=1}^\infty\biggl(\sum_{n=0}^l
		\alpha_{n,l-n}(\mr t(r))e^{i\vartheta(2n-l)}\biggr)r^l\bigg|
		\preceq \sum_{l=1}^\infty\biggl(\sum_{n=0}^l|\alpha_{n,l-n}(\mr t(r))|\biggr)r^l
		\\
		&\preceq \biggl(\sum_{l=1}^\infty(l+1)m^+(\mr t(r))^lr^l\biggr)M^+(\mr t(r))
		= \biggl(\sum_{l=1}^\infty(l+1)\eta^l\biggr)M^+(\mr t(r)).
	\end{align*}
	We have
	\begin{equation}\label{Y17}
		\sum_{l=1}^\infty(l+1)\eta^l
		= \sum_{k=1}^\infty k\eta^{k-1}-1
		= \frac{1}{(1-\eta)^2}-1 = \sigma,
	\end{equation}
	and this implies that
	\begin{align*}
		\big|a(\mr t(r),re^{i\vartheta})-m_2(\mr t(r))\big| &\le \sigma m_2(\mr t(r)), \\[1ex]
		\big|b(\mr t(r),re^{i\vartheta})-m_3(\mr t(r))\big| &\le \sigma\sqrt{(m_1m_2)(\mr t(r))}
		= \frac{\sigma\eta}{2r}.
	\end{align*}
	Remembering that $m_3(t)^2\le(m_1m_2)(t)$, we therefore obtain
	\begin{align}
		& (1-\sigma)m_2(\mr t(r)) \le a\bigl(\mr t(r),re^{i\vartheta}\bigr)
		\le(1+\sigma)m_2(\mr t(r)),
		\label{Y72}
		\\[1ex]
		& \big|b\bigl(\mr t(r),re^{i\vartheta}\bigr)\big| \le \frac{(1+\sigma)\eta}{2r},
		\qquad
		\Im b\bigl(\mr t(r),re^{i\vartheta}\bigr) \le \frac{\sigma\eta}{2r}.
		\nonumber
	\end{align}
	Note that the assumption $\eta\in(0,1-\frac{1}{\sqrt{2}\,})$
	implies that $\sigma\in(0,1)$.
	Now it follows from \thref{Y51} that
	(recall from \eqref{Y36} that $A(r)=\frac{\eta}{2} \frac{1}{rm_2(\mr t(r)}$)
	\begin{align*}
		\big|q_H(re^{i\vartheta})\big|
		&\le \frac{|b(\mr t(r),re^{i\vartheta})|}{a(\mr t(r),re^{i\vartheta})}
		+ 2\cdot\frac 1{2(r\sin\vartheta)a(\mr t(r),re^{i\vartheta})}
		\\[0.5ex]
		&\le \frac{ \frac{(1+\sigma)\eta}{2r}}{(1-\sigma)m_2(\mr t(r))}
		+ \frac 1{(r\sin\vartheta)(1-\sigma)m_2(\mr t(r))}
		\\[0.5ex]
		&= \frac{(1+\sigma)\eta+\frac{2}{\sin\vartheta}}{2(1-\sigma)}\cdot\frac{1}{rm_2(\mr t(r))}
		= \frac{1+\sigma+\frac{2}{\eta\cdot \sin\vartheta}}{1-\sigma}\cdot A(r),
		\displaybreak[0]\\[2ex]
		\big|\Re \eta_H(re^{i\vartheta})\big|
		&\le \frac{\Re b(\mr t(r),re^{i\vartheta})}{a(\mr t(r),re^{i\vartheta})}
		+ \frac 1{2(r\sin\vartheta)a(\mr t(r),re^{i\vartheta})}
		\\[0.5ex]
		&\le \frac{ \frac{(1+\sigma)\eta}{2r}}{(1-\sigma)m_2(\mr t(r))}
		+ \frac 1{2(r\sin\vartheta)(1-\sigma)m_2(\mr t(r))}
		\\[0.5ex]
		&= \frac{(1+\sigma)\eta+\frac{1}{\sin\vartheta}}{2(1-\sigma)}\cdot\frac{1}{rm_2(\mr t(r))}
		= \frac{1+\sigma+\frac{1}{\eta\cdot \sin\vartheta}}{1-\sigma}\cdot A(r),
		\displaybreak[0]\\[2ex]
		\Im q_H(re^{i\vartheta})
		&\le \frac{\Im b(\mr t(r),re^{i\vartheta})}{a(\mr t(r),re^{i\vartheta})}
		+ 2\cdot\frac 1{2(r\sin\vartheta)a(\mr t(r),re^{i\vartheta})}
		\\[0.5ex]
		&\le \frac{\frac{\sigma\eta}{2r}}{(1-\sigma)m_2(\mr t(r))}
		+ \frac 1{(r\sin\vartheta)(1-\sigma)m_2(\mr t(r))}
		\\[0.5ex]
		&= \frac{\sigma\eta+\frac{2}{\sin\vartheta}}{2(1-\sigma)}\cdot\frac{1}{rm_2(\mr t(r))}
		= \frac{\sigma+\frac{2}{\eta\cdot\sin\vartheta}}{1-\sigma}\cdot A(r).
		\qedhere
	\end{align*}
\end{proof}

\medskip

\begin{proof}[Proof of Theorem~\textup{\ref{Y98}}\, (lower bound for the absolute value)]
	Let the data be given as in the formulation of the theorem.
	We use the already established upper bound for the Hamiltonian
	\[
		\wt H\DE -JHJ = \begin{pmatrix} h_2 & -h_3 \\[1ex] -h_3 & h_1 \end{pmatrix}.
	\]
	With the obvious notation we have
	\[
		q_{\wt H} = -\frac{1}{q_H}, \qquad
		\wt m_1 = m_2, \quad \wt m_2 = m_1, \qquad
		\mr{\wt t}(r) = \mr t(r).
	\]
	Hence
	\[
		\wt A(r)
		= \sqrt{\frac{\wt m_1\bigl(\mr{\wt t}(r)\bigr)}{\wt m_2\bigl(\mr{\wt t}(r)\bigr)}\,}
		= \sqrt{\frac{m_2\bigl(\mr t(r)\bigr)}{ m_1\bigl(\mr t(r)\bigr)}\,}
		= \frac{1}{A(r)},
	\]
	and the upper bound from \eqref{Y35} gives
	\[
		\big|q_H\bigl(re^{i\vartheta}\bigr)\big|=\frac{1}{|q_{\wt H}(re^{i\vartheta})|}
		\ge \Big(\frac{1+\sigma+\frac{2}{\eta\sin\vartheta}}{1-\sigma}\Big)^{-1}A(r).
		\qedhere
	\]
\end{proof}

\subsection{The lower bound for $\bm{\Im q_H}$}
\label{Y42}

The bound for $\alpha_{n,m}(t)$ given in \thref{Y15} puts absolute values everywhere
and does not take care of possible cancellations.
Proving the lower bound of $\Im q_H$ asserted in \eqref{Y96} requires more
delicate coefficient estimates.

We start with some preliminary computations.

\begin{lemma}\thlab{Y16}
	We have
	\begin{equation}\label{Y05}
	\begin{aligned}
		\int_a^t W_n(s)H(s)W_m(s)^*\DD s
		&= W_n(t)W_1(t)JW_m(t)^*
		+ \int_a^t W_n(s)W_1(s)H(s)W_{m-1}(s)^*\DD s
		\\[1ex]
		&\quad +\int_a^t W_{n-1}(s)H(s)W_1(s)^*W_m(s)^*\DD s,
		\hspace*{10ex} n,m\ge 1,
	\end{aligned}
	\end{equation}
	and
	\begin{equation}\label{Y04}
	\begin{aligned}
		W_n(t) &= W_{n-1}(t)W_1(t)+\bigl(\det M(t)\bigr)W_{n-2}(t)
		+ \int_a^t \bigl(\det M(s)\bigr)W_{n-3}(s)H(s)J\DD s
		\\[1ex]
		&\quad -\int_a^t W_{n-2}(s)JW_1(s)^*\cdot JH(s)J\DD s,
		\qquad n\ge 3.
	\end{aligned}
	\end{equation}
\end{lemma}

\begin{proof}
	To show \eqref{Y05}, first note that, by \eqref{Y03}, we have
	\begin{align*}
		\frac{\RD}{\RD t}\bigl(W_n(t)JW_1(t)^*\bigr)
		&= W_{n-1}(t)H(t)(-J)JW_1(t)^*+W_n(t)J(-H(t)J)^*
		\\
		&= W_{n-1}(t)H(t)W_1(t)^*-W_n(t)H(t),
		\qquad n\ge 1.
	\end{align*}
	Using integration by parts, \eqref{Y12} and \eqref{Y11} we obtain
	\begin{align*}
		& \int_a^t W_n(s)H(s)W_m^*(s)\DD s
		\\
		&= \int_a^t\Bigl[W_{n-1}(s)H(s)W_1(s)^*
		-\frac{\RD}{\RD s}\bigl(W_n(s)JW_1^*(s)\bigr)\Bigr]W_m^*(s)\DD s
		\\
		&= \int_a^t W_{n-1}(s)H(s)W_1^*(s)W_m^*(s)\DD s-W_n(t)JW_1^*(t)W_m^*(t)
		\\
		&\quad + \int_a^t W_n(s)JW_1^*(s)\cdot\bigl(-W_{m-1}(s)H(s)J\bigr)^*\DD s
		\\
		&= \int_a^t W_{n-1}(s)H(s)W_1^*(s)W_m^*(s)\DD s+W_n(t)W_1(t)JW_m^*(t)
		\\
		&\quad + \int_a^t W_n(s)W_1(s)H(s)W_{m-1}(s)\DD s,
	\end{align*}
	which is \eqref{Y05}.
	To show \eqref{Y04}, we use again integration by parts, \eqref{Y05}
	with $(n,m)$ replaced by $(n-2,1)$, and also \eqref{Y38} and \eqref{Y11}:
	\begin{align*}
		W_n(t) &= \int_a^t W_{n-1}(s)\big(-H(s)J\big)\DD s
		\\[0.5ex]
		&= W_{n-1}(t)W_1(t) - \int_a^t \big(-W_{n-2}(s)H(s)J\big)W_1(s)\DD s
		\\[0.5ex]
		&= W_{n-1}(t)W_1(t) - \int_a^t W_{n-2}(s)H(s)W_1^*(s)J\DD s
		\displaybreak[0]\\[1ex]
		&= W_{n-1}(t)W_1(t) - W_{n-2}(t)W_1(t)JW_1^*(t)J
		\\
		&\quad - \int_a^t W_{n-2}(s)W_1(s)H(s)J\DD s
		- \int_a^t W_{n-3}(s)H(s)\big(W_1(s)W_1(s)\big)^*J\DD s
		\displaybreak[0]\\[0.5ex]
		&= W_{n-1}(t)W_1(t) - W_{n-2}(t)\cdot (\det M(t))J\cdot J
		\\
		&\quad - \int_a^t W_{n-2}(s)JW_1^*(s)\cdot JH(s)J\DD s
		- \int_a^t W_{n-3}(s)H(s)\big(-\det M(s)\big) I\cdot J\DD s.
		\qedhere
	\end{align*}
\end{proof}

\bigskip

\noindent
Set
\[
	\beta_{n,m}(t) \DE W_n(t)JW_m(t)^*, \qquad n,m\ge 0,
\]
and note that the symmetry relation $\beta_{m,n}(t)=-\beta_{n,m}(t)^*$ holds.

\begin{proposition}\thlab{Y02}
\hspace*{-1ex}\footnote{%
	Seeking a finer estimate of the form \eqref{Y07} was already
	suggested in \cite{winkler:2000b} as a potential way to progress.
}
	For $k,l\ge 0$, $(k,l)\ne(0,0)$ we have
	\begin{equation}\label{Y07}
		|\beta_{2k+1,2l+1}(t)| \preceq (\det M(t))\cdot\bigl(1+3(k+l)\bigr)
		\cdot m^+(t)^{2(k+l)-1}M^+(t).
	\end{equation}
\end{proposition}

\begin{proof}
We divide the proof into three steps.
\begin{Steps}
\item
	Since the matrix on the right-hand side of \eqref{Y07} is symmetric,
	the stated assertion is symmetric in $k$ and $l$.
	This means, it holds for a pair $(k,l)$ if and only if it holds for $(l,k)$.
\item
	We proceed by induction on $k+l$. First, assume that $k+l=1$,
	i.e.\ $(k,l)\in\{(1,0),(0,1)\}$.
	By symmetry, it is enough to consider the case $(k,l)=(1,0)$.
	Using \eqref{Y04}, \eqref{Y38}, \eqref{Y06} and Lemma~\ref{Y01} we compute
	\begin{align*}
		|\beta_{3,1}(t)| &= |W_3(t)\cdot JW_1(t)^*|
		\\
		&= \Big|\Big(W_2(t)W_1(t)+(\det M(t))W_1(t)
		+ \int_a^t (\det M(s))H(s)J\DD s
		\\
		&\quad - \int_a^t W_1(s)JW_1(s)^*\cdot JH(s)J\DD s\Big)JW_1(t)^*\Big|
		\\[1ex]
		&\preceq |W_2(t)|\cdot|\underbrace{W_1(t)JW_1(t)^*}_{=(\det M(t))J}|+(\det M(t))|W_1(t)|\cdot|J|\cdot|W_1(t)|^*
		\\
		&\quad + \int_a^t (\det M(s))|H(s)|\DD s\cdot|W_1(t)|^*
		\\
		&\quad + \int_a^t |\underbrace{W_1(s)JW_1(s)^*}_{=(\det M(s))J}|\cdot|J|\cdot|H(s)|\DD s\cdot|W_1(t)|^*
		\displaybreak[0]\\[1ex]
		&\preceq (\det M(t))|W_2(t)||J|+(\det M(t))|W_1(t)||J||W_1(t)|^*
		\\
		&\quad + (\det M(t))\cdot \int_a^t|H(s)|\DD s\cdot |W_1(t)|^*
		\\
		&\quad + (\det M(t))\cdot \int_a^t|H(s)|\DD s\cdot |W_1(t)|^*
		\displaybreak[0]\\[1ex]
		&\preceq (\det M(t))\Big(m^+(t)M^+(t)|J|\cdot|J|+M^+(t)|J|\cdot|J|\cdot|J|M^+(t)
		\\
		&\quad + 2M^+(t)\cdot|J|\cdot M^+(t)\Big)
		\\[1ex]
		&= 4(\det M(t)) m^+(t)M^+(t),
	\end{align*}
	which is \eqref{Y07} for $(k,l)=(1,0)$.
\item
	Let $(k,l)$ with $k+l\ge 2$ be given, and assume that \eqref{Y07}
	holds for all $(k',l')$ with $k'+l'<k+l$.
	By symmetry, we can assume, w.l.o.g., that $k\ge l$. Then, certainly, $k\ge 1$.
	It follows from \eqref{Y04} that
	\begin{align*}
		\beta_{2k+1,2l+1}(t) &= W_{2k+1}(t)JW_{2l+1}^*(t)
		\\[1ex]
		&= W_{2k}(t)\cdot W_1(t)JW_{2l+1}^*(t) + (\det M(t))W_{2k-1}(t)JW_{2l+1}^*(t)
		\\
		&\quad -\int_a^t (\det M(s))W_{2k-2}(s)H(s)\DD s \cdot W_{2l+1}^*(t)
		\\
		&\quad +\int_a^t W_{2k-1}(s)JW_1^*(s)\cdot JH(s)\DD s \cdot W_{2l+1}^*(t).
	\end{align*}
	Let us estimate each term on the right-hand side separately.
	Using \eqref{Y06}, the induction hypothesis, \eqref{Y38} and Lemma~\ref{Y01}(ii)
	we obtain
	\begin{align*}
		&\hspace*{-3ex} |W_{2k}(t)\cdot W_1(t)JW_{2l+1}(t)^*|
		\preceq |W_{2k}(t)|\cdot|W_1(t)JW_{2l+1}(t)^*|
		\\[1ex]
		&\preceq
		m^+(t)^{2k-1}M^+(t)|J|\cdot
		\begin{cases}
			(\det M(t))(1+3l)m^+(t)^{2l-1}M^+(t), & l>0
			\\[0.5ex]
			(\det M(t))|J|, & l=0
		\end{cases}
		\\[1ex]
		&=
		(\det M(t))(1+3l)\cdot
		\begin{cases}
			m^+(t)^{2(k+l)-2}M^+(t)|J|M^+(t), & l>0
			\\[0.5ex]
			m^+(t)^{2k-1}M^+(t)|J||J|, & l=0
		\end{cases}
		\\[1ex]
		&= (\det M(t))(1+3l)m^+(t)^{2(k+l)-1}M^+(t).
	\end{align*}
	Moreover, \eqref{Y06} and Lemma~\ref{Y01} imply that
	\begin{align*}
		&\hspace*{-3ex} \big|(\det M(t))W_{2k-1}(t)JW_{2l+1}(t)^*\big|
		\\[0.5ex]
		&\preceq (\det M(t))m^+(t)^{2k-2}M^+(t)|J|\cdot|J|\cdot m^+(t)^{2l}|J|M^+(t)
		\\[0.5ex]
		&= (\det M(t))m^+(t)^{2(k+l)-2}M^+(t)|J|M^+(t)
		\\[0.5ex]
		&= (\det M(t))m^+(t)^{2(k+l)-1}M^+(t)
	\end{align*}
	and
	\begin{align*}
		&\hspace*{-3ex} \bigg|\int_a^t (\det M(s))W_{2k-2}(s)H(s)\DD s\cdot W_{2l+1}(t)^*\bigg|
		\\[1ex]
		&\preceq \int_a^t (\det M(s))|W_{2k-2}(s)|\cdot |H(s)|\DD s\cdot |W_{2l+1}(t)|^*
		\\[1ex]
		&\preceq \int_a^t (\det M(s))\cdot
		\left\{\hspace*{-1ex}\begin{array}{ll}
			m^+(s)^{2k-3}M^+(s)|J|, & k>1
			\\[0.5ex]
			I, & k=1
		\end{array}\hspace*{-1ex}\right\}
		\cdot|H(s)|\DD s
		\\
		&\quad \cdot m^+(t)^{2l}|J|M^+(t)
		\displaybreak[0]\\[1ex]
		&\preceq (\det M(t))\cdot
		\left\{\hspace*{-1ex}\begin{array}{ll}
			m^+(t)^{2k-3}M^+(t)|J|, & k>1
			\\[0.5ex]
			I, & k=1
		\end{array}\hspace*{-1ex}\right\}
		\cdot \underbrace{\int_a^t |H(s)|\DD s}_{\preceq M^+(t)}
		\\
		&\quad \cdot m^+(t)^{2l}|J|M^+(t)
		\displaybreak[0]\\[1ex]
		&\preceq (\det M(t))\cdot
		\begin{cases}
			m^+(t)^{2(k+l)-3}(M^+(t)|J|)^2M^+(t), & k>1
			\\[0.5ex]
			m^+(t)^{2l}M^+(t)|J|M^+(t), & k=1
		\end{cases}
		\\[1ex]
		&= (\det M(t))m^+(t)^{2(k+l)-1}M^+(t).
	\end{align*}
	Finally, we use again the induction hypothesis to show
	\begin{align*}
		&\hspace*{-3ex} \bigg|\int_a^t W_{2k-1}(s)JW_1(s)^*
		\cdot JH(s)\DD s\cdot W_{2l+1}(t)^*\bigg|
		\\[1ex]
		&\preceq \int_a^t |W_{2k-1}(s)JW_1(s)^*|\cdot|J|\cdot|H(s)|\DD s\cdot |W_{2l+1}(t)|^*
		\displaybreak[0]\\[1ex]
		&\preceq \int_a^t
		\left\{\hspace*{-1ex}\begin{array}{ll}
			(\det M(s))(1+3(k-1))m^+(s)^{2(k-1)-1}M^+(s), & k>1
			\\[0.5ex]
			(\det M(s))|J|, & k=1
		\end{array}\hspace*{-1ex}\right\}
		\\[0.5ex]
		&\quad \cdot|J|\cdot|H(s)|\DD s\cdot m^+(t)^{2l}|J|M^+(t)
		\displaybreak[0]\\[1ex]
		&\preceq (\det M(t))\cdot
		\left\{\hspace*{-1ex}\begin{array}{ll}
			(1+3(k-1))m^+(t)^{2(k-1)-1}M^+(t), & k>1
			\\[0.5ex]
			|J|, & k=1
		\end{array}\hspace*{-1ex}\right\}
		\\[0.5ex]
		&\quad \cdot|J|\cdot \underbrace{\int_a^t |H(s)|\DD s}_{\preceq M^+(t)}
		\cdot m^+(t)^{2l}|J|M^+(t)
		\displaybreak[0]\\[1ex]
		&\preceq (\det M(t))(1+3(k-1))
		\\[0.5ex]
		&\quad \cdot
		\begin{cases}
			m^+(t)^{2(k-1+l)-1}(M^+(t)|J|)^2M^+(t), & k>1
			\\[0.5ex]
			m^+(t)^{2l}M^+(t)|J|M^+(t), & k=1
		\end{cases}
		\\[1ex]
		&= (\det M(t))(1+3(k-1))m^+(t)^{2(k+l)-1}M^+(t).
	\end{align*}
	Combining all estimates we obtain
	\begin{align*}
		&\hspace*{-3ex} |\beta_{2k+1,2l+1}(t)| = |W_{2k+1}(t)JW_{2l+1}(t)^*|
		\\[1ex]
		&\preceq (\det M(t))m^+(t)^{2(k+l)-1}M^+(t)\cdot
		\big[\underbrace{(1+3l)+1+1+(1+3(k-1))}_{=1+3(k+l)}\big],
	\end{align*}
	\end{Steps}
	which finishes the proof.
\end{proof}

\noindent
We also need the following formulae that relate $\alpha_{n,m}$ to $\beta_{k,l}$.

\begin{lemma}\thlab{Y90}
	We have
	\begin{alignat*}{2}
		& \alpha_{n,m+1}(t)-\alpha_{n+1,m}(t) = \beta_{n+1,m+1}(t), \qquad && n,m\ge 0,
		\\[1ex]
		& \alpha_{n,0}(t) = \beta_{n+1,0}(t),\quad \alpha_{0,n}(t) = -\beta_{0,n+1}(t),
		\qquad && n\ge 0.
	\end{alignat*}
\end{lemma}

\begin{proof}
	The first line follows since
	\[
		\frac{\RD}{\RD t}\bigl(W_{n+1}(t)JW_{m+1}(t)^*\bigr)
		= W_n(t)H(t)W_{m+1}(t)^*-W_{n+1}(t)H(t)W_m(t)^*
	\]
	and $W_n(a)=0$, $n\ge 1$.  The second line is just the recurrence relation \eqref{Y03}.
\end{proof}

\begin{proof}[Proof of Theorem~\textup{\ref{Y98}}\,(lower bound for the imaginary part)]
	Let the data be given as in the formulation of the theorem.
	We first estimate $\Im q_H(z)$ along the imaginary axis.
	Let $t\in [a,b)$ and $r>0$ and consider \eqref{Y94},
	which, for $\vartheta=\pi/2$, becomes
	\[
		\nabla(t,ir) = \sum_{l=0}^\infty \biggl(\sum_{n=0}^l \alpha_{n,l-n}(t)(-1)^n\biggr)
		(-i)^l r^l.
	\]
	The coefficient of $r^1$ is
	\[
		\alpha_{0,1}(t)i^{-1}+\alpha_{1,0}(t)i^1 = -i\beta_{1,1}(t)
		= -iW_1(t)JW_1(t)^* = -i(\det M(t))J
	\]
	by \eqref{Y38}, and hence
	\begin{align*}
		\nabla(t,ir)-M(t)+i(\det M(t))J\cdot r
		&= \sum_{\substack{l=2 \\[0.2ex] l\text{ even}}}^\infty
		\Big(\sum_{n=0}^l\alpha_{n,l-n}(t)(-1)^n\Big)(-1)^{\frac l2}r^l
		\\
		&\quad + i\sum_{\substack{l=3 \\[0.2ex] l\text{ odd}}}^\infty
		\Big(\sum_{n=0}^l\alpha_{n,l-n}(t)(-1)^n\Big)(-1)^{\frac{l+1}2}r^l.
	\end{align*}
	The imaginary part of the right-hand side can be estimated with the help of \thref{Y02};
	namely, using \thref{Y90} to join consecutive summands we obtain
	\begin{align*}
		&\hspace*{-2ex} \bigg|\sum_{\substack{l=3\\ l\text{ odd}}}^\infty
		\Big(\sum_{n=0}^l\alpha_{n,l-n}(t)(-1)^n\Big)(-1)^{\frac{l+1}2}r^l\bigg|
		\\
		&= \bigg|\sum_{\substack{l=3 \\[0.2ex] l\text{ odd}}}^\infty
		\biggl(\sum_{k=0}^{\frac{l-1}2}\big[\alpha_{2k,l-2k}(t)-\alpha_{2k+1,l-2k-1}(t)\big]\biggr)
		(-1)^{\frac{l+1}2}r^l\bigg|
		\\
		&\preceq \sum_{\substack{l=3 \\[0.2ex] l\text{ odd}}}^\infty
		\biggl(\sum_{k=0}^{\frac{l-1}2}\big|\beta_{2k+1,l-2k}(t)\big|\biggr)r^l
		= \sum_{\substack{l=3 \\[0.2ex] l\text{ odd}}}^\infty
		\biggl(\sum_{k=0}^{\frac{l-1}2}\big|\beta_{2k+1,\,2(\frac{l-1}{2}-k)+1}(t)\big|\biggr)r^l
		\displaybreak[0]\\
		&\preceq \sum_{\substack{l=3 \\[0.2ex] l\text{ odd}}}^\infty
		\biggl(\sum_{k=0}^{\frac{l-1}2}(\det M(t))\Bigl[1+3\Bigl(k+\frac{l-1}{2}-k\Bigr)\Bigr]
		\cdot m^+(t)^{2(k+\frac{l-1}{2}-k)-1}M^+(t)r^l
		\\
		&= \det M(t)\cdot\biggl(\sum_{\substack{l=3 \\[0.2ex] l\text{ odd}}}^\infty
		\frac{l+1}{2}\cdot\frac{3l-1}{2}m^+(t)^{l-2}r^l\biggr)M^+(t).
	\end{align*}
	Using the definition of $a(t,z)$ and $b(t,z)$ in \thref{Y51} we obtain
	\begin{align*}
		&\hspace*{-3ex} \big|\Im b(t,ir)-(\det M(t))r\big|
		= \bigg|\Im\biggl(\binom10^*\nabla(t,ir)\binom01\biggr)-(\det M(t))r\bigg|
		\\[1ex]
		&= \bigg|\Im\binom10^*\Bigl(\nabla(t,ir)-M(t)+i(\det M(t))J\cdot r\Bigr)\binom01\bigg|
		\\[1ex]
		&\le \binom10^*\det M(t)\cdot\biggl(\sum_{\substack{l=3 \\[0.2ex] l\text{ odd}}}^\infty
		\frac{l+1}{2}\cdot\frac{3l-1}{2}m^+(t)^{l-2}r^l\biggr)M^+(t)\binom01
		\\[1ex]
		&\le (\det M(t))r\cdot \sum_{\substack{l=3 \\[0.2ex] l\text{ odd}}}^\infty
		\frac{(l+1)(3l-1)}{8}\bigl(m^+(t)r\bigr)^{l-1}
		\displaybreak[0]\\[1ex]
		&\le (\det M(t))r\cdot \sum_{k=1}^\infty
		\frac{(2k+2)\bigl(3(2k+1)-1\bigr)}{8}\bigl(m^+(t)r\bigr)^{2k}
		\\[1ex]
		&= (\det M(t))r\cdot \sum_{k=1}^\infty (k+1)\frac{3k+1}{2}\bigl(m^+(t)r\bigr)^{2k}.
	\end{align*}
	Recall that $m^+(\mr t(r))r=\eta$.  Moreover, it is easy to check that $\eta\le\frac12$ implies
	\begin{equation}\label{Y18}
		\frac{3k+1}{2}\eta^k \le 1, \qquad k\in\bb N.
	\end{equation}
	Continuing the above chain of inequalities with $t=\mr t(r)$
	and using \eqref{Y18} and \eqref{Y17} we obtain
	\begin{align*}
		&\hspace*{-3ex} \big|\Im b(\mr t(r),ir)-(\det M(\mr t(r)))r\big|
		\le (\det M(\mr t(r)))r\cdot \sum_{k=1}^\infty (k+1)\frac{3k+1}{2}\eta^{2k}
		\\
		&\le (\det M(\mr t(r)))r\cdot \sum_{k=1}^\infty (k+1)\eta^k
		= \sigma(\det M(\mr t(r)))r,
	\end{align*}
	and hence $\Im b(\mr t(r),ir)\ge(1-\sigma)(\det M(\mr t(r)))r$.
	This, together with \eqref{Y72} and \thref{Y51}, implies that
	\[
		\Im q_H(ir) \ge \frac{\Im b(\mr t(r),ir)}{a(\mr t(r),ir)}
		\ge \frac{(1-\sigma)\det M(\mr t(r))r}{(1+\sigma)m_2(\mr t(r))}
		= \frac{1-\sigma}{1+\sigma}\cdot\frac{\eta}{2}L(r),
	\]
	which is the lower bound in \eqref{Y96} for $\vartheta=\pi/2$.
	Having an estimate along the imaginary axis, \eqref{Y96} for general $\vartheta$
	follows using a standard property of the Poisson kernel.
	Using the Herglotz integral representation \eqref{Y156} of $q_H$ we obtain,
	for each $\vartheta\in(0,\pi)$, that
	\begin{align*}
		\Im q_H(re^{i\vartheta})
		&= \beta_H r\sin\vartheta+\int_{\bb R}
		\frac{r\sin\vartheta}{|t-re^{i\vartheta}|^2}\DD\mu_H(t)
		\\[1ex]
		&\ge \beta_H r\sin\vartheta+\int_{\bb R}
		\frac{r\sin\vartheta}{(t^2+r^2)(1+|\cos\vartheta|)}\DD\mu_H(t)
		\\[1ex]
		&\ge \frac{\sin\vartheta}{1+|\cos\vartheta|}
		\biggl(\underbrace{\beta_H r+\int_{\bb R}\frac{r}{t^2+r^2}
		\DD\mu_H(t)}_{=\Im q_H(ir)}\biggr),
	\end{align*}
	which finishes the proof of \thref{Y98}.
\end{proof}

\subsection{The comparison result}
\label{Y29}

Let $H$ and $\widetilde H$ be Hamiltonians on $[a,b)$ such that none of their
diagonal entries vanishes a.e.
We use the freedom in the choice of the parameter $\eta$ in \thref{Y98},
and therefore also make a notational distinction: set
\[
	\mr r_{(\eta)}(t) \DE \frac{\eta}{2}\bigl(m_1(t)m_2(t)\bigr)^{-\frac 12}, \qquad
	\mr t_{(\eta)}(r) \DE r_{(\eta)}^{-1}(r), \qquad
	A_{(\eta)}(r) \DE \sqrt{\frac{m_1(\mr t_{(\eta)}(r))}{m_2(\mr t_{(\eta)}(r))}\,}.
\]
for $t\in(\mr a,b)$ and $r>0$, where $\mr a$ is defined in \eqref{Y196}.
Moreover, recall that $A_{(\eta)}(r)$ can be written as
\begin{equation}\label{Y119}
	A_{(\eta)}(r) = \frac{2r}{\eta}\cdot m_1\bigl(\mr t_{(\eta)}(r)\bigr)
	= \frac{\eta}{2r}\cdot\frac{1}{m_2(\mr t_{(\eta)}(r))}.
\end{equation}
Analogous notation applies to $\widetilde H$.

\begin{lemma}\thlab{Y20}
	Let $c,\gamma>0$ and let $t\in(\mr a,b)$.  Assume that $t\le\mr{\tilde a}$ or that
	\begin{equation}\label{Y24}
		\tr\tildM(t) \le c\tr M(t),\qquad
		\widetilde m_2(t) \le \gamma m_2(t),\qquad
		m_1(t)\ge \frac 12\tr M(t).
	\end{equation}
	Further, let $\eta\in(0,\frac{1}{\sqrt{2c\gamma}\,}[1-\frac{1}{\sqrt{2}\,}])$, and set
	\[
		\tilde\eta \DE \eta\cdot\sqrt{2c\gamma},\quad
		\tilde\sigma \DE \frac{1}{(1-\tilde\eta)^2}-1,\quad
		\tilde\delta \DE \bigg(\frac{1+\tilde\sigma+\frac 2{\tilde\eta}}{1-\tilde\sigma}\bigg)^{-1}
		\frac{2}{\,\tilde\eta\,}.
	\]
	Then
	\[
		\frac{1}{\mr r_{(\eta)}(t)}\big|q_{\widetilde H}\bigl(i\mr r_{(\eta)}(t)\bigr)\big|
		\ge \tilde\delta\cdot\widetilde m_1(t).
	\]
\end{lemma}

\begin{proof}
	First assume that $t>\mr{\tilde a}$.  Then, by assumption, \eqref{Y24} holds,
	from which we obtain
	\begin{gather*}
		\widetilde m_1(t) \le \tr\tildM(t) \le c\cdot\tr M(t) \le 2c\cdot m_1(t),
		\\[1ex]
		\widetilde m_1(t)\widetilde m_2(t) \le 2c\gamma\cdot m_1(t)m_2(t),
	\end{gather*}
	and hence
	\[
		\mr r_{(\eta)}(t) = \frac{\eta}{2}\bigl(m_1(t)m_2(t)\bigr)^{-\frac 12}
		\le \frac{\eta\sqrt{2c\gamma}\,}{2}\bigl(\widetilde m_1(t)\widetilde m_2(t)\bigr)^{-\frac 12}
		= \mr{\tilde r}_{(\tilde\eta)}(t).
	\]
	Since $\mr{\tilde t}_{(\tilde\eta)}$ is decreasing, it follows that
	\[
		\mr{\tilde t}_{(\tilde\eta)}\bigl(\mr r_{(\eta)}(t)\bigr)
		\ge \mr{\tilde t}_{(\tilde\eta)}\bigl(\mr{\tilde r}_{(\eta)}(t)\bigr)
		= t.
	\]
	In the case when $t\le\mr{\tilde a}$ we have
	\[
		\mr{\tilde t}_{(\tilde\eta)}\bigl(\mr r_{(\eta)}(t)\bigr)
		> \mr{\tilde a} \ge t.
	\]
	In both cases we can apply \thref{Y98} to estimate
	\begin{align*}
		\frac{1}{\mr r_{(\eta)}(t)}\big|q_{\widetilde H}\bigl(i\mr r_{(\eta)}(t)\bigr)\big|
		&\ge \biggl(\frac{1+\tilde\sigma+\frac 2{\tilde\eta}}{1-\tilde\sigma}\biggr)^{-1}
		\frac{1}{\mr r_{(\eta)}(t)}\tilde A_{(\tilde\eta)}\bigl(\mr r_{(\eta)}(t)\bigr)
		\\[1ex]
		&= \biggl(\frac{1+\tilde\sigma+\frac{2}{\tilde\eta}}{1-\tilde\sigma}\biggr)^{-1}
		\frac{2}{\,\tilde\eta\,}
		\cdot \widetilde m_1\Bigl(\mr{\tilde t}_{(\tilde\eta)}\bigl(\mr r_{(\eta)}(t)\bigr)\Bigr)
		\ge \bigg(\frac{1+\tilde\sigma+\frac{2}{\tilde\eta}}{1-\tilde\sigma}\bigg)^{-1}
		\frac{2}{\,\tilde\eta\,}\cdot \widetilde m_1(t).
	\end{align*}
\end{proof}

\begin{proof}[Proof of \thref{Y09}]
	Let the data be given as in the statement of the proposition
	and assume first that $a'<b$.
	Choose $\eta>0$ such that
	\[
		\max\bigl\{\eta,\,\eta\sqrt{2c_1\gamma_1},\,\eta\sqrt{2c_2\gamma_2}\bigr\}
		< 1-\frac{1}{\sqrt{2}\,},
	\]
	and set
	\begin{equation}\label{Y31}
		r_0 \DE \max\bigl\{\mr r_{(\eta)}(a'),\mr{\tilde r}_{(\eta)}(a')\bigr\}.
	\end{equation}
	Further, let $r>r_0$.
	We distinguish three cases.
	\begin{Steps}
	\item
		Assume that $A_{(\eta)}(r)\ge 1$. \\[0.5ex]
		Set $t\DE\mr t_{(\eta)}(r)$; then $t\in(\mr a,a')$.  Since $A_{(\eta)}(r)\ge 1$,
		we have $m_1(t)\ge m_2(t)$, and hence $m_1(t)\ge \frac 12\tr M(t)$.
		Set
		\[
			\eta_2 \DE \eta\cdot\sqrt{2c_2\gamma_2}, \qquad
			\sigma_2 \DE \frac{1}{(1-\eta_2)^2}-1, \qquad
			\delta_2 \DE \bigg(\frac{1+\sigma_2+\frac{2}{\eta_2}}{1-\sigma_2}\bigg)^{-1}
			\frac{2}{\eta_2}.
		\]
		The assumptions of \thref{Y20} are satisfied with $c=c_2$, $\gamma=\gamma_2$,
		and hence
		\begin{align}
			\frac 1r|q_{\widetilde H}(ir)|
			&\ge \delta_2\cdot\widetilde m_1(t)\ge\frac{\delta_2}{\gamma_1}\cdot m_1(t)
			= \frac{\delta_2}{\gamma_1}\cdot\frac{\eta}{2r}A_{(\eta)}(r)
			\nonumber
			\\
			&\ge
			\frac{\delta_2}{\gamma_1}\cdot\frac{\eta}{2r}\cdot
			\bigg(\frac{1+\sigma+\frac{2}{\eta}}{1-\sigma}\bigg)^{-1}\cdot|q_H(ir)|.
			\label{Y57}
		\end{align}
	\item
		Assume that $\widetilde A_{(\eta)}(r)\le 1$. \\[0.5ex]
		We proceed similarly.
		Set $t\DE\mr{\tilde t}_{(\eta)}(r)$; then $t\in(\mr{\tilde a},a')$.  Moreover, we have
		$\widetilde m_1(t)\le\widetilde m_2(t)$, and hence
		$\widetilde m_2(t)\ge\frac 12\tr\tildM(t)$.
		We apply \thref{Y20} with $-J\widetilde HJ$ in place of $H$
		and $-JHJ$ in place of $\widetilde H$.
		The assumptions of the lemma are now satisfied with $c=c_1$, $\gamma=\gamma_1$.
		Set
		\[
			\eta_1 \DE \eta\cdot\sqrt{2c_1\gamma_1}, \qquad
			\sigma_1 \DE \frac{1}{(1-\eta_1)^2}-1, \qquad
			\delta_1 \DE \biggl(\frac{1+\sigma_1+\frac{2}{\eta_1}}{1-\sigma_1}\biggr)^{-1}
			\frac{2}{\eta_1};
		\]
		then
		\begin{align}
			\frac 1r\Big|\frac {-1}{q_H(ir)}\Big|
			&= \frac 1r\big|q_{-JHJ}(ir)\big|
			\ge \delta_1\cdot m_2(t)\ge\frac{\delta_1}{\gamma_2}\cdot\widetilde m_2(t)
			= \frac{\delta_1}{\gamma_2}\cdot\frac{\eta}{2r}\big[\widetilde A_{(\eta)}(r)\big]^{-1}
			\nonumber
			\\
			&\ge \frac{\delta_1}{\gamma_2}\cdot\frac{\eta}{2r}\cdot
			\biggl[\biggl(\frac{1+\sigma+\frac{2}{\eta}}{1-\sigma}\biggr)
			\cdot|q_{\widetilde H}(ir)|\biggr]^{-1}.
			\label{Y58}
		\end{align}
	\item
		Assume that $A_{(\eta)}(r)<1$ and $\widetilde A_{(\eta)}(r)>1$. \\[0.5ex]
		Then
		\[
			|q_H(ir)|<\frac{1+\sigma+\frac{2}{\eta}}{1-\sigma},\qquad
			|q_{\widetilde H}(ir)|>\biggl(\frac{1+\sigma+\frac{2}{\eta}}{1-\sigma}\biggr)^{-1},
		\]
		and we see that
		\begin{equation}\label{Y59}
			|q_H(ir)| \le \biggl(\frac{1+\sigma+\frac{2}{\eta}}{1-\sigma}\biggr)^2
			\cdot|q_{\widetilde H}(ir)|.
		\end{equation}
	\end{Steps}
	The above cases together cover the whole ray $(r_0,\infty)$.
	From \eqref{Y57}, \eqref{Y58} and \eqref{Y59} we obtain that,
	for all $r$ on this ray, $|q_H(ir)|\le C|q_{\widetilde H}(ir)|$
	where $C$ is the maximum of the three terms
	\[
		\frac{\gamma_1}{\delta_2}\cdot\frac{2}{\eta}
		\biggl(\frac{1+\sigma+\frac{2}{\eta}}{1-\sigma}\biggr),
		\qquad
		\frac{\gamma_2}{\delta_1}\cdot\frac{2}{\eta}
		\biggl(\frac{1+\sigma+\frac{2}{\eta}}{1-\sigma}\biggr),
		\qquad
		\biggl(\frac{1+\sigma+\frac{2}{\eta}}{1-\sigma}\biggr)^2.
	\]
	Observe that $C$ is independent of $H,\widetilde H$ and $a'$.
	Finally, the statement that we can choose $r_0=0$ when $a'=b$ follows from the fact
	that $C$ is independent of $a'$ and from \eqref{Y31}, which shows
	that $r_0\to0$ as $a'\to b$.
\end{proof}

\section{Asymptotic behaviour of the spectral measure}
\label{Y43}

Having available the estimates of the Weyl coefficient $q_H$ from \thref{Y98},
we employ Abelian--Tauberian-type theorems to translate knowledge about the growth
of $\Im q_H(ir)$ for $r\to\infty$ to knowledge about the growth of the
distribution function of the spectral measure $\mu_H$ towards infinity.
This enables us to give conditions for weighted integrability
and boundedness of tails of $\mu_H$ relative to suitable comparison functions.
Thereby we use functions of regular variation (see Appendix~\ref{Y50}) to compare with.
This class of regularly varying functions provides a much finer scale 
than just the class of power functions; see, e.g.\ Examples~\ref{Y205} and \ref{Y206}.

In this section we assume that neither $h_1$ nor $h_2$ vanishes in a neighbourhood
of the left endpoint $a$, i.e.\ we assume that $\mr a=a$,
where $\mr a$ is defined in \eqref{Y196}.
The reason is the following.  If $h_1=0$ in a neighbourhood of $a$, then the
measure $\mu_H$ is finite and therefore the growth of the distribution function
of $\mu_H$ trivial.
If $h_2=0$ in a neighbourhood of $a$, then the linear term $\beta_Hz$ in \eqref{Y156}
dominates the integral when $z\to\infty$ non-tangentially.
Hence the growth of $\Im q_H(ir)$ does not determine the growth of the
distribution function of the spectral measure $\mu_H$.
These two special cases are considered in more detail in \S\ref{Y127}.

Throughout this section we use the following notation.

\begin{definition}\thlab{Y141}
	If $\mu$ is a positive Borel measure on the real line, then we denote
	by $\tilde\mu$ the push-forward measure of $\mu$ under the map $t\mapsto|t|$,
	and by $\dblarrow\mu$ the distribution function of $\tilde\mu$.
	Explicitly, this means
	\begin{equation}\label{Y140}
		\tilde\mu([0,r)) = \mu((-r,r)) = \dblarrow\mu(r),\;\; r\ge 0,\qquad
		\tilde\mu((-\infty,0)) = 0.
	\end{equation}
\end{definition}

\subsection{Membership of Kac classes}
\label{Y190}

Recall that a Nevanlinna function $q$ is said to belong to the \emph{Kac class}
with index $\alpha\in(0,2)$ if the measure
$\mu$ in its Herglotz integral representation (cf.\ \eqref{Y156}) satisfies
\[
	\int_{\bb R}\frac{\DD\mu(r)}{1+|r|^\alpha} < \infty
\]
and no linear term is present.
These classes have been investigated for a long time because of their role
in the spectral theory of the string;
see \cite[11.6$^\circ$]{kac.krein:1968a}, \cite{kac.krein:1968,kac:1959}.
They are also known to play a role in a broader operator-theoretic context
\cite{winkler:2002,hassi.snoo.winkler:2007}.
More general classes occur in \cite{kac:1982} where weighted integrability conditions
for the spectral measure of a Krein string are characterised in terms of
the mass distribution function of the string.
Kac's result is formulated in a way to allow arbitrary non-decreasing comparison functions,
while we prefer to give more explicit conditions on the cost of restricting the class
of comparison functions to regularly varying functions.
See Appendix~\ref{Y50} for the definition and some properties of regularly varying functions.

\begin{definition}\thlab{Y88}
	Let $\ms g$ be a continuous, regularly varying function with index $\alpha\le 2$
	and $\lim_{r\to\infty}\ms g(r)=\infty$.
	We denote by $\mc M_{\ms g}$ the set of all positive Borel measures $\mu$ on $\bb R$
	such that
	\[
		\int_{[1,\infty)} \frac{\DD\tilde\mu(r)}{\ms g(r)} < \infty.
	\vspace*{-3mm}
	\]
\end{definition}

\begin{remark}\thlab{Y55}
	The assumption that $\alpha\le 2$ is natural.
	Namely, our aim is to relate the spectral measure of a canonical system with the
	Hamiltonian of the system. Such measures are always Poisson integrable,
	i.e.\ $\int_{[1,\infty)} r^{-2}\DD\tilde\mu(r)<\infty$.
	
	Contrasting this, the assumption that $\lim_{r\to\infty}\ms g(r)=\infty$ is a restriction.
	The necessity to impose this assumption comes from the fact that finite measures $\mu$
	correspond to Hamiltonians which start with an interval where $h_1=0$.
	Some cases of finite measures $\mu$ can be reduced to the case of an infinite measure.
	Explicit formulae that relate the Hamiltonians with spectral measures $\mu$
	and $t^2\DD\mu(t)$ are known; see \cite[Rule~6]{winkler:1995a}.
	Iterating these formulae and combining them with our results below one can obtain
	corresponding results for a class of finite measures;
	however, these formulae will be very lengthy (and presumably hard to apply in practice).
	Hence, we do not go into further details in this respect.
\end{remark}

\noindent
\thref{Y86} below is a generalisation of \cite[Theorem]{kac:1982} to the case of a
non-diagonal Hamiltonian; the connection is worked out in detail in Section~\ref{Y89}.
In the formulation of the theorem we use the following notation:
for a regularly varying function $\ms g$ with index $\alpha\le 2$ set
\begin{equation}\label{Y46}
	\ms g_\star(r)\DE \int_1^r\frac t{\ms g(t)}\DD t,\quad r \ge 1.
\end{equation}
By Karamata's theorem (see \thref{Y48}(i)) the function $\ms g_\star$ is
regularly varying with index $2-\alpha$, and
\begin{equation}\label{Y137}
	\ms g_\star(r)\;
	\begin{cases}
		\; \asymp\frac{r^2}{\ms g(r)}, & \alpha<2,
		\\[2ex]
		\; \gg\frac{r^2}{\ms g(r)}, & \alpha=2.
	\end{cases}
\end{equation}
For the following, it turns out to be more convenient to use a slight
modification of the class $\mc M_{\ms g}$.

\begin{definition}\thlab{Y146}
	Let $\ms g$ be a regularly varying function with index $\alpha\le2$
	and $\lim_{r\to\infty}\ms g(r)=\infty$.
	We denote by $\hatcalM_{\ms g}$ the set of all positive Borel measures $\mu$
	on $\bb R$ such that
	\[
		\int_1^\infty \dblarrow\mu(r)\frac{\ms g_\star(r)}{r^3}\DD r<\infty,
	\]
	where $\dblarrow{\mu}$ and $\ms g_\star$ are defined in \eqref{Y140}
	and \eqref{Y46} respectively.
\end{definition}

\noindent
For $\alpha\in(0,2)$ the classes $\mc M_{\ms g}$ and $\hatcalM_{\ms g}$
coincide as the following proposition shows.

\begin{proposition}\thlab{Y147}
	Let $\ms g$ be a regularly varying function with index $\alpha\le2$
	and $\lim_{r\to\infty}\ms g(r)=\infty$.
	\begin{Enumerate}
	\item
		If $\alpha\in(0,2)$, then $\hatcalM_{\ms g}=\mc M_{\ms g}$.
	\item
		If $\alpha=0$ or $\alpha=2$, then $\hatcalM_{\ms g}\subseteq\mc M_{\ms g}$.
	\item
		If $\alpha\in[0,2)$, then
		\[
			\mu\in\hatcalM_{\ms g} \;\;\Leftrightarrow\;\;
			\int_1^\infty \dblarrow\mu(r)\frac{\DD r}{r\ms g(r)}<\infty.
		\]
	\item
		If $\alpha\in(0,2]$, then
		\[
			\mu\in\mc M_{\ms g} \;\;\Leftrightarrow\;\;
			\int_1^\infty \dblarrow\mu(r)\frac{\DD r}{r\ms g(r)}<\infty.
		\]
	\end{Enumerate}
\end{proposition}

\noindent
Before we prove \thref{Y147} we formulate a lemma about integration by parts
in a measure-theoretic form, see e.g.\ \cite[Lemma~2]{kac:1965}.

\begin{lemma}\thlab{Y49}
	Let $-\infty<a<b\le\infty$ and let $\mu$ and $\nu$ be positive Borel measures
	on $[a,b)$.  Then
	\begin{equation}\label{Y142}
		\int_{[a,b)}\mu([a,t))\DD\nu(t)=\int_{[a,b)}\nu((t,b))\DD\mu(t).
	\end{equation}
	If these integrals are finite, then $\lim\limits_{t\to b}\mu([a,t))\nu([t,b))=0$.
\end{lemma}

\begin{proof}
	If $\nu([a,b))=\infty$, then either both sides are zero (when $\mu$ is the zero measure)
	or both sides are infinity (otherwise).
	In the case when $\nu$ is a finite measure, relation \eqref{Y142} follows
	from Fubini's theorem.
	To show the last assertion, assume that both sides of \eqref{Y142} are finite
	and let $x\in(a,b)$.  Then \eqref{Y142} applied to $[a,x)$ instead of $[a,b)$ yields
	\[
		\int_{[a,x)}\mu([a,t))\DD\nu(t)
		= \int_{[a,x)}\nu((t,x))\DD\mu(t)
		= \int_{[a,x)}\nu((t,b))\DD\mu(t) - \mu([a,x))\nu([x,b)).
	\]
	Letting $x\to b$ we obtain the claimed limit relation.
\end{proof}

\begin{proof}[Proof of \thref{Y147}]
	If $\int_1^\infty\frac{\RD t}{t\ms g(t)}<\infty$, then
	\begin{equation}\label{Y145}
		\int_r^\infty\frac{\RD t}{t\ms g(t)}\;
		\begin{cases}
			\; \asymp\frac{1}{\ms g(r)}, & \alpha>0,
			\\[2ex]
			\; \gg\frac{1}{\ms g(r)}, & \alpha=0,
		\end{cases}
	\end{equation}
	by \thref{Y48}(ii).  Using \eqref{Y137}, \thref{Y49} and \eqref{Y145}
	we then obtain the following implications:
	\begin{alignat*}{2}
		\mu\in\hatcalM_{\ms g}
		&\quad\Leftrightarrow\quad
		\int_1^\infty\frac{\dblarrow\mu(r)}{r^3}\ms g_\star(r)\DD r<\infty
		\hspace*{-5ex}
		\\[1ex]
		& \biggl\{\!
		\begin{array}{c}
			\scalebox{0.5}{if $\alpha<2$} \\[-1.7ex]
			\scalebox{0.7}{$\Longleftrightarrow$} \\[-0.4ex]
			\scalebox{0.8}{$\Longrightarrow$} \\[-1.8ex]
			\scalebox{0.5}{if $\alpha=2$}
		\end{array}\!\biggr\}
		\quad
		\int_1^\infty\frac{\dblarrow\mu(r)}{r\ms g(r)}\DD r<\infty
		\quad&&\Leftrightarrow\quad
		\int\limits_{[1,\infty)}\biggl(\int_r^\infty\frac 1{t\ms g(t)}\DD t\biggr)\DD\tilde\mu(r)<\infty
		\\[1ex]
		& \biggl\{\!
		\begin{array}{c}
			\scalebox{0.5}{if $\alpha>0$} \\[-1.7ex]
			\scalebox{0.7}{$\Longleftrightarrow$} \\[-0.4ex]
			\scalebox{0.8}{$\Longrightarrow$} \\[-1.8ex]
			\scalebox{0.5}{if $\alpha=0$}
		\end{array}\!\biggr\}
		\quad
		\int\limits_{[1,\infty)}\frac{\DD\tilde\mu(r)}{\ms g(r)}<\infty
		\quad&&\Leftrightarrow\quad \mu\in\mc M_{\ms g},
	\end{alignat*}
	which shows all assertions.
\end{proof}

\noindent
The following proposition contains the core of the argument in the proof
of \thref{Y86} below.
However, it is more flexible and is also used in Section~\ref{Y89}.

\begin{proposition}\thlab{Y158}
	Let $H$ be a Hamiltonian defined on some interval $[a,b)$, and assume that
	neither $h_1$ nor $h_2$ vanishes on a neighbourhood of the left endpoint $a$.
	Let $\ms f$ be a continuous, non-decreasing, regularly varying function,
	and denote by $\mu_H$ the spectral measure of $H$.

	Then the statements
	\begin{Enumerate}
	\item
		$\exists a'\in(a,b)$ such that
		\[
			\int_{a}^{a'}
			h_1(x)\cdot\ms f\bigl((m_1m_2)(t)^{-\frac12}\bigr)\DD t<\infty;
		\]
	\item
		\begin{equation}\label{Y161}
			\int_1^\infty \dblarrow\mu_H(r)\frac{\ms f(r)}{r^3}\DD r<\infty;
		\end{equation}
	\item
		$\exists a'\in(a,b)$ such that
		\[
			\int_{a}^{a'}
			\frac{1}{m_2(t)^2}\binom{m_2(t)}{-m_3(t)}^*H(t)\binom{m_2(t)}{-m_3(t)}
			\cdot\ms f\bigl((m_1m_2)(t)^{-\frac12}\bigr)\DD t<\infty.
		\]
	\end{Enumerate}
	satisfy \textup{(i)}\,$\Rightarrow$\,\textup{(ii)}\,$\Rightarrow$\,\textup{(iii)}.

	If, in addition, $\ms f$ is differentiable and $\ms f'$ is regularly varying,
	then
	\begin{alignat*}{3}
		& \text{\textup{(i)}} \quad&&\Leftrightarrow\quad &
		& \exists a'\in(a,b)\DP
		\int_{a}^{a'}m_1(t)\ms f'\bigl((m_1m_2)(t)^{-\frac12}\bigr)
		\frac{(m_1m_2)'(t)}{(m_1m_2)(t)^{\frac32}}\DD t < \infty,
		\\[1ex]
		& \text{\textup{(iii)}} \quad&&\Leftrightarrow\quad &
		& \exists a'\in(a,b)\DP
		\int_{a}^{a'}\frac{\det M(t)}{m_2(t)}\ms f'\bigl((m_1m_2)(t)^{-\frac12}\bigr)
		\frac{(m_1m_2)'(t)}{(m_1m_2)(t)^{\frac32}}\DD t < \infty.
	\end{alignat*}
\end{proposition}

\medskip

\begin{remark}\thlab{Y163}
	It can be seen from the proof below that, in the first part, instead of assuming
	that $\ms f$ is regularly varying,
	it is sufficient to assume that $\ms f\in\mathop{OR}$,
	i.e.\ for every $\lambda>0$ there exist $c_1,c_2>0$
	such that $c_1\le\frac{\ms f(\lambda r)}{\ms f(r)}\le c_2$, $r\in[1,\infty)$;
	for the latter definition see, e.g.\ \cite[\S2.0.2]{bingham.goldie.teugels:1989}.
\end{remark}

\noindent
For the proof of \thref{Y158} we use a simple Abelian--Tauberian-type theorem
for the Poisson integral of a positive measure.
This is folklore; an explicit proof can be found in, e.g.\
\cite[Lemma~4]{kac:1982}\footnote{%
	For the case when $\ms g(r)=r^\alpha$ it goes back at least to \cite[Theorema~1]{kac:1956}.
}.

\begin{lemma}\thlab{Y83}
	Let $\mu$ be a positive Borel measure on $\bb R$
	with $\int_{\bb R}\frac{\DD\mu(t)}{1+t^2}<\infty$,
	let $r_0>0$, and let $\xi$ be a positive Borel measure on $[r_0,\infty)$.
	Define $\dblarrow\mu$ as in \eqref{Y140}, set
	\[
		\vec\xi(r) \DE \xi([r_0,r)),\quad r \ge r_0,
	\]
	and let $\Poi{\mu}(z)$ be the Poisson integral
	\begin{equation}\label{Y144}
		\Poi{\mu}(z) \DE \int_{\bb R}\Im\frac{1}{t-z}\DD\mu(t),\quad z\in\bb C^+,
	\end{equation}
	of $\mu$.  Then
	\[
		\int_{[r_0,\infty)}\frac 1r\Poi{\mu}(ir)\DD\xi(r)<\infty
		\quad\Leftrightarrow\quad
		\int_{[r_0,\infty)}\frac{\dblarrow\mu(r)\vec\xi(r)}{r^3}\DD r<\infty.
	\]
\end{lemma}

\medskip

\begin{proof}[Proof of \thref{Y158}]
	First note that finiteness of the integrals in the proposition clearly
	does not depend on $a'\in(a,b)$.

	Let $\xi$ be the measure on $[1,\infty)$ such that $\ms f(r)=\xi([1,r))$, $r\ge1$.
	It follows from \thref{Y83} that
	\[
		\int_{[1,\infty)}\frac{\Poi{\mu_H}(ir)}{r}\DD\xi(r) < \infty
		\quad\Leftrightarrow\quad
		\int_1^\infty \frac{\dblarrow\mu_H(r)\ms f(r)}{r^3}\DD r < \infty.
	\]
	Fix $\eta\in(0,1-\frac{1}{\sqrt{2}\,})$ and let $\mr r$ be as in \eqref{Y124}.
	By \thref{Y98} with $A(r)$ and $L(r)$ in the form of \eqref{Y36} and \eqref{Y120},
	we have
	\[
		\frac{\det M(\mr t(r))}{m_2(\mr t(r))}
		\lesssim \frac{\Poi{\mu_H}(ir)}{r}
		\lesssim m_1(\mr t(r)).
	\]
	Hence
	\begin{equation}\label{Y162}
		\int_{[1,\infty)}m_1\bigl(\mr t(r)\bigr)\DD\xi(r) < \infty
		\;\;\Rightarrow\;\;
		\text{(\ref{Y161})}
		\;\;\Rightarrow\;\;
		\int_1^\infty\frac{\det M\bigl(\mr t(r)\bigr)}{m_2\bigl(\mr t(r)\bigr)}\DD\xi(r)
		< \infty.
	\end{equation}
	Let $\nu$ be the measure on $(0,\infty)$ such that $\nu((r,\infty))=m_1(\mr t(r))$, $r>0$,
	and let $\mr t_*\nu$ be the push-forward measure of $\nu$ under the mapping $\mr t$.
	For $t\in(a,b)$ we have
	$\mr t_*\nu((a,t))=\nu((\mr t^{-1}(t),\infty))=m_1(t)$.
	Moreover, recall that $\mr r$, defined in \eqref{Y124}, is the inverse function
	of $\mr t$.
	Hence, with \thref{Y49} we can rewrite the first integral in \eqref{Y162} as follows:
	\begin{align*}
		& \int_{[1,\infty)}m_1\bigl(\mr t(r)\bigr)\DD\xi(r)
		= \int_{[1,\infty)}\nu\bigl((r,\infty)\bigr)\DD\xi(r)
		= \int_{[1,\infty)}\vec\xi(r)\DD\nu(r)
		= \int_{[1,\infty)}\ms f(r)\DD\nu(r)
		\\
		&= \int_{(a,\mr t(1)]}\ms f\bigl(\mr r(t)\bigr)\DD(\mr t_*\nu)(t)
		= \int_{(a,\mr t(1)]}\ms f\bigl(\mr r(t)\bigr)\DD m_1(t)
		= \int_a^{\mr t(1)}\ms f\bigl(\mr r(t)\bigr)h_1(t)\DD t.
	\end{align*}
	Since $\ms f$ is regularly varying, the last integral is finite
	if and only the integral in (i) is finite.

	In a similar way one can rewrite the last integral in \eqref{Y162}
	by using a measure $\nu$ such that
	$\nu((r,\infty))=\frac{\det M(\mr t(r))}{m_2(\mr t(r))}$, $r>0$,
	which is possible since $t\mapsto \frac{\det M(t)}{m_2(t)}$ is
	non-decreasing by \eqref{Y71}.

	For the last part let us assume that $\ms f$ is differentiable and that
	$\ms f'$ is regularly varying.  Using a substitution we can rewrite
	the first integral in \eqref{Y162} differently:
	\begin{align*}
		\int_{[1,\infty)}m_1\bigl(\mr t(r)\bigr)\DD\xi(r)
		&= \int_1^\infty m_1\bigl(\mr t(r)\bigr)\ms f'(r)\DD r
		= \int_{\mr t(1)}^{a} m_1(t)\ms f'\bigl(\mr r(t)\bigr)\mr r'(t)\DD t
		\\
		&= \frac{\eta}{4}\int_{a}^{\mr t(1)}m_1(t)\ms f'\bigl(\mr r(t)\bigr)
		\frac{(m_1m_2)'(t)}{(m_1m_2)(t)^{\frac32}}\DD t.
	\end{align*}
	Since $\ms f'$ is assumed to be regularly varying, the last integral
	is finite if and only if it is finite with $\ms f'(\mr r(t))$ replaced
	by $\ms f'((m_1m_2)(t)^{-\frac12})$.
	In exactly the same way one can rewrite the last integral in \eqref{Y162}.
\end{proof}

\noindent
The following theorem is the main result of this subsection.
It provides, in particular, information when the spectral measure $\mu_H$
belongs to the class $\hatcalM_{\ms g}$.

\begin{theorem}\thlab{Y86}
	Let $H$ be a Hamiltonian defined on some interval $[a,b)$, and assume that
	neither $h_1$ nor $h_2$ vanishes on a neighbourhood of the left endpoint $a$.
	Let $\ms g$ be a continuous, regularly varying function with index $\alpha\le 2$
	and $\lim_{r\to\infty}\ms g(r)=\infty$, let $\ms g_\star$ be as in \eqref{Y46},
	and denote by $\mu_H$ the spectral measure of $H$ as in \eqref{Y156}.
	For every $a'\in(a,b)$ the following statements
	\begin{flalign*}
	\hspace*{2ex}
	\begin{alignedat}{2}
		&{\rm(i)} \;\; &&
		\int_{a}^{a'}
		h_1(t)\cdot\ms g_\star\bigl((m_1m_2)(t)^{-\frac 12}\bigr)\DD t<\infty,
		\\[1ex]
		&{\rm(i)'} \;\; &&
		\int_{a}^{a'}m_1(t)
		\cdot\frac{(m_1m_2)'(t)}{(m_1m_2)(t)^2\ms g\bigl((m_1m_2)(t)^{-\frac 12}\bigr)}\DD t
		< \infty,
		\\[1ex]
		&{\rm(i)''} \;\; &&
		\int_{a}^{a'}h_1(t)
		\cdot\frac{\DD t}{(m_1m_2)(t)\ms g\bigl((m_1m_2)(t)^{-\frac 12}\bigr)} < \infty,
		\\[1ex]
		&{\rm(ii)} \;\; &&
		\mu_H\in\hatcalM_{\ms g},
		\\[1ex]
		&{\rm(ii)'} \;\; &&
		\mu_H\in\mc M_{\ms g},
		\\[1ex]
		&{\rm(iii)} \;\; &&
		\int_{a}^{a'}
		\frac{1}{m_2(t)^2}\binom{m_2(t)}{-m_3(t)}^*H(t)\binom{m_2(t)}{-m_3(t)}
			\cdot\ms g_\star\bigl((m_1m_2)(t)^{-\frac 12}\bigr)\DD t<\infty,
		\\[2ex]
		&{\rm(iii)'} \;\; &&
			\int_{a}^{a'}\frac{\det M(t)}{m_2(t)}
			\cdot\frac{(m_1m_2)'(t)}{(m_1m_2)(t)^2\ms g\bigl((m_1m_2)(t)^{-\frac 12}\bigr)}\DD t
			< \infty,
		\\[2ex]
		&{\rm(iii)''} \;\; &&
			\int_{a}^{a'}\mkern-10mu
			\frac{1}{m_2(t)^2}\binom{m_2(t)}{-m_3(t)}^*\mkern-5mu H(t)\binom{m_2(t)}{-m_3(t)}
			\frac{\DD t}{(m_1m_2)(t)\ms g\bigl((m_1m_2)(t)^{-\frac 12}\bigr)} < \infty
	\end{alignedat}
	&&
	\end{flalign*}
	satisfy the relations:
	\begin{alignat*}{5}
		&{\rm(i)} \quad &&\Longleftrightarrow\quad && {\rm(i)'} \quad&&
		\begin{array}{c}
			\scalebox{0.6}{if $\alpha<2$} \\[-1.1ex]
			\scalebox{1.}{$\Longleftrightarrow$} \\[-0.8ex]
			\scalebox{1.}{$\Longrightarrow$} \\[-1.7ex]
			\scalebox{0.6}{if $\alpha=2$}
		\end{array}
		\quad && {\rm(i)''}
		\\[-1.5ex]
		&\Downarrow
		\\[-1.5ex]
		&{\rm(ii)} \quad&&
		\begin{array}{c}
			\scalebox{0.6}{if $\alpha\in(0,2)$} \\[-1.1ex]
			\scalebox{1.0}{$\Longleftrightarrow$} \\[-0.8ex]
			\scalebox{1.0}{$\Longrightarrow$} \\[-1.7ex]
			\scalebox{0.6}{if $\alpha\in\{0,2\}$}
		\end{array}
		\quad && {\rm(ii)'}
		\\[-1.5ex]
		&\Downarrow
		\\[-1.5ex]
		&{\rm(iii)} \quad &&\Longleftrightarrow\quad && {\rm(iii)'} \quad&&
		\begin{array}{c}
			\scalebox{0.6}{if $\alpha<2$} \\[-1.1ex]
			\scalebox{1.0}{$\Longleftrightarrow$} \\[-0.8ex]
			\scalebox{1.0}{$\Longrightarrow$} \\[-1.7ex]
			\scalebox{0.6}{if $\alpha=2$}
		\end{array}
		\quad && {\rm(iii)''}.
	\end{alignat*}
\end{theorem}

\begin{proof}
	The implications (i)\,$\Rightarrow$\,(ii)\,$\Rightarrow$\,(iii) and the
	equivalences (i)\,$\Leftrightarrow$\,(i)$'$ and (iii)\,$\Leftrightarrow$\,(iii)$'$
	follow directly from \thref{Y158} with $\ms f=\ms g_\star$.
	The relations between (i) and (i)$''$ and between (iii) and (iii)$''$
	follow from \eqref{Y137}.
	Finally, \thref{Y147} implies the relations between (ii) and (ii)$'$.
\end{proof}

\begin{remark}\thlab{Y204}
	Let us consider the case of the regularly varying function $\ms g(r)=r^\alpha(\log r)^\beta$ 
	with $\alpha\in[0,2]$ and $\beta\in\bb R$.
	It is easy to check that (i)$''$ is equivalent to
	\[
		\int_a^{a'}\frac{h_1(t)\DD t}{\bigl[(m_1m_2)(t)\bigr]^{1-\frac{\alpha}{2}}
		\big|\log\bigl((m_1m_2)(t)\bigr)\big|^\beta} < \infty.
	\]
\end{remark}

\medskip

\noindent
The following corollary shows that for diagonally-dominant Hamiltonians we
obtain a characterisation when $\mu_H$ belongs to $\hatcalM_{\ms g}$.

\begin{corollary}\thlab{Y47}
	Consider the situation from \thref{Y86}, and assume, in addition, that
	$\limsup_{t\to a}\frac{m_3(t)^2}{(m_1m_2)(t)}<1$
	\textup{(}this holds in particular if $H$ is diagonal\textup{)}.
	Then, for every $a'\in(a,b)$, also
	\textup{(i)}\,$\Leftrightarrow$\,\textup{(ii)}\,$\Leftrightarrow$\,\textup{(iii)}.
\end{corollary}

\begin{proof}
	The additional hypothesis just means that $\frac{\det M(t)}{m_2(t)}\asymp m_1(t)$,
	which implies that (i)$'\,\Leftrightarrow\,$(iii)$'$.
\end{proof}

\begin{example}\thlab{Y205}
	Let $\rho_1,\rho_2>0$ and let $H$ be a Hamiltonian on $[0,b)$ with $0<b\le\infty$ 
	such that $h_1(t)\asymp t^{\rho_1-1}$ and $m_2(t)\asymp t^{\rho_2}$ as $t\to0$
	and $\limsup_{t\to0}\frac{m_3(t)^2}{(m_1m_2)(t)}<1$.
	Let us consider the regularly function $\ms g(r)=r^\alpha(\log r)^\beta$ 
	with $\alpha\in(0,2)$ and $\beta\in\bb R$.
	It follows from \thref{Y86} and \thref{Y204} that $\mu_H\in\mc M_{\ms g}$
	if and only if
	\[
		\int_0^{a'}\frac{t^{\rho_1-1}}{t^{(\rho_1+\rho_2)(1-\frac{\alpha}{2})}|\log t|^\beta}\DD t < \infty,
	\]
	which, in turn, is equivalent to
	\[
		\alpha>\frac{2\rho_2}{\rho_1+\rho_2} \qquad\text{or}\qquad
		\biggl(\alpha=\frac{2\rho_2}{\rho_1+\rho_2} \quad\text{and}\quad \beta>1\biggr).
	\]
\end{example}

\subsection{Limit superior conditions}

In this section we investigate $\limsup$-conditions for the
quotient $\frac{\dblarrow\mu_H(r)}{\ms g(r)}$ instead of integrability conditions.
Let us introduce the corresponding classes of measures.

\begin{definition}\thlab{Y77}
	Let $\ms g(r)$ be a regularly varying function with index $\alpha\le 2$
	and $\lim_{r\to\infty}\ms g(r)=\infty$.
	Then we set
	\[
		\mc F_{\ms g} \DE \big\{\mu\DS \dblarrow\mu(r)=\BigO(\ms g(r))\big\},\qquad
		\mc F_{\ms g}^0 \DE \big\{\mu\DS \dblarrow\mu(r)=\Smallo(\ms g(r))\big\},
	\]
	where again $\dblarrow\mu(r)\DE\mu((-r,r))$.
\end{definition}

\noindent
Clearly, we have $\mc F_{\ms g}^0 \subseteq \mc F_{\ms g}$.
In the next proposition relations among $\mc F_{\ms g}$, $\mc F_{\ms g}^0$
and the classes $\mc M_{\ms g}$ and $\hatcalM_{\ms g}$ from Section~\ref{Y190} are discussed.

\begin{proposition}\thlab{Y76}
	Let	$\ms g$, $\ms g_1$ and $\ms g_2$ be continuous, regularly varying functions
	with indices $\alpha,\alpha_1,\alpha_2\le 2$ respectively, such that
	$\ms g(r),\ms g_1(r),\ms g_2(r)\to\infty$ as $r\to\infty$.
	\begin{Enumerate}
	\item
		If $\ms g$ is non-decreasing, then $\mc M_{\ms g} \subseteq \mc F_{\ms g}^0$.
	\item
		Assume that
		\begin{equation}\label{Y164}
			\alpha_2<2 \qquad\text{and}\qquad
			\int_1^\infty\frac{\ms g_1(r)}{r\ms g_2(r)}\DD r<\infty
		\end{equation}
		or that $\alpha_1<\alpha_2$.  Then $\mc F_{\ms g_1}\subseteq \hatcalM_{\ms g_2}$.
	\end{Enumerate}
\end{proposition}

\begin{proof}
	\phantom{}

	(i) Let $\mu\in\mc M_{\ms g}$, i.e.\
	$\int_1^\infty\frac 1{\ms g(t)}\DD\tilde\mu(t)<\infty$, where $\tilde\mu$
	is as in \eqref{Y140}.
	\thref{Y49} (with a measure $\nu$ such that $\nu((t,\infty))=1/\ms g(t)$) yields
	\[
		\lim_{r\to\infty}\frac{\dblarrow\mu(r)}{\ms g(r)}
		= \lim_{r\to\infty}\nu\bigl((r,\infty)\bigr)\tilde\mu\bigl([0,r)\bigr) = 0.
	\]

	(ii) Let $\mu\in\mc F_{\ms g_1}$.
	Then
	\begin{equation}\label{Y166}
		\dblarrow\mu(r)\frac{(\ms g_2)_\star(r)}{r^3}
		\lesssim \ms g_1(r)\frac{(\ms g_2)_\star(r)}{r^3}\,.
	\end{equation}
	First, we consider the case when \eqref{Y164} is satisfied.
	It follows from \eqref{Y137} that
	\[
		\ms g_1(r)\frac{(\ms g_2)_\star(r)}{r^3}
		\asymp \ms g_1(r)\frac{r^2}{\ms g_2(r)}\cdot\frac{1}{r^3}\,.
	\]
	This, together with \eqref{Y166} and the second relation in \eqref{Y164}, implies that
	$\int_1^\infty \dblarrow\mu(r)\frac{(\ms g_2)_\star(r)}{r^3}\DD r<\infty$
	and hence $\mu\in\hatcalM_{\ms g_2}$.
	Let us now assume that $\alpha_1<\alpha_2$.
	It follows from the sentence around \eqref{Y137}
	that $(\ms g_2)_\star$ is regularly varying with index $2-\alpha_2$.
	Hence the right-hand side of \eqref{Y166} is regularly varying with
	index $\alpha_1+2-\alpha_2-3<-1$ and therefore integrable by \thref{Y114}.
	It follows again that $\mu\in\hatcalM_{\ms g_2}$.
\end{proof}

\begin{remark}\thlab{Y165}
	\rule{0ex}{1ex}
	\begin{Enumerate}
	\item
		If $\alpha>0$ in \thref{Y76}\,(i), then the assumption that $\ms g$ is non-decreasing
		is not necessary because, by \thref{Y115},
		there exists a non-decreasing, regularly varying function $\bar{\ms g}$
		such that $\bar{\ms g}(r)\sim\ms g(r)$ as $r\to\infty$.
	\item
		The inclusion in \thref{Y76}\,(i) is the analogue on the level of measures
		to the fact that an entire function of convergence class is necessarily of
		minimal type.
	\item
		Let $\ms g_1,\ms g_2$ be non-decreasing functions as in \thref{Y76}\,(ii).
		We can combine items (i) and (ii) of \thref{Y76} with \thref{Y147}
		to obtain the following chain of inclusions:
		\begin{equation}\label{Y143}
			\mc M_{\ms g_1} \subseteq \mc F_{\ms g_1}^0 \subseteq \mc F_{\ms g_1}
			\subseteq \hatcalM_{\ms g_2} \subseteq \mc M_{\ms g_2}.
		\end{equation}
		This is satisfied, in particular, when $\ms g_i(r)=r^{\alpha_i}$, $i=1,2$,
		with $\alpha_1<\alpha_2$, or when $\ms g_i(r)=r^\alpha(\log r)^{\beta_i}$
		with $\beta_2>\beta_1+1$.
	\end{Enumerate}
\vspace*{-3ex}
\end{remark}

\noindent
The next theorem can be viewed as a generalisation of \cite[Theorem~4]{kasahara:1975}
to the case of a non-diagonal Hamiltonian; the connection is worked out in detail in
Section~\ref{Y91}.

\begin{theorem}\thlab{Y74}
	Let $H$ be a Hamiltonian defined on some interval $[a,b)$, and assume that
	neither $h_1$ nor $h_2$ vanishes on a neighbourhood of the left endpoint $a$.
	Let $\ms g(r)$ be a regularly varying function with index $\alpha\le 2$
	and $\lim_{r\to\infty}\ms g(r)=\infty$,
	and denote by $\mu_H$ the spectral measure of $H$.
	Then the following implications hold:
	\begin{Enumerate}
	\item
		${\displaystyle
		\limsup_{t\to a}\frac{m_1(t)}{(m_1m_2)(t)\ms g\bigl((m_1m_2)(t)^{-\frac12}\bigr)}<\infty
		\;\;\Rightarrow\;\;
		\mu_H\in\mc F_{\ms g}
		}$;
	\item
		${\displaystyle
		\lim_{t\to a}\frac{m_1(t)}{(m_1m_2)(t)\ms g\bigl((m_1m_2)(t)^{-\frac12}\bigr)}=0
		\;\;\Rightarrow\;\;
		\mu_H\in\mc F_{\ms g}^0
		}$.
	\end{Enumerate}
	If, in addition, $\alpha<2$, then
	\begin{Enumerate}
	\setcounter{enumi}{2}
	\item
		${\displaystyle
		\mu_H\in\mc F_{\ms g}
		\;\;\Rightarrow\;\;
		\limsup_{t\to a}\frac{\det M(t)\big/m_2(t)}{(m_1m_2)(t)\ms g\bigl((m_1m_2)(t)^{-\frac12}\bigr)}<\infty
		}$;
	\item
		${\displaystyle
		\mu_H\in\mc F_{\ms g}^0
		\;\;\Rightarrow\;\;
		\lim_{t\to a}\frac{\det M(t)\big/m_2(t)}{(m_1m_2)(t)\ms g\bigl((m_1m_2)(t)^{-\frac12}\bigr)}=0
		}$.
	\end{Enumerate}
\end{theorem}

\noindent
The proof of \thref{Y74} relies on the following Abelian--Tauberian-type result.
Most probably this fact is folklore, but we do not know an explicit reference and
therefore provide the proof.

\begin{lemma}\thlab{Y75}
	Let $\mu$ be a positive Borel measure on $\bb R$
	with $\int_{\bb R}\frac{\DD\mu(t)}{1+t^2}<\infty$,
	define $\dblarrow\mu(r)$ as in \eqref{Y140}, denote by $\Poi{\mu}(z)$
	the Poisson integral of $\mu$ as in \eqref{Y144},
	and let $\ms g(r)$ be regularly varying with index $\alpha\in[0,2]$.
	\begin{Enumerate}
	\item
		We have
		\begin{equation}\label{Y25}
			\limsup_{r\to\infty}\biggl(\frac r{\ms g(r)}\Poi{\mu}(ir)\biggr)
			\ge \Bigl(1-\frac{\alpha}{2}\Bigr)^{1-\frac{\alpha}{2}}
			\Bigl(\frac{\alpha}{2}\Bigr)^{\frac{\alpha}{2}}
			\cdot\limsup_{r\to\infty}\frac{\dblarrow\mu(r)}{\ms g(r)},
		\end{equation}
		where the constant in front of $\limsup$ on the right-hand side is
		interpreted as $1$ if $\alpha$ equals $0$ or $2$.
	\item
		If $\alpha<2$ and $\lim_{r\to\infty}\ms g(r)=\infty$, then
		\begin{equation}\label{Y26}
			\limsup_{r\to\infty}\biggl(\frac r{\ms g(r)}\Poi{\mu}(ir)\biggr)
			\le \B\Bigl(1+\frac{\alpha}2,1-\frac{\alpha}2\Bigr)
			\cdot\limsup_{r\to\infty}\frac{\dblarrow\mu(r)}{\ms g(r)},
		\end{equation}
		where $\B$ denotes Euler's beta function.
	\end{Enumerate}
\end{lemma}

\begin{proof}
	For the proof of \Enumref{1} observe that, for every $x>0$ and $r>0$,
	\begin{align*}
		&\hspace*{-5ex} \frac r{\ms g(r)}\Poi{\mu}(ir)
		= \frac r{\ms g(r)}\int_{\bb R}\frac r{t^2+r^2}\DD\mu(t)
		\ge \frac r{\ms g(r)}\int_{(-xr,xr)}\frac r{t^2+r^2}\DD\mu(t)
		\\[1ex]
		&\ge \frac r{\ms g(r)}\int_{(-xr,xr)}\frac r{(1+x^2)r^2}\DD\mu(t)
		= \frac{1}{1+x^2}\cdot\underbrace{\frac{\ms g(xr)}{\ms g(r)}}_{\to x^{\alpha}}
		\cdot\frac{\mu\big((-xr,xr)\big)}{\ms g(xr)}\,.
	\end{align*}
	If $\alpha\in(0,2)$, then the function $x\mapsto\frac{x^{\alpha}}{1+x^2}$
	attains the maximum at $x_0=\sqrt{\alpha/(2-\alpha)}$;
	with this $x_0$ the above inequality yields \eqref{Y25}.
	For $\alpha=0$ we use arbitrarily small $x$,
	and for $\alpha=2$ we use $x$ that are arbitrarily close to 2.

	We come to the proof of \Enumref{2}.
	Let $\tilde\mu$ be the measure defined in \eqref{Y140}.
	For every $r_0>0$ we estimate as follows (where we use \thref{Y49}):
	\begin{align*}
		\Poi{\mu}(ir) &= \int_{\bb R}\frac r{t^2+r^2}\DD\mu(t)
		= \int_{[0,\infty)}\frac r{t^2+r^2}\DD\tilde\mu(t)
		\\[1ex]
		&= r\int_0^\infty\underbrace{\tilde\mu\big([0,t)\big)}_{=\dblarrow\mu(t)}\cdot\frac{2t}{(t^2+r^2)^2}\DD t
		\\
		&= r\int_0^{r_0}\dblarrow\mu(t)\cdot\frac{2t}{(t^2+r^2)^2}\DD t
		+ r\int_{r_0}^\infty\frac{\dblarrow\mu(t)}{\ms g(t)}\cdot\frac{2t\ms g(t)}{(t^2+r^2)^2}\DD t
		\\[1ex]
		&\le \frac{2r_0}{r^3}\int_0^{r_0}\dblarrow\mu(t)\DD t
		+ 2r\Big(\sup_{t\ge r_0}\frac{\dblarrow\mu(t)}{\ms g(t)}\Big)
		\cdot\int_{r_0}^\infty\frac{t\ms g(t)}{(t^2+r^2)^2}\DD t.
	\end{align*}
	The first summand tends to $0$ when multiplied by $\frac r{\ms g(r)}$
	and hence does not contribute to the limit superior on the left-hand side
	of \eqref{Y26}.
	The integral in the second summand is estimated by
	\[
		\int_{r_0}^\infty\frac{t\ms g(t)}{(t^2+r^2)^2}\DD t
		\le \int_0^\infty\frac{t\ms g(t)}{(t^2+r^2)^2}\DD t
		= r^{\alpha-2}\int_0^\infty\frac{x^{1+\alpha}}{(x^2+1)^2}
		\cdot\frac{\ms g(rx)}{(rx)^{\alpha}}\DD x.
	\]
	Since $\alpha<2$, we can apply \cite[Theorems~2.6, 2.7]{seneta:1976} and obtain
	\[
		\lim_{r\to\infty}\bigg(
		\raisebox{2pt}{${\displaystyle
		\int_0^\infty\frac{x^{1+\alpha}}{(x^2+1)^2}
		\cdot\frac{\ms g(rx)}{(rx)^{\alpha}}\DD x
		}$}
		\bigg/
		\raisebox{-2pt}{${\displaystyle
		\frac{\ms g(r)}{r^{\alpha}}
		}$}
		\bigg)
		= \int_0^\infty\frac{x^{1+\alpha}}{(x^2+1)^2}\DD x
		= \frac 12\B\Bigl(1+\frac{\alpha}2,1-\frac{\alpha}2\Bigr).
	\]
	Putting these estimates together we obtain
	\begin{align*}
		\limsup_{r\to\infty}\biggl(\frac r{\ms g(r)}\Poi{\mu}(ir)\biggr)
		&\le \Bigl(\sup_{t\ge r_0}\frac{\dblarrow\mu(t)}{\ms g(t)}\Bigr)
		\limsup_{r\to\infty}\biggl[\frac{2r^2}{\ms g(r)}
		\int_{r_0}^\infty \frac{t\ms g(t)}{(t^2+r^2)^2}\DD t\biggr]
		\\[1ex]
		&\le \Bigl(\sup_{t\ge r_0}\frac{\dblarrow\mu(t)}{\ms g(t)}\Bigr)
		\limsup_{r\to\infty}\biggl[\frac{2r^{\alpha}}{\ms g(r)}
		\int_0^\infty \frac{x^{1+\alpha}}{(x^2+1)^2}
		\cdot\frac{\ms g(rx)}{(rx)^{\alpha}}\DD x\biggr]
		\\[1ex]
		&= \Bigl(\sup_{t\ge r_0}\frac{\dblarrow\mu(t)}{\ms g(t)}\Bigr)
		\B\Bigl(1+\frac{\alpha}2,1-\frac{\alpha}2\Bigr);
	\end{align*}
	since $r_0$ can be chosen arbitrarily large, \eqref{Y26} follows.
\end{proof}

\noindent
Note that for $\alpha>2$ both $\limsup$ appearing in \thref{Y75} are equal to $0$, and
for $\alpha<0$ both are equal to $+\infty$ (unless $\mu=0$).

\begin{proof}[Proof of \thref{Y74}]
	\thref{Y98} with $A(r)$ in the form \eqref{Y36}, together with \eqref{Y25},
	implies that there exist $c_1,c_2>0$ such that
	\[
		\limsup_{r\to\infty}\frac{\dblarrow\mu_H(r)}{\ms g(r)}
		\le c_1 \limsup_{r\to\infty}\biggl(\frac{r}{\ms g(r)}\Im q_H(ir)\biggr)
		\le c_2 \limsup_{r\to\infty}\biggl(\frac{r^2}{\ms g(r)}m_1\bigl(\mr t(r)\bigr)\biggr).
	\]
	With the substitution $t=\mr t(r)$ the last $\limsup$ can be rewritten as
	\begin{align*}
		\limsup_{r\to\infty}\biggl(\frac{r^2}{\ms g(r)}m_1\bigl(\mr t(r)\bigr)\biggr)
		&= \limsup_{t\to a}
		\frac{\bigl(\frac{\eta}{2}(m_1m_2)(t)^{-\frac12}\bigr)^2}{\ms g\bigl(\frac{\eta}{2}(m_1m_2)(t)^{-\frac12}\bigr)}m_1(t)
		\\
		&= \frac{(\frac{\eta}{2})^2}{(\frac{\eta}{2})^{\alpha}}
		\limsup_{t\to a}\frac{m_1(t)}{(m_1m_2)(t)\ms g\bigl((m_1m_2)(t)^{-\frac12}\bigr)}\,.
	\end{align*}
	Items \Enumref{3} and \Enumref{4} are shown in a similar way,
	when \thref{Y98} with $L(r)$ in the form \eqref{Y120} and \eqref{Y26} are used.
\end{proof}

\noindent
Analogously to \thref{Y47}, we obtain the obvious corollary.

\begin{corollary}\thlab{Y70}
	Consider the situation from \thref{Y74}, and assume in addition that
	$\limsup_{t\to a}\frac{m_3(t)^2}{(m_1m_2)(t)}<1$
	\textup{(}this holds in particular if $H$ is diagonal\textup{)}.
	Moreover, let $\alpha\in[0,2)$.  Then
	\begin{align*}
		\mu_H\in\mc F_{\ms g}
		& \;\;\Leftrightarrow\;\;
		\limsup_{t\to a}\frac{m_1(t)}{(m_1m_2)(t)\ms g\bigl((m_1m_2)(t)^{-\frac 12}\bigr)}
		< \infty,
		\\[1ex]
		\mu_H\in\mc F_{\ms g}^0
		& \;\;\Leftrightarrow\;\;
		\lim_{t\to a}\frac{m_1(t)}{(m_1m_2)(t)\ms g\bigl((m_1m_2)(t)^{-\frac 12}\bigr)}=0.
	\end{align*}
\end{corollary}

\noindent
We finish this section with two examples, the first of which is also used in Section~\ref{Y194}.

\begin{example}\thlab{Y167}
	Let $\rho_1,\rho_2>0$ and let $H$ be a Hamiltonian on $[0,b)$ with $0<b\le\infty$
	such that $m_j(t)\asymp t^{\rho_j}$, $j=1,2$, as $t\to 0$
	and $\limsup_{t\to 0}\frac{m_3(t)^2}{(m_1m_2)(t)}<1$.
	Let us consider $\ms g_\alpha(r)\DE r^\alpha$ with $\alpha\in(0,2)$.  Then
	\[
		\frac{m_1(t)}{(m_1m_2)(t)\ms g_\alpha\bigl((m_1m_2)(t)^{-\frac 12}\bigr)}
		= m_1(t)^{\frac{\alpha}{2}}m_2(t)^{\frac{\alpha}{2}-1}
		\asymp t^{\rho_1\frac{\alpha}{2}+\rho_2(\frac{\alpha}{2}-1)}.
	\]
	It follows from \thref{Y70} that
	\begin{align*}
		\mu_H\in\mc F_{\ms g_\alpha}\setminus\mc F_{\ms g_\alpha}^0
		\quad&\Leftrightarrow\quad
		0<\limsup_{t\to 0}
		\frac{m_1(t)}{(m_1m_2)(t)\ms g_\alpha\bigl((m_1m_2)(t)^{-\frac 12}\bigr)}
		<\infty
		\\[1ex]
		&\Leftrightarrow\quad \rho_1\frac{\alpha}{2}+\rho_2\Bigl(\frac{\alpha}{2}-1\Bigr)=0
		\quad\Leftrightarrow\quad \alpha=\frac{2\rho_2}{\rho_1+\rho_2}\,.
	\end{align*}
	Now \eqref{Y143} implies that
	\[
		\mu_H \in \mc M_{\ms g_\alpha} \quad\Leftrightarrow\quad
		\alpha > \frac{2\rho_2}{\rho_1+\rho_2}\,.
	\]
	Moreover, we have $\mu_H\in\mc F_{\ms g_\alpha}^0$
	if $\alpha>\frac{2\rho_2}{\rho_1+\rho_2}$, and
	$\mu_H\notin\mc F_{\ms g_\alpha}$ if $\alpha<\frac{2\rho_2}{\rho_1+\rho_2}$.
\end{example}

\medskip

\begin{example}\thlab{Y206}
	Let $\rho>0$ and let $H$ be a Hamiltonian on $[0,b)$ with $0<b\le\infty$
	such that $m_1(t)\asymp e^{-\frac{1}{t}}$ and $m_2(t)\asymp t^\rho$ as $t\to0$
	and $\limsup_{t\to 0}\frac{m_3(t)^2}{(m_1m_2)(t)}<1$.
	We choose the slowly varying function $\ms g(r) = (\log r)^\beta$ with $\beta\in\bb R$.
	Since
	\[
		\frac{m_1(t)}{(m_1m_2)(t)\ms g\bigl((m_1m_2)(t)^{-\frac 12}\bigr)}
		\asymp \frac{e^{-\frac{1}{t}}}{e^{-\frac{1}{t}}t^\rho\big|\log\bigl(e^{-\frac{1}{t}}t^\rho\bigr)\big|^\beta}
		\sim t^{\beta-\rho}
	\]
	as $t\to0$, it follows from \thref{Y70} that $\mu_H\in\mc F_{\ms g}\setminus\mc F_{\ms g}^0$
	if and only if $\beta=\rho$.
\end{example}

\section{Further discussion of the main theorem}
\label{Y34}
\subsection{A monotonicity property}
\label{Y78}

In many situations one cannot determine the solution $\mr t(r)$ of \eqref{Y30}
exactly.  Having a lower bound can often be sufficient to obtain
estimates for $A(r)$ and $L(r)$.
With the following lemma we can easily prove \thref{Y121} below.

\begin{lemma}\thlab{Y79}
	Let $H$ be as in \thref{Y98}, and let $r>0$.  If $\hat t\in(\mr a,\mr t(r))$, then
	\[
		\frac{2}{\eta}\cdot rm_1(\hat t) \le A(r) \le \frac{\eta}{2}\cdot \frac 1{rm_2(\hat t)},\qquad
		\frac{2}{\eta}\cdot \frac{r\det M(\hat t)}{m_2(\hat t)} \le L(r).
	\]
	If $\check t\in(\mr t(r),b)$, then
	\[
		\frac{\eta}{2}\cdot \frac{1}{rm_2(\check t)} \le A(r) \le \frac{2}{\eta}\cdot rm_1(\check t),\qquad
		L(r) \le \frac{2}{\eta}\cdot \frac{r\det M(\check t)}{m_2(\check t)}.
	\]
\end{lemma}

\begin{proof}
	Since $m_1$ and $m_2$ are non-decreasing functions, the assertions involving $A(r)$
	are obvious from \eqref{Y36}.
	To show the assertion involving $L(r)$, we use the middle term in \eqref{Y120}
	and the monotonicity of $\frac{\det M}{m_2}$, shown in \eqref{Y71}.
\end{proof}

\noindent
The following statement is now obvious.

\begin{corollary}\thlab{Y121}
	Let $H$ be a Hamiltonian defined on some interval $[a,b)$, and assume that
	neither $h_1=0$ a.e.\ nor $h_2=0$ a.e.

	Let $\hat t\DF(0,\infty)\to(\mr a,b)$ be a function such that for all $r>0$
	\begin{equation}\label{Y149}
		\frac{\eta}{2}(m_1m_2)^{-\frac 12}\bigl(\hat t(r)\bigr) \ge r.
	\end{equation}
	Then, for each $\vartheta\in(0,\pi)$,
	\begin{align*}
		rm_1(\hat t(r))
		\lesssim \big|q_H\bigl(re^{i\vartheta}\bigr)\big|
		&\lesssim \frac{1}{rm_2(\hat t(r))},
		\\
		\big|\Re q_H\bigl(re^{i\vartheta}\bigr)\big|
		&\lesssim \frac 1{rm_2(\hat t(r))},
		\\
		\frac{r\det M(\hat t(r))}{m_2(\hat t(r))}\lesssim
		\Im q_H\bigl(re^{i\vartheta}\bigr)
		&\lesssim \frac 1{rm_2(\hat t(r))}.
	\end{align*}
	The various constants in ``\,$\lesssim$'' depend on $\vartheta$ but not on $H$.
\end{corollary}

\noindent
Note that the validity of \eqref{Y149} just means that the points $(r,\hat t(r))$
belong to the dotted region in Figure~\ref{Y37}.

\subsection{Hamiltonians starting with an indivisible interval of type $\bm{0}$ or $\bm{\pi/2}$}
\label{Y127}

For $\phi\in\bb R$ we set
\[
	\xi_\phi\DE\binom{\cos\phi}{\sin\phi}.
\]

\begin{definition}\thlab{Y125}
	Let $H$ be a Hamiltonian defined on some interval $[a,b)$.
	An interval $(c,d)\subseteq(a,b)$ is called \emph{$H$-indivisible}
	if there exists $\phi\in\bb R$ such that
	\[
		H(t) = \tr H(t)\cdot\xi_\phi\xi_\phi^*,\qquad t\in(c,d)\text{ a.e.}
	\]
	Equivalently, one may say that $J\xi_\phi=(-\sin\phi,\cos\phi)^*\in\ker H(t)$
	for $t\in(c,d)$ a.e.

	The number $\phi$ is uniquely determined up to integer multiples of $\pi$;
	it will always be understood modulo $\pi$,
	and is called the \emph{type} of the indivisible interval $(c,d)$.
\end{definition}

\noindent
The property that $\mr a>a$ can now be formulated as:
$H$ does not start with an indivisible interval of type $0$ or $\frac\pi2$.

Although we can apply \thref{Y98} in the situation when $\mr a>a$,
we can also split off an interval of type $0$ or $\frac\pi2$ adjacent to $a$
to get an asymptotic expansion with a leading-order term, which is a power of $z$
with exponent $1$ or $-1$, and an estimate for the remainder term.

\begin{remark}\thlab{Y54}
	Assume that $H$ starts with an indivisible interval of type $0$, i.e.\
	let $(a,\mr a)$ with $\mr a>a$ be the maximal interval such that the
	Hamiltonian $H$ is of the form
	\[
		H(t) = \begin{pmatrix} h_1(t) & 0 \\ 0 & 0 \end{pmatrix},
		\qquad t\in(a,\mr a)\text{ a.e.}
	\]
	Then
	\begin{equation}\label{Y197}
		\lim_{r\to\infty}\frac{1}{ir}q_H(ir) = m_1(\mr a)
		= \int_a^{\mr a}h_1(t)\DD t = \beta_H,
	\end{equation}
	where $\beta_H$ is as in \eqref{Y156}; see, e.g.\ \cite{kac.krein:1968a,winkler:1995}.
	Nevertheless, \thref{Y98} provides additional information as it yields
	explicit bounds.  Clearly, \eqref{Y197} implies that $A(r)\asymp 1$, $r\to\infty$,
	which can also be seen directly from the behaviour of the functions $m_i$.
	Further, we also obtain a good lower bound for the imaginary part, which is
	seen as follows.  For $t>\mr a$ we have
	\begin{align*}
		m_3(t)^2 &= \bigg|\int_{\mr a}^t h_3(s)\DD s\bigg|^2
		\le \biggl[\int_{\mr a}^t |h_3(s)|\DD s\biggr]^2
		\le \biggl[\int_{\mr a}^t \sqrt{h_1(s)h_2(s)}\DD s\biggr]^2
		\\[1ex]
		&\le \int_{\mr a}^t h_1(s)\DD s\cdot \int_{\mr a}^t h_2(s)\DD s
		= \bigl(m_1(t)-m_1(\mr a)\bigr)m_2(t),
	\end{align*}
	which implies
	\begin{align*}
		1 &\ge \limsup_{t\searrow\mr a}\frac{\det M(t)}{m_1(t)m_2(t)}
		\ge \liminf_{t\searrow\mr a}\frac{\det M(t)}{m_1(t)m_2(t)}
		\\[1ex]
		&\ge \liminf_{t\searrow\mr a}
		\frac{m_1(t)m_2(t)-\bigl(m_1(t)-m_1(\mr a)\bigr)m_2(t)}{m_1(t)m_2(t)}
		= \liminf_{t\searrow\mr a}\frac{m_1(\mr a)}{m_1(t)} = 1.
	\end{align*}
	From this we obtain $L(r)\sim A(r)$, $r\to\infty$.

	On the other hand, we can also split off the indivisible interval;
	namely, one can show that
	\[
		q_H(z) = m_1(\mr a)z + q_{\wh H}(z)
	\]
	where $\wh H=H|_{(\mr a,b)}$, and then apply \thref{Y98} to obtain an estimate
	for the remainder term.

	Note that the case when $h_2=0$ a.e.\ on $(a,\mr a)$ with $\mr a>a$
	corresponds to $\mu_H$ including a ``point mass at infinity''.
\end{remark}

\begin{remark}\thlab{Y126}
	Assume now that $H$ starts with an indivisible interval of type $\frac\pi2$,
	i.e.\ let $(a,\mr a)$ with $\mr a>a$ be the maximal interval such that the
	Hamiltonian $H$ is of the form
	\[
		H(t) = \begin{pmatrix} 0 & 0 \\ 0 & h_2(t) \end{pmatrix},
		\qquad t\in(a,\mr a)\text{ a.e.}
	\]
	Then the representation in \eqref{Y156} can be rewritten
	as $q_H(z)=\int_{\bb R}\frac{1}{t-z}\DD\mu_H(t)$ with the finite measure $\mu_H$,
	and
	\[
		\lim_{r\to\infty}\bigl(-irq_H(ir)\bigr) = \frac{1}{m_2(\mr a)}
		= \mu_H(\bb R).
	\]
	Again \thref{Y98} can be applied to obtain explicit bounds, and as in \thref{Y54},
	we have $L(r)\sim A(r)$.  One can also split off the indivisible interval,
	namely, $q_H$ can be rewritten as
	\[
		q_H(z) = -\frac{1}{m_2(\mr a)z - \frac{1}{q_{\check H}(z)}\,},
	\]
	where $\check H=H|_{(\mr a,b)}$, and one can apply \thref{Y98} to $\check H$
	in order to obtain an estimate for the remainder term.
\end{remark}

\noindent
Assume we are given a Hamiltonian $H$ which starts with a finite number of
consecutive indivisible intervals whose types alternate between $0$ and $\frac\pi2$.
Iterating the splitting-off procedure from Remarks~\ref{Y54} and \ref{Y126}
we obtain a representation of $q_H$ as a continued fraction of finite length
and a certain remainder term to which \thref{Y98} can be applied.
Thus, we obtain information about the size of the remainder.

\subsection{A rotation transformation}
\label{Y173}

Sometimes it is possible to improve the bounds \eqref{Y96} by applying a
transformation to $H$.
In this subsection we consider the situation when $q_H$ has a real, non-zero limit.
More specifically, let us assume that
\begin{equation}\label{Y152}
	\lim_{r\to\infty}q_H(ir) = \cot\phi
\end{equation}
with some $\phi\in(0,\frac\pi2)\cup(\frac\pi2,\pi)$.
Obviously we have
\[
	A(r)\asymp|q_H(ir)|\asymp 1,\qquad
	L(r)\lesssim\Im q_H(ir)\ll 1,
\]
and hence \eqref{Y96} is certainly not strong enough to fully (i.e.\ up to universal constants)
determine $\Im q_H(ir)$.
The following lemma shows that in this situation a transformation, where the
dependent variable in \eqref{Y99} is rotated in the two-dimensional space $\bb C^2$,
strictly improves the upper bound for the imaginary part of $q_H$ and that
the lower bound does not get worse.
In \thref{Y168} below the transformation is made more explicit in terms
of the Hamiltonian $H$.

\begin{lemma}\thlab{Y62}
	Let $H$ be a Hamiltonian such that \eqref{Y152}
	with $\phi\in(0,\frac\pi2)\cup(\frac\pi2,\pi)$ is satisfied.  Set
	\begin{equation}\label{Y170}
		Q \DE \begin{pmatrix} \sin\phi & -\cos\phi \\[0.5ex]
		\cos\phi & \sin\phi \end{pmatrix},\qquad
		\widetilde H\DE QHQ^T.
	\end{equation}
	Then
	\begin{equation}\label{Y198}
		\Im q_{\widetilde H}(ir) \asymp \Im q_H(ir),\qquad
		\widetilde A(r) \ll A(r), \qquad
		\widetilde L(r) \gtrsim L(r)
	\end{equation}
	as $r\to\infty$.
\end{lemma}

\begin{proof}
	The Weyl coefficients of $H$ and $\widetilde H$ are related via
	\[
		q_{\widetilde H}(z) = \frac{\sin\phi\cdot q_H(z)-\cos\phi}{\cos\phi\cdot q_H(z)+\sin\phi}\,,
		\qquad
		\Im q_{\widetilde H}(z)=\frac{\Im q_H(z)}{|\cos\phi\cdot q_H(z)+\sin\phi|^2};
	\]
	see, e.g.\ \cite[(3.20)]{gesztesy.tsekanovskii:2000}.
	Thus we have
	\[
		\lim_{r\to\infty}q_{\widetilde H}(ir)=0,\qquad
		\lim_{r\to\infty}\frac{\Im q_{\widetilde H}(ir)}{\Im q_H(ir)}=\sin^2\phi>0,
	\]
	and therefore, in particular,
	\[
		\Im q_{\widetilde H}(ir)\asymp\Im q_H(ir),\qquad
		\widetilde A(r)\asymp|q_{\widetilde H}(ir)|\ll|q_H(ir)|\asymp A(r),
	\]
	which proves the first two relations in \eqref{Y198}.

	The proof of the third relation in \eqref{Y198} requires slightly more effort.
	Let us first consider the case when $H$ starts with an indivisible interval,
	say $(a,c)$, which must be of type $\phi$.  Since $\phi\notin\{0,\pi/2\}$,
	we have $\mr a=a$, where $\mr a$ is as in \eqref{Y196}.
	Clearly, $\det M(t)=0$ for $t\in[a,c]$, which shows that $L(r)=0$
	for large enough $r$; so the third relation in \eqref{Y198}
	is trivially satisfied in this case.

	Now let us assume that $H$ does not start with an indivisible interval.
	Then $\widetilde H$ does not start with an indivisible interval either,
	and hence $\mr{\tilde a}=a$, where $\mr{\tilde a}$ is as in \eqref{Y196}
	for $H$ replaced by $\widetilde H$.
	First, note that
	\[
		\det\tildM(t)=\det M(t),\qquad \tr\tildM(t)=\tr M(t),
	\]
	and, since $A(r)\asymp 1$ and $\widetilde A(r)\ll 1$, we have
	\[
		m_1(t)\asymp m_2(t),\qquad \widetilde m_1(t)\ll\widetilde m_2(t).
	\]
	Together, it follows that
	\[
		m_1(t) \asymp m_2(t) \asymp \tr M(t) \asymp \widetilde m_2(t) \gg \widetilde m_1(t),
	\]
	and hence
	\[
		\mr r(t) = \frac{\eta}{2}\bigl(m_1(t)m_2(t)\bigr)^{-\frac 12}
		\ll \frac{\eta}{2}\bigl(\widetilde m_1(t)\widetilde m_2(t)\bigr)^{-\frac 12}
		= \mr{\tilde r}(t),
	\]
	where $\mr r$ and $\mr{\tilde r}$ are as in \eqref{Y124}.
	In particular, $\mr r(t)\le\mr{\tilde r}(t)$ for small enough $t$,
	and therefore $\mr t(r)\le\mr{\tilde t}(r)$ for large enough $r$,
	say, $r\ge r_0$.

	Now we use the monotonicity property in \eqref{Y71} and the representation \eqref{Y120}
	of $L(r)$ and $\widetilde L(r)$ to obtain
	\[
		L(r) = \frac{2r}{\eta}\cdot\frac{\det M(\mr t(r))}{m_2(\mr t(r))}
		\le \frac{2r}{\eta}\cdot\frac{\det M(\mr{\tilde t}(r))}{m_2(\mr{\tilde t}(r))}
		\asymp \frac{2r}{\eta}\cdot\frac{\det\tildM(\mr{\tilde t}(r))}{\widetilde m_2(\mr{\tilde t}(r))}
		= \widetilde L(r)
	\]
	for $r\ge r_0$.
\end{proof}

\noindent
According to \cite[Corollary~3.2]{eckhardt.kostenko.teschl:2018}, relation \eqref{Y152}
holds if and only if
\begin{equation}\label{Y191}
	\lim_{t\to a}\frac{1}{m_1(t)+m_2(t)}
	\begin{pmatrix} m_1(t) & m_3(t) \\[0.5ex] m_3(t) & m_2(t) \end{pmatrix}
	= \begin{pmatrix} c_1 & c_3 \\[0.5ex] c_3 & c_2 \end{pmatrix}
\end{equation}
with $c_1,c_2>0$, $c_3\in\bb R$ such that $c_1c_2-c_3^2=0$ and $\frac{c_3}{c_2}=\cot\phi$
(note that the Weyl coefficient is defined slightly differently
in \cite{eckhardt.kostenko.teschl:2018}).
We can use this fact to obtain the following proposition, which provides a
transformation that is in terms of the asymptotic behaviour of $M$ at $a$.

\begin{proposition}\thlab{Y168}
	Let $H$ be a Hamiltonian defined on some interval $[a,b)$, let $M$
	be as in \eqref{Y08}, and assume that \eqref{Y191} holds with
	$c_1,c_2>0$, $c_3\in\bb R$ such that $c_1c_2-c_3^2=0$.
	Then there exist $Q$ and $\widetilde H$ as in \eqref{Y170} such that
	$\tildM(t)\DE\int_a^t \widetilde H(s)\DD s$ satisfies
	\begin{equation}\label{Y171}
		\tildM =
		\begin{pmatrix}
			c_2m_1+c_1m_2-2c_3m_3 & c_3(m_1-m_2)+(c_2-c_1)m_3
			\\[1ex]
			c_3(m_1-m_2)+(c_2-c_1)m_3 & c_1m_1+c_2m_2+2c_3m_3
		\end{pmatrix}.
	\end{equation}
	Hence
	\begin{equation}\label{Y172}
		L(r) \lesssim \widetilde L(r) \lesssim \Im q_H(ir) \asymp \Im q_{\wt H}(ir)
		\lesssim \widetilde A(r) \ll A(r).
	\end{equation}
\end{proposition}

\begin{proof}
	Let $\phi\in(0,\frac\pi2)\cup(\frac\pi2,\pi)$ such that $\frac{c_3}{c_2}=\cot\phi$.
	It follows from \eqref{Y191} that $c_1+c_2=1$.
	Hence
	\[
		c_2^2(1-\sin^2\phi) = c_2^2\cos^2\phi = c_3^2\sin^2\phi = c_2(1-c_2)\sin^2\phi,
	\]
	which yields $c_2=\sin^2\phi$.
	Since $\sin\phi>0$, we arrive at $\sin\phi=\sqrt{c_2}$
	and $\cos\phi=\frac{c_3}{\sqrt{c_2}\,}$.
	Applying the transformation \eqref{Y170} to $H$ we obtain $\widetilde H$,
	whose primitive $\tildM$ satisfies
	\begin{align*}
		\tildM &= QMQ^T
		=
		\begin{pmatrix}
			\sqrt{c_2} & -\frac{c_3}{\sqrt{c_2}\,}
			\\[1ex]
			\frac{c_3}{\sqrt{c_2}\,} & \sqrt{c_2}
		\end{pmatrix}
		\begin{pmatrix}
			m_1 & m_3 \\[1ex] m_3 & m_2
		\end{pmatrix}
		\begin{pmatrix}
			\sqrt{c_2} & \frac{c_3}{\sqrt{c_2}\,}
			\\[1ex]
			-\frac{c_3}{\sqrt{c_2}\,} & \sqrt{c_2}
		\end{pmatrix}
		\\[1ex]
		&=
		\begin{pmatrix}
			c_2m_1-2c_3m_3+\frac{c_3^2}{c_2}m_2
			& c_3m_1-\frac{c_3^2}{c_2}m_3+c_2m_3-c_3m_2
			\\[1ex]
			c_3m_1+c_2m_3-\frac{c_3^2}{c_2}m_3-c_3m_2
			& \frac{c_3^2}{c_2}m_1+2c_3m_3+c_2m_2
		\end{pmatrix}.
	\end{align*}
	Using again $c_1c_2=c_3^2$ we see that this is equal to the right-hand side
	of \eqref{Y171}.
\end{proof}

\noindent
Let us consider an example where this transformation trick is beneficial.

\begin{example}\thlab{Y65}
	Let $H$ be a Hamiltonian on $[0,\infty)$ whose primitive is
	\[
		M(t)
		=
		\begin{pmatrix}
			4t+t^\gamma & 2t+t^\gamma \\[1ex] 2t+t^\gamma & t+t^\gamma
		\end{pmatrix}
	\]
	with some $\gamma>1$.
	Let us first determine the bounds we obtain when we apply \thref{Y98} directly.
	Since $(m_1m_2)(t)\sim 4t^2$, we have $\mr t(r)\asymp \frac1r$.
	Further,
	\[
		\sqrt{\frac{m_1(t)}{m_2(t)}} \to 2,
		\qquad
		\frac{\det M(t)}{(m_1m_2)(t)}
		= \frac{t^{\gamma+1}}{4t^2+5t^{\gamma+1}+t^{2\gamma}}
		\sim \frac{1}{4}t^{\gamma-1},
	\]
	and hence
	\[
		A(r) \asymp 1, \qquad
		L(r) \asymp \frac{\det M(\mr t(r))}{(m_1m_2)(\mr t(r))}
		\asymp r^{-(\gamma-1)}.
	\]
	Let us now apply the transformation in \thref{Y168}.
	Since $m_1(t)+m_2(t)\sim 5t$, we obtain $c_1=\frac45$, $c_2=\frac15$ and $c_3=\frac25$.
	According to \eqref{Y171} the primitive of the transformed Hamiltonian
	$\widetilde H$ is
	\begin{align*}
		\tildM
		&=
		\begin{pmatrix}
			\frac15(4t+t^\gamma)+\frac45(t+t^\gamma)-\frac45(2t+t^\gamma) \hspace*{-2ex}
			& \frac25\!\times\!3t-\frac35(2t+t^\gamma)
			\\[2ex]
			\frac25\!\times\!3t-\frac35(2t+t^\gamma)
			& \hspace*{-2ex} \frac45(4t+t^\gamma)+\frac15(t+t^\gamma)+\frac45(2t+t^\gamma)
		\end{pmatrix}
		\\[1ex]
		&=
		\begin{pmatrix}
			\frac{1}{5}t^\gamma & -\frac{3}{5}t^\gamma
			\\[2ex]
			-\frac{3}{5}t^\gamma & 5t+\frac{9}{5}t^\gamma
		\end{pmatrix}.
	\end{align*}
	Since $(\wt m_1\wt m_2)\sim t^{\gamma+1}$, we have
	$\mr{\tilde t}(r)\asymp r^{-\frac{2}{\gamma+1}}$.
	Further, the relations
	\[
		\sqrt{\frac{\wt m_1(t)}{\wt m_2(t)}} \sim \frac{1}{5}t^{\frac{\gamma-1}{2}},
		\qquad
		\frac{\det\tildM(t)}{(\wt m_1\wt m_2)(t)}
		= \frac{t^{\gamma+1}}{t^{\gamma+1}+\frac{9}{25}t^{2\gamma}} \to 1
	\]
	and \eqref{Y172} yield
	\[
		\Im q_H(ir) \asymp \wt A(r) \asymp \wt L(r) \asymp r^{-\frac{\gamma-1}{\gamma+1}}.
	\]
	Hence, in this example the actual asymptotic behaviour of $\Im q_H(ir)$ lies
	strictly between $L(r)$ and $A(r)$:
	\[
		\underset{\scriptsize\begin{array}{c}\mkern8mu\vphantom{\rule{0pt}{7pt}}\begin{rotate}{90}$\asymp$\end{rotate} \\
		r^{-(\gamma-1)}\end{array}}{L(r)}
		\ll
		\underset{\scriptsize\begin{array}{c}\mkern8mu\vphantom{\rule{0pt}{7pt}}\begin{rotate}{90}$\asymp$\end{rotate} \\
		r^{-\frac{\gamma-1}{\gamma+1}}\end{array}}{\Im q_H(ir)}
		\ll
		\underset{\scriptsize\begin{array}{c}\mkern8mu\vphantom{\rule{0pt}{7pt}}\begin{rotate}{90}$\asymp$\end{rotate} \\
		1\end{array}}{A(r)}.
	\]
\end{example}

\begin{remark}\thlab{Y68}
	\thref{Y65} also shows that the lower bound for $\Im q_H(ir)$ from \eqref{Y96}
	may be far too small.  However, this particular example is in a sense not proper
	since it occurred only because the Hamiltonian is ``turned in the wrong direction''
	(thinking of matrices $Q$ as above as rotation matrices).
	It indicates that the crux for finding out whether or not
	we may have $L(r)\ll\Im q_H(ir)$ in an intrinsic and essential manner,
	is to understand the situation when $q_H(ir)$ tends to $0$.

	So far we have no example of a Hamiltonian $H$ such that $\lim_{r\to\infty}q_H(ir)=0$
	and $L(r)\ll\Im q_H(ir)\ll A(r)$.
	We expect that there exist large classes of Hamiltonians with these properties.
\end{remark}

\subsection{Examples}
\label{Y193}

Let us consider more examples, in particular, such where the Hamiltonian oscillates
in a neighbourhood of the left endpoint.

\begin{example}\thlab{Y44}
	Let $\phi\in(0,\frac\pi2)$, let $(0,\infty)=I_+\cup I_-$ be a partition
	into two disjoint measurable sets, and consider on $(0,\infty)$ the Hamiltonian
	(recall the notation $\xi_\phi\DE(\cos\phi,\sin\phi)^*$)
	\begin{equation}\label{Y52}
		H(t)\DE
		\begin{cases}
			\xi_\phi\xi_\phi^*, \quad & t \in I_+,
			\\[2mm]
			\xi_{-\phi}\xi_{-\phi}^*, \quad & t \in I_-.
		\end{cases}
	\end{equation}
	Let $\lambda$ denote the Lebesgue measure, and set
	\[
		l_+(t) \DE \lambda\bigl(I_+\cap(0,t)\bigr),\qquad
		l_-(t) \DE \lambda\bigl(I_-\cap(0,t)\bigr)
	\]
	for $t>0$.  Note that $l_+(t)+l_-(t)=t$.  It follows easily that
	\[
		m_1(t) = t\cos^2\phi, \quad m_2(t) = t\sin^2\phi, \quad
		m_3(t) = \bigl(l_+(t)-l_-(t)\bigr)\cos\phi\sin\phi.
	\]
	Let $\eta\in(0,1-\frac1{\sqrt2\,})$.  Then
	\[
		A(r) = \sqrt{\frac{m_1(\mr t(r))}{m_2(\mr t(r))}\,}
		= \cot\phi\in(0,\infty).
	\]
	Further, we have
	\[
		\mr r(t) = \frac{\eta}{2}(t\cos\phi\sin\phi)^{-1}
		= \frac{\alpha}{t},
	\]
	with
	\begin{equation}\label{Y150}
		\alpha \DE \frac{\eta}{2\cos\phi\sin\phi}.
	\end{equation}
	and hence
	\[
		\mr t(r) = \frac{\alpha}{r}.
	\]
	From this and
	\begin{align*}
		\frac{\det M(t)}{m_1(t)m_2(t)}
		&= 1-\frac{m_3(t)^2}{m_1(t)m_2(t)}
		= 1-\biggl(\frac{l_+(t)}{t}-\frac{l_-(t)}{t}\biggr)^2
		\\[1ex]
		&= \biggl(1+\frac{l_+(t)}{t}-\frac{l_-(t)}{t}\biggr)
		\biggl(1-\frac{l_+(t)}{t}+\frac{l_-(t)}{t}\biggr)
		= 4\frac{l_+(t)}{t}\cdot\frac{l_-(t)}{t}
	\end{align*}
	we obtain
	\begin{equation}\label{Y151}
		L(r) = 4\cot\phi\cdot\frac{l_+(\mr t(r))}{\mr t(r)}
		\cdot\frac{l_-(\mr t(r))}{\mr t(r)}
		= \frac{4\cot\phi}{\alpha^2}\cdot r^2l_+\Bigl(\frac{\alpha}{r}\Bigr)
		l_-\Bigl(\frac{\alpha}{r}\Bigr).
	\end{equation}
	Hence, if the sets $I_+$ and $I_-$ both have positive density at $0$ in the sense that
	\[
		\liminf_{t\to 0}\frac{l_+(t)}{t} > 0
		\qquad\text{and}\qquad
		\liminf_{t\to 0}\frac{l_-(t)}{t} > 0,
	\]
	then $L(r)\asymp A(r)\asymp1$ as $r\to\infty$.
\end{example}

\noindent
Specialising the sets $I_+,I_-$ in \thref{Y44} we obtain an example where $|q_H|$
and $\Im q_H$ oscillate between the bounds given by \thref{Y98}.

\begin{example}\thlab{Y60}
	Let $\phi\in(0,\frac\pi2)$, set
	\[
		I_+ \DE \bigcup_{n=-\infty}^\infty\bigl(2^{2n-1},2^{2n}\bigr],\qquad
		I_- \DE \bigcup_{n=-\infty}^\infty\bigl(2^{2n},2^{2n+1}\bigr],
	\]
	and let $H$ be as in \eqref{Y52}.  We have $2\cdot(I_+\cap(0,t))=I_-\cap(0,2t)$,
	and hence $\frac 1{2t}l_-(2t)=\frac 1t l_+(t)$.
	Analogously, we find $\frac 1{2t}l_+(2t)=\frac 1t l_-(t)$.
	Putting this together we obtain
	\[
		L(2r) = L(r),\quad r>0.
	\]
	Let $r\in[\alpha,2\alpha]$ where $\alpha$ is as in \eqref{Y150};
	then $\mr t(r)=\frac{\alpha}{r}\in[\frac12,1]$.
	Since $l_-(t)=\frac{1}{3}$ for $t\in[\frac12,1]$, it follows from \eqref{Y151}
	that
	\[
		L(r) = 4\cot\phi\cdot\biggl(1-\frac{l_-(\mr t(r))}{\mr t(r)}\biggr)
		\cdot\frac{l_-(\mr t(r))}{\mr t(r)}
		= 4\cot\phi\cdot\Bigl(1-\frac{r}{3\alpha}\Bigr)\cdot\frac{r}{3\alpha},
		\quad r\in[\alpha,2\alpha],
	\]
	which is quadratic in $r$ and satisfies $L(\alpha)=L(2\alpha)=\frac{8\cot\phi}{9}$
	and $L(\frac{3\alpha}{2})=\cot\phi$.
	The Hamiltonian $H$ satisfies
	\[
		H\Bigl(\frac 12 x\Bigr) = \begin{pmatrix} 1 & 0 \\ 0 & -1 \end{pmatrix}
		H(x)\begin{pmatrix} 1 & 0 \\ 0 & -1 \end{pmatrix}.
	\]
	Since the Weyl coefficient for the Hamiltonian on the left-hand side
	of the latter equation is $q_H(2z)$ and the Weyl coefficient
	for the Hamiltonian on the right-hand side is $-q_H(-z)$, we obtain
	that $q_H(2z)=-q_H(-z)$.  From this we see that
	\[
		|q_H(i\cdot 2r)|=|q_H(ir)|,\quad \Im q_H(i\cdot 2r)=\Im q_H(ir),
		\quad \Re q_H(i\cdot 2r)=-\Re q_H(ir),
	\]
	for $r>0$.
	Of course, $q_H$ is not constant since $H$ is not constant.
	Hence the limit $\lim_{r\to\infty}q_H(ir)$ does not exist.

	Let us collect what we have computed.
	First, the absolute value $|q_H(ir)|$ is a non-constant function which oscillates
	between the constant bounds \eqref{Y35} and is $2$-periodic on a logarithmic scale.
	Second, the imaginary part $\Im q_H(ir)$ is a non-constant ``$2$-periodic'' function
	which lies in between the constant upper bound and the ``$2$-periodic''
	lower bound from \eqref{Y96}.
\end{example}

\noindent
In the above example it seems that the lower bound $L(r)$ mimics the behaviour
of $|q_H|$ and $\Im q_H$ better than the upper bound $A(r)$ in the sense
that it oscillates with the same period.
For $|q_H|$ this is not always the case as the following example shows in a striking way.
For the imaginary part it is not so clear how well $L(r)$ describes its behaviour.

\begin{example}\thlab{Y63}
	Choose a sequence $(\phi_n)_{n=1}^\infty$ which oscillates between $+1$ and $-1$
	with decaying step width so that $\{\phi_n:n\in\bb N\}$ is dense in $[-1,1]$.
	For example, let
	\[
		(\phi_n)_{n=1}^\infty
		\DE \biggl(1,\frac 12,0,-\frac 12,-1,-\frac 23,-\frac 13,0,
		\frac 13,\frac 23,1,\frac 34,\ldots\biggr).
	\]
	Further, set $t_n\DE 2^{1-n^2}$ and consider the Hamiltonian
	\[
		H(t) \DE
		\begin{cases}
			\dfrac 12\begin{pmatrix} 1 & \sin(\frac\pi 2\phi_n) \\
			\sin(\frac\pi 2\phi_n) & 1 \end{pmatrix},
			\quad & t\in(t_{n+1},t_n],
			\\[3ex]
			\begin{pmatrix} 0 & 0 \\ 0 & 1 \end{pmatrix},
			\quad & t\in(1,\infty).
		\end{cases}
	\]
	This definition is made so that \cite[Theorem~4.1]{pruckner.woracek:limp-arXiv} is applicable.
	It follows that
	\begin{align*}
		& \lim_{r\to\infty}|q_H(ir)|=1
		\\[1ex]
		& \;\forall\alpha\in[0,1]\DQ\exists r_n>0\DP
		\lim_{n\to\infty}r_n=\infty \quad\text{and}\quad
		\lim_{n\to\infty}\Im q_H(ir_n)=\alpha.
	\end{align*}
	Loosely speaking we may say that $\Im q_H(ir)$ oscillates between $0$ and $|q_H(ir)|$.
	On the other hand, $m_1(t)=m_2(t)$, and hence $A(r)=1$.
	Let us consider $L(r)$.  For $n\in\bb N$ we have
	\begin{align*}
		m_3(t_n)
		&= \frac12\biggl[t_n\sin\Bigl(\frac\pi2\phi_n\Bigr)
		- t_{n+1}\sin\Bigl(\frac\pi2\phi_n\Bigr)
		+ \sum_{k=n+1}^\infty \bigl(t_k-t_{k+1}\bigr)\sin\Bigl(\frac\pi2\phi_k\Bigr)\biggr]
		\\[1ex]
		&= \frac12 t_n\sin\Bigl(\frac\pi2\phi_n\Bigr) + \rho_n
	\end{align*}
	where $|\rho_n|\le\frac12(t_{n+1}+t_{n+1})=t_{n+1}$.
	Since $(m_1m_2)(t)=\frac14r^2$, the relation $\mr t(r)=\frac{\eta}{r}$ holds with $\eta$ as
	in \thref{Y98}.  With $r_n\DE\mr r(t_n)=\frac{\eta}{t_n}\to\infty$ we obtain
	\begin{align*}
		L(r_n) &= \frac{\det M(t_n)}{(m_1m_2)(t_n)}
		= \frac{\frac14t_n^2-\bigl(\frac12t_n\sin(\frac\pi2\phi_n)+\rho_n\bigr)^2}{\frac14t_n^2}
		\\[1ex]
		&= \cos^2\Bigl(\frac\pi2\phi_n\Bigr)
		- 4\frac{\rho_n}{t_n}\sin\Bigl(\frac\pi2\phi_n\Bigr)
		- 4\Bigl(\frac{\rho_n}{t_n}\Bigr)^2
		= \cos^2\Bigl(\frac\pi2\phi_n\Bigr) + \BigO\Bigl(\frac{t_{n+1}}{t_n}\Bigr)
		\\[1ex]
		&= \cos^2\Bigl(\frac\pi2\phi_n\Bigr) + \BigO\bigl(4^{-n}\bigr),
	\end{align*}
	which oscillates between $0$ and $1$.
\end{example}

\subsection{$\bm{A(r)}$ in terms of the associated string}
\label{Y107}

A Krein string is a pair $\String Lm$ where
\begin{Itemize}
\item
	$L$ is a number in $[0,\infty]$;
\item
	$\ms m$ is a non-decreasing, left-continuous, $[0,\infty)$-valued function
	on $[0,L)$ with $\ms m(0)=0$.
\end{Itemize}
The number $L$ is called the length of the string, and the function $\ms m$
its mass distribution function.
The string equation can be written as an integro-differential equation,
\[
	y_+'(x) + z\int_{[0,x]}y(t)\DD\ms m(t) = 0,
\]
where $z\in\bb C$ is the spectral parameter and $y_+'$ denotes the right-hand
derivative of $y$; see, e.g.\ \cite{kac.krein:1968}.
Further, set $\ms m(L-)\DE\lim_{x\nearrow L}\ms m(x)$
and $\ell\DE\sup\{x\in[0,L):\ms m(x)<\ms m(L-)\}$;
then a string is called \emph{regular} if both $\ell$ and $\ms m(L-)$ are finite.

Given a string, one can construct a function $\qS$, the
\emph{principal Titchmarsh--Weyl coefficient} of the string;
see \cite{kac.krein:1968}.
This function belongs to the Stieltjes class, i.e.\ it is analytic
on $\bb C\setminus[0,\infty)$, has non-negative imaginary part in
the upper half-plane, and takes positive values on $(-\infty,0)$.
It is a fundamental theorem proved by M.\,G.~Krein that the assignment
\[
	\String Lm\mapsto \qS
\]
sets up a bijection between the set of all strings and the Stieltjes class.

Given a Hamiltonian $H$ on some interval $[a,b)$, we can define a string by setting
\begin{equation}\label{Y153}
	L \DE \lim_{t\to b}m_1(t),\qquad
	\ms m(x) \DE (m_2\circ m_1^-)(x),\ x\in[0,L),
\end{equation}
where $m_1^-$ denotes the generalised inverse of $m_1$; see \thref{Y109}.
Note that $\ms m$ is well defined and satisfies the properties above
by \thref{Y116}\,(i)--(iii).
Further, this string does not depend on the off-diagonal entry $h_3$.

The correspondence between strings and canonical systems is studied in detail in
\cite{kaltenbaeck.winkler.woracek:bimmel}.
In our present context, we need two additions, which are given in the following lemma,
namely, that the definition of the string associated with $H$ as above
does not depend on the parameterisation of $H$, and a characterisation
when the string is regular.
Recall from \thref{Y125} that an interval $(c,d)$ is $H$-indivisible of
type $0$ (respectively type $\frac\pi2$) if and only if $h_2(t)=0$ (respectively $h_1(t)=0$)
for a.e.\ $t\in(c,d)$.
Further, set
\begin{equation}\label{Y199}
	\mr a_i \DE \sup\bigl\{t\in[a,b):m_i(t)=0\bigr\}, \qquad i\in\{1,2\}.
\end{equation}
Then $H$ starts with an indivisible interval of type $0$ (respectively type $\frac\pi2$)
at the left endpoint $a$ if and only if $\mr a_2>a$ (respectively $\mr a_1>a$).

\begin{lemma}\thlab{Y117}
	Let $H$ be a Hamiltonian defined on $[a,b)$, let $\String Lm$ be
	the string associated with $H$ via \eqref{Y153} and let $\mr a_i$ be as in \eqref{Y199}.
	\begin{Enumerate}
	\item
		The following relations hold:
		\begin{align}
			m_1(\mr a_2) &= x_0 \DE \sup\{x\in[0,L):\ms m(x)=0\},
			\label{Y202}
			\\[1ex]
			m_2(\mr a_1) &= \ms m(0+) \DE \lim_{x\searrow0}\ms m(x).
			\label{Y203}
		\end{align}
		In particular, the following equivalences are true:
		\begin{alignat}{3}
			\mr a_2 &> a \quad&&\Leftrightarrow\quad && x_0>0,
			\label{Y200}
			\\[0.5ex]
			\mr a_1 &> a \quad&&\Leftrightarrow\quad && \ms m(0+)>0.
			\label{Y201}
		\end{alignat}
	\item
		The string associated with $H$ via \eqref{Y153} is regular
		if and only if there exists $c\in[a,b)$ such that
		$(c,b)$ is $H$-indivisible of type $0$ or of type $\frac\pi2$.
	\item
		Let $\wt H$ be a second Hamiltonian defined on $[\tilde a,\tilde b)$
		and assume that $H$ and $\wt H$ are related via
		\[
			\wt H = (H\circ\varphi)\cdot\varphi'\quad\text{a.e.}
		\]
		where $\varphi\DF[\tilde a,\tilde b)\to[a,b)$ is an increasing bijection
		such that $\varphi$ and $\varphi^{-1}$ are absolutely continuous.
		Then the strings associated with $H$ and $\widetilde H$ coincide.
	\end{Enumerate}
\end{lemma}

\begin{proof}
	(i)
	Since $m_1$ is continuous, it follows from \thref{Y116}\,(vi)
	that, for $x\in[0,L)$,
	\[
		\ms m(x) = 0 \quad\Leftrightarrow\quad
		m_1^-(x) \le \mr a_2 \quad\Leftrightarrow\quad
		x \le m_1(\mr a_2),
	\]
	which proves \eqref{Y202} and the equivalence in \eqref{Y200}.

	Next let us prove \eqref{Y203}.  Since $m_1$ is continuous, \thref{Y116}\,(iv)
	implies that, for $x\in(0,L)$, we have $m_1(m_1^-(x))=x>0$ and
	hence $m_1^-(x)>\mr a_1$, which shows $\lim_{x\searrow0}m_1^-(x)\ge\mr a_1$.
	Further, there exist $t_n\in(\mr a_1,b)$, $n\in\bb N$, such that $t_n\searrow\mr a_1$
	and $t_n$ is not right endpoint of an interval where $m_1$ is constant.
	Now \thref{Y16}\,(iii) yields $m_1^-(m_1(t_n))=t_n$.
	Since $\lim_{t\to\infty}m_1(t_n)=0$, we obtain $\lim_{x\searrow0}m_1^-(x)=\mr a_1$.
	Together with the continuity of $m_2$, this yields \eqref{Y203},
	and the equivalence in \eqref{Y201} follows.

	(ii)
	If $m_1\equiv0$, then $L=0$ and the string is regular.
	From now on assume that $m_1\not\equiv0$ and let $\ell$ be defined as above.
	We have $\ell<L$ if and only if $m_2$ is constant
	on an interval of the form $[c,b)$ with $c\in[a,b)$,
	i.e.\ $(c,b)$ is an indivisible interval of type $0$.
	If $\ell<L$, then $\ms m(L-)=\ms m(x)<\infty$ for $x\in(\ell,L)$
	and hence the string is regular.

	Now assume that $\ell=L$ and set $t_0\DE\sup\{t\in[a,b):m_1(t)<L\}$.
	By \thref{Y116}\,(vii) we have $\lim_{x\nearrow L}m_1^-(x)=t_0$.
	Hence it follows from the continuity of $m_1$ and \thref{Y116}\,(iv) that
	\begin{align*}
		L+\ms m(L-) &= \lim_{x\nearrow L}\bigl(x+\ms m(x)\bigr)
		= \lim_{x\nearrow L}\bigl[m_1(m_1^-(x))+m_2(m_1^-(x))\bigr]
		\\[1ex]
		&= \lim_{x\nearrow L}\int_a^{m_1^-(x)} \tr H(t)\DD t
		= \int_a^{t_0}\tr H(t)\DD t.
	\end{align*}
	Since we assumed that $H$ is in the limit point case at $b$,
	the right-hand side is finite if and only if $t_0<b$, which, in turn,
	is equivalent to the fact that $h_1(t)=0$ for a.e.\ $t\in(c,b)$ for
	some $c\in[a,b)$.

	(iii)
	Let the notation $\widetilde m_j$ correspond to $\widetilde H$.  Then we have
	\[
		\widetilde m_j(u) = \int_{a'}^u h_j\bigl(\varphi(v)\bigr)\varphi'(v)\DD v
		= m_j(\varphi(u)).
	\]
	By \thref{Y116}\,\Enumref{9} we therefore have
	\[
		\wt m_2\circ\wt m_1^- = (m_2\circ\varphi)\circ(\varphi^-\circ m_1^-)
		= m_2\circ m_1^-.
		\qedhere
	\]
\end{proof}

\noindent
The next lemma contains the relation between the Weyl coefficient $q_H$ of
the Hamiltonian and the principal Titchmarsh--Weyl coefficient $\qS$ of the string.
We also state a relation between the corresponding spectral measures,
which is used in \S\ref{Y89}.
Since $\qS$ belongs to the Stieltjes class, it has the following representation:
\begin{equation}\label{Y154}
	\qS(z) = \alphaS + \int_{[0,\infty)}\frac{1}{t-z}\DD\muS(t),
	\qquad z\in\bb C\setminus[0,\infty)
\end{equation}
with $\alphaS\ge0$ and $\muS$ a measure supported on $[0,\infty)$
satisfying $\int_{[0,\infty)}\frac{\DD\muS(t)}{1+t}<\infty$.
According to M.\,G.~Krein the measure $\muS$ is called
\emph{principal spectral measure} of the string $\String Lm$.

\begin{lemma}\thlab{Y118}
	Let $H$ be a Hamiltonian defined on some interval $[a,b)$,
	and assume that $H$ is diagonal, i.e.\ $h_3=0$.
	Moreover, let $\String Lm$ be its associated string.
	Let $q_H$, $\qS$, $\mu_H$ and $\muS$ be the Weyl coefficients
	and spectral measures of $H$ and $\String Lm$ respectively,
	and let $\alpha_H,\beta_H$ and $\alphaS$ be the constants in \eqref{Y156}
	and \eqref{Y154} respectively.
	Define the mapping $\tau:\bb R\to\bb R,t\mapsto t^2$
	and let $\tau_*\mu_H$ be the push-forward measure of $\mu_H$.
	Then
	\begin{equation}\label{Y155}
		q_H(z) = z\qS(z^2), \qquad
		\muS = \tau_*\mu_H, \qquad
		\alphaS = \beta_H \qquad\text{and}\qquad
		\alpha_H = 0.
	\end{equation}
\end{lemma}

\begin{proof}
	The first relation in \eqref{Y155} is shown
	in \cite[Theorem~4.2]{kaltenbaeck.winkler.woracek:bimmel}
	for trace-normed Hamiltonians (i.e.\ $\tr H=1$ a.e.).
	Every Hamiltonian can be reparameterised to a trace-normed Hamiltonian,
	and this changes neither its Weyl coefficient nor the associated string
	by \thref{Y117}.
	The second relation in \eqref{Y155} can either be deduced from
	\cite[Theorem~2.1]{kaltenbaeck.winkler.woracek:bimmel} or follows from
	the following considerations.
	Since $H$ is diagonal, $q_H$ is an odd function and hence $\alpha_H=0$
	in the representation \eqref{Y156} of $q_H$, and $\mu_H$ is a symmetric measure;
	see \cite[Lemma~2.2]{kaltenbaeck.winkler.woracek:bimmel}.
	Hence
	\begin{align*}
		q_H(z) &= \beta_Hz
		+ \int_{\bb R}\Bigl(\frac{1}{t-z}-\frac{t}{1+t^2}\Bigr)\DD\mu_H(t)
		\\[1ex]
		&= \beta_Hz + \frac{1}{2}\biggl[\int_{\bb R}\Bigl(\frac{1}{t-z}-\frac{t}{1+t^2}\Bigr)\DD\mu_H(t)
		+ \int_{\bb R}\Bigl(\frac{1}{-t-z}+\frac{t}{1+t^2}\Bigr)\DD\mu_H(t)\biggr]
		\\[1ex]
		&= z\biggl[\beta_H + \int_{\bb R}\frac{1}{t^2-z^2}\DD\mu_H(t)\biggr]
		= z\biggl[\beta_H + \int_{[0,\infty)}\frac{1}{s-z^2}\DD(\tau_*\mu_H)(s)\biggr].
	\end{align*}
	Now the first relation in \eqref{Y155} and the uniqueness of the
	integral representation imply that $\muS=\tau_*\mu_H$.
	The relations among the constants $\alpha_H,\beta_H$ and $\alphaS$ are clear.
\end{proof}

\noindent
We can now give a formula which represents $A(r)$ in term of the associated string.

\begin{proposition}\thlab{Y122}
	Let $H$ be a Hamiltonian defined on some interval $[a,b)$ and assume that
	neither $h_1=0$ a.e.\ nor $h_2=0$ a.e.
	Let $\String Lm$ be the string associated with $H$, and set
	\begin{equation}\label{Y169}
		\ms f(x) \DE
		\begin{cases}
			x\ms m(x) & \text{if} \ x\in[0,L), \\[0.5ex]
			L\ms m(L-) & \text{if $L<\infty$ and $x=L$} , \\[0.5ex]
			\infty & \text{if $L<\infty$ and $x\in(L,\infty)$}.
		\end{cases}
	\end{equation}
	Moreover, let $\eta\in(0,1-\frac{1}{\sqrt{2}\,})$.  Then
	\begin{equation}\label{Y123}
		A(r) = \frac{2r}{\eta}\ms f^-\Bigl(\frac{\eta^2}{4r^2}\Bigr),
		\qquad r>0.
	\end{equation}
\end{proposition}

\begin{proof}
We divide the proof into three steps.
\begin{Steps}
\item
	Let $\mr a$ and $\mr a_i$ be as in \eqref{Y196} and \eqref{Y199} respectively.
	It follows from \thref{Y117}\,(i) that, for $x\in[0,L)$,
	\[
		\ms m(x) = 0 \quad\Leftrightarrow\quad \ms f(x) = 0 \quad\Leftrightarrow\quad
		x \le m_1(\mr a_2).
	\]
	Hence $\ms f$ is strictly increasing on $[m_1(\mr a),L)$.

	Let us consider the continuity of $\ms f^-$.  Let $y\in(0,L\ms m(L-))$; then
	\begin{equation}\label{Y105}
		\ms f^-(y) = \inf\bigl\{x\in[0,L):\ms f(x)\ge y\bigr\} \ge m_1(\mr a_2),
	\end{equation}
	and it follows from \thref{Y116}\,(v) that $\ms f^-$ is continuous at $y$.
	The definition of $\ms f^-$ shows that
	\begin{equation}\label{Y129}
		\ms f^-(y)=L, \qquad y\in\bigl[L\ms m(L-),\infty\bigr).
	\end{equation}
	Since $\ms f^-$ is left-continuous at $L\ms m(L-)$ by \thref{Y116}\,(ii),
	$\ms f^-$ is continuous on $(0,\infty)$.

\item
	Our aim is to show
	\begin{equation}\label{Y178}
		m_1(t)=\ms f^-\bigl(m_1(t)m_2(t)\bigr)
	\end{equation}
	for $t\in(\mr a,b)$.
	Set $t_0\DE\sup\{t\in[a,b):m_1(t)<L\}$, which is the left endpoint
	of the maximal interval of the form $(c,b)$ on which $h_1$ vanishes
	if such an interval exists, and is equal to $b$ otherwise.

	First, let $t\in(\mr a,t_0]\cap(\mr a,b)$ be such that $t$ is not right endpoint of
	an interval where $m_1$ is constant.  Then \thref{Y116}\,(iii) implies that
	\begin{equation}\label{Y106}
		\ms f(m_1(t)) = m_1(t)m_2\bigl(m_1^-(m_1(t))\bigr) = m_1(t)m_2(t).
	\end{equation}
	Since $t>\mr a$ and $t$ is not right endpoint of an interval where $m_1$
	is constant, we have $m_1(t)>m_1(\mr a)\ge m_1(\mr a_2)$,
	and hence $m_1(t)$ is not right endpoint of an interval where $\ms f$ is constant.
	Now \thref{Y116}\,(iii), together with \eqref{Y106}, implies that
	\[
		m_1(t) = \ms f^-\bigl(\ms f(m_1(t))\bigr)
		= \ms f^-\bigl(m_1(t)m_2(t)\bigr).
	\]
	
	Next, let $[c,d]\subseteq[\mr a,t_0)$ with $c<d$ be a maximal interval
	where $m_1$ is constant.
	There exist $d_n\in(d,t_0)$, $n\in\bb N$, such that $d_n\searrow d$ and $d_n$ is not
	right endpoint of an interval where $m_1$ is constant.
	Hence \eqref{Y178} holds for $t=d_n$.
	Since $m_1$ and $m_2$ are continuous at $d$ and $\ms f^-$ is continuous
	at $m_1(d)m_2(d)>0$, we can take the limit as $n\to\infty$,
	and therefore \eqref{Y178} holds also for $t=d$.
	Now let $s\in(c,d]$.
	If $c>\mr a$, then \eqref{Y178} holds also for $t=c$, which yields
	\[
		m_1(s) = m_1(c) = \ms f^-\bigl(m_1(c)m_2(c)\bigr)
		\le \ms f^-\bigl(m_1(s)m_2(s)\bigr).
	\]
	If $c=\mr a$, then we must have $\mr a=\mr a_2>a$, and \eqref{Y105} implies
	\[
		m_1(s) = m_1(c) = m_1(\mr a_2) \le \ms f^-\bigl(m_1(s)m_2(s)\bigr).
	\]
	In both cases we obtain
	\[
		m_1(s) \le \ms f^-\bigl(m_1(s)m_2(s)\bigr)
		\le \ms f^-\bigl(m_1(d)m_2(d)\bigr) = m_1(d) = m_1(s).
	\]
	Since we must have equality everywhere, it follows that \eqref{Y178}
	holds for $s\in[c,d]\cap(\mr a,t_0)$.

	Finally, let $t\in[t_0,b)$.  It follows from \eqref{Y106} that
	\[
		m_1(t)m_2(t) \ge m_1(t_0)m_2(t_0) = \ms f\bigl(m_1(t_0)\bigr)
		= \ms f(L) = L\ms m(L-).
	\]
	From \eqref{Y129} we obtain $m_1(t)=L=\ms f^-(m_1(t)m_2(t))$.
	This finishes the proof of \eqref{Y178} for all $t\in(\mr a,b)$.

\item
	Now \eqref{Y36} and \eqref{Y178} yield
	\[
		A(r) = \frac{2r}{\eta}m_1\bigl(\mr t(r)\bigr)
		= \frac{2r}{\eta}\ms f^-\Bigl(m_1\bigl(\mr t(r)\bigr)m_2\bigl(\mr t(r)\bigr)\Bigr)
		= \frac{2r}{\eta}\ms f^-\biggl(\frac{\eta^2}{4r^2}\biggr)
	\]
	for $r>0$.
\end{Steps}
\end{proof}

\section{Relation to previous work}
\label{Y81}
\subsection{Strings and I.\,S.~Kac's theorem}
\label{Y89}

Let $\String Lm$ be a Krein string where the length $L$ and the
mass distribution function $\ms m$ are as at the beginning of \S\ref{Y107}.
The aim of the current section is to establish a relation between
the asymptotic behaviour of the spectral measure of the string at infinity
and the asymptotic behaviour of the mass distribution function at $0$.
To this end, let $H$ be a diagonal Hamiltonian that is related to the string
as in \cite[\S4]{kaltenbaeck.winkler.woracek:bimmel},
and that, in particular, \eqref{Y153} is satisfied.
For a given string, such a Hamiltonian can be constructed as follows.
Set
\[
	\tildmsm(x) \DE
	\begin{cases}
		x+\ms m(x) & \text{if} \ x\in[0,L), \\[0.5ex]
		L+\ms m(L-) & \text{if $L<\infty$ and $x=L$}, \\[0.5ex]
		\infty & \text{if $L<\infty$ and $x\in(L,\infty)$}.
	\end{cases}
\]
The generalised inverse $\tildmsm^-$ of $\tildmsm$ is defined on $[0,\infty)$
(it is even defined on $[0,\infty]$ if $L<\infty$) and absolutely continuous;
see Appendix~\ref{Y108} for the definition of the generalised inverse.
A trace-normed Hamiltonian $H$ on $[0,\infty)$ that satisfies \eqref{Y153} is then given by
\begin{equation}\label{Y157}
	H(t) \DE
	\begin{pmatrix}
		\frac{\RD\widetilde{\ms m}^{\!-\!}(t)}{\RD t} & 0
		\\[1ex]
		0 & 1-\frac{\RD\widetilde{\ms m}^{\!-\!}(t)}{\RD t}
	\end{pmatrix},
	\qquad t\in[0,\infty);
\end{equation}
see \cite[\S4]{kaltenbaeck.winkler.woracek:bimmel}
and \cite[\S5]{eckhardt.kostenko.teschl:2018}.

Using this diagonal Hamiltonian we can now prove the following characterisation
of the asymptotic behaviour of the spectral measure in the sense of
an integrability condition.
This essentially reproves the main theorem in \cite{kac:1982} by I.\,S.~Kac
for the case when no potential is present.  The theorem in \cite{kac:1982}
is stated only for regular strings although it is stated that it can be carried
over to singular strings.
In order to apply the results from \S\ref{Y43}, we have to assume that the corresponding
Hamiltonian $H$ does not start with an indivisible interval of type $0$ or $\frac\pi2$,
which, by \thref{Y117}\,(i), is equivalent to the fact that $\ms m(0+)=0$
and $\ms m(x)>0$ for $x\in(0,L)$.

\begin{corollary}\thlab{Y28}
	Let $\String Lm$ be a Krein string such that $\ms m(0+)=0$ and $\ms m(x)>0$
	for every $x\in(0,L)$.
	Further, let $\muS$ the principal spectral measure of the string as in \eqref{Y154},
	set $\vec\muS(r)\DE\muS([0,r))$, $r>0$,
	and let $\ms g$ be a continuous, non-decreasing, regularly varying function.
	Then
	\begin{equation}\label{Y159}
		\int_1^\infty \vec\muS(s)\frac{\ms g(s)}{s^2}\DD s < \infty
		\quad\Leftrightarrow\quad
		\exists x_0\in(0,L)\DP
		\int_0^{x_0}\ms g\biggl(\frac{1}{\int_0^x\ms m(t)\DD t}\biggr)\DD x < \infty.
	\end{equation}
\end{corollary}

\medskip

\noindent
Before we prove this corollary, we need the following lemma.

\begin{lemma}\thlab{Y27}
	Let $F:(0,\infty)\to(0,\infty)$ be a non-increasing function and let $c\in(0,b)$.  Then
	\begin{align*}
		\frac{1}{2}\int_0^{2m_1(c)} F\biggl(\int_0^x \ms m(\xi)\DD\xi\biggr)\DD x
		&\le \int_0^c F\bigl((m_1m_2)(t)\bigr)h_1(t)\DD t
		\\
		&\le \int_0^{m_1(c)} F\biggl(\int_0^x \ms m(\xi)\DD\xi\biggr)\DD x.
	\end{align*}
\end{lemma}

\begin{proof}
	To show the second inequality,
	we apply \cite[Proposition~1]{falkner.teschl:2012} to obtain
	\begin{align*}
		\int_0^c F\bigl((m_1m_2)(t)\bigr)h_1(t)\DD t
		&= \int_0^c F\bigl(m_1(t)m_2(t)\bigr)\DD m_1(t)
		\\
		&= \int_0^{m_1(c)} F\Bigl[m_2\bigl(m_1^-(x)\bigr)
		\underbrace{m_1\bigl(m_1^-(x)\bigr)}_{=x}\Bigr]\DD x
		\\
		&= \int_0^{m_1(x)} F\biggl[\ms m(x)\int_0^x \DD\xi\biggr]\DD x
		\le \int_0^{m_1(c)} F\biggl[\int_0^x \ms m(\xi)\DD\xi\biggr]\DD x,
	\end{align*}
	where in the last step the monotonicity of $\ms m$ and $F$ was used.

	For the first inequality we use \cite[Proposition~1]{falkner.teschl:2012} again
	\begin{align*}
		& \int_0^c F\bigl((m_1m_2)(t)\bigr)h_1(t)\DD t
		= \frac{1}{2}\int_0^c F\bigl(m_1(t)m_2(t)\bigr)\DD(2m_1)(t)
		\\[1ex]
		&= \frac{1}{2}\int_0^{2m_1(c)} F\biggl[m_2\biggl(m_1^-\Bigl(\frac{x}{2}\Bigr)\biggr)
		\underbrace{m_1\biggl(m_1^-\Bigl(\frac{x}{2}\Bigr)\biggr)}_{=\frac{x}{2}}\biggr]\DD x
		\\[1ex]
		&= \frac{1}{2}\int_0^{2m_1(c)} F\biggl[\ms m\Bigl(\frac{x}{2}\Bigr)
		\int_{x/2}^x \DD\xi\biggr]\DD x
		\ge \frac{1}{2}\int_0^{2m_1(c)} F\biggl[\int_{x/2}^x \ms m(\xi)\DD\xi\biggr]\DD x
		\\[1ex]
		&\ge \frac{1}{2}\int_0^{2m_1(c)} F\biggl[\int_0^x \ms m(\xi)\DD\xi\biggr]\DD x,
	\end{align*}
	where in the penultimate step the monotonicity of $\ms m$ and $F$ was used.
\end{proof}

\begin{proof}[Proof of Corollary~\ref{Y28}]
	Let $H$ be the Hamiltonian in \eqref{Y157}, and let $\tau$ and $\tau_*\mu_H$
	be as in \thref{Y118}.
	From \thref{Y118} we obtain that
	\[
		\vec\muS(r^2) = \muS\bigl([0,r^2)\bigr) = \tau_*\mu_H\bigl([0,r^2)\bigr)
		= \mu_H\bigl((-r,r)\bigr) = \dblarrow\mu_H(r).
	\]
	Define $\ms f(r)\DE\ms g(r^2)$, $r>0$, which is also a continuous, non-decreasing,
	regularly varying function.  A substitution yields
	\[
		\int_1^\infty\vec\muS(s)\frac{\ms g(s)}{s^2}\DD s
		= 2\int_1^\infty\vec\muS(r^2)\frac{\ms g(r^2)}{r^3}\DD r
		= 2\int_1^\infty\dblarrow\mu_H(r)\frac{\ms f(r)}{r^3}\DD r.
	\]
	The assumptions on $\ms m$ imply that neither $h_1$ nor $h_2$ vanish in
	a neighbourhood of $0$; see \thref{Y117}\,(i).  Hence we can apply \thref{Y158}, which,
	for the diagonal Hamiltonian $H$, gives the second equivalence in the following chain:
	\begin{align*}
		\int_1^\infty\vec\muS(s)\frac{\ms g(s)}{s^2}\DD s < \infty
		\quad&\Leftrightarrow\quad
		\int_1^\infty\dblarrow\mu_H(r)\frac{\ms f(r)}{r^3}\DD r < \infty
		\\[1ex]
		&\Leftrightarrow\quad
		\exists c>0\DP
		\int_0^c h_1(t)\ms f\bigl((m_1m_2)(t)^{-\frac{1}{2}}\bigr)\DD t < \infty
		\\[1ex]
		&\Leftrightarrow\quad
		\exists c>0\DP
		\int_0^c h_1(t)\ms g\Bigl(\frac{1}{(m_1m_2)(t)}\Bigr)\DD t < \infty.
	\end{align*}
	It follows from \thref{Y27} with $F(u)=\ms g(\frac{1}{u})$ that the last condition
	in the above chain is equivalent to the condition on the right-hand side
	of \eqref{Y159}.
\end{proof}

\subsection{The work of Y.~Kasahara}
\label{Y91}

Y.~Kasahara's paper \cite{kasahara:1975} deals with Krein strings as discussed
in \S\ref{Y107}.  It was a milestone in the study of high-energy asymptotics of the
principal Titchmarsh--Weyl coefficient\footnote{%
	Kasahara works with the function $\qS(-z)$ and calls it the ``characteristic function
	of the string''.}.
Its recent successor \cite{kasahara.watanabe:2010} extends the results
to so-called Kotani strings.
In these papers also some theorems about pointwise relations between the
mass distribution function of the string and its principal Titchmarsh--Weyl coefficient
are proved;
see \cite[Theorem~4]{kasahara:1975}, \cite[Section~3]{kasahara.watanabe:2010}.

Since Krein strings can be considered as diagonal canonical systems, we can use
our present theorems to obtain quite precise information about the
principal Titchmarsh--Weyl coefficient; see \thref{Y136} below.
Concerning its content, our proposition is a variant of Kasahara's Theorem~4.
It is weaker than \cite[Theorem~4]{kasahara:1975} in the sense that we have
worse universal constants in the estimates, but stronger in the sense that
we do not restrict ourselves to invertible comparison
functions\footnote{%
	This assumption is not explicitly stated in \cite[Theorem~4]{kasahara:1975}.
	However, it is needed in the proof.}.
This allows us to give a formulation which is similar
to \cite[Theorem~3.4]{kasahara.watanabe:2010}.  That result, however, is not
comparable with ours since on the one hand it deals with a more general type of strings
(for which it is not clear how we could treat them) but on the other hand restricts
to the particular situation when $\qS$ oscillates around a regularly varying function.

\begin{proposition}\thlab{Y136}
	Let $\String Lm$ be a string with $L>0$ and $\ms m\not\equiv0$.
	Denote by $\qS$ its principal Titchmarsh--Weyl coefficient,
	and define $\ms f$ as in \eqref{Y169}.
	Then
	\[
		\qS(-y) \asymp \ms f^-\Bigl(\frac 1y\Bigr),
		\qquad y\in(0,\infty).
	\]
	The constants in ``\,$\asymp$'' are independent of the string.
\end{proposition}

\begin{proof}
	Let $H$ be a diagonal Hamiltonian associated with the string $\String Lm$,
	e.g.\ as in \eqref{Y157}.
	By assumption, neither $h_1=0$ a.e.\ nor $h_2=0$ a.e.
	Moreover, fix $\eta\in(0,1-\frac{1}{\sqrt{2}\,})$.
	By \thref{Y98} and \thref{Y122} we have
	\[
		|q_H(ir)| \asymp A(r) = \frac{2r}{\eta}\ms f^-\Bigl(\frac{\eta^2}{4r^2}\Bigr).
	\]
	The values $q_H(ir)$ are purely imaginary since $H$ is diagonal,
	and \thref{Y118} gives $q_H(ir)=(ir)\qS(-r^2)$.
	Setting $r=\sqrt y$ we obtain
	\[
		\qS(-y) = \frac{|q_H(i\sqrt y)|}{\sqrt y}
		\asymp \ms f^-\Bigl(\frac{\eta^2}{4y}\Bigr)
		\asymp \ms f^-\Bigl(\frac 1y\Bigr),
	\]
	where in the last step we used \thref{Y128}.
\end{proof}

\subsection{Sturm--Liouville equations}

In this subsection we consider Sturm--Liouville equations of the form
\begin{equation}\label{Y181}
	-(py')'+qy = \lambda wy
\end{equation}
on an interval $(a,b)$, where $p(x),w(x)>0$, $q(x)\in\bb R$ for a.e.\ $x\in(a,b)$,
$\frac{1}{p},q,w\in L^1_{\text{loc}}([a,b))$,
and $\lambda$ is the spectral parameter.  Further, we assume that the equation
is in the limit-point case at the right endpoint $b$.
The Titchmarsh--Weyl coefficient corresponding to a Dirichlet boundary condition at $a$
is defined as follows.  Let $\theta(\cdot;\lambda), \phi(\cdot;\lambda)$ be solutions
of \eqref{Y181} that satisfy the initial conditions
\begin{alignat*}{2}
	\theta(a;\lambda) &= 1, \qquad & (p\theta')(a;\lambda) &= 0, \\
	\phi(a;\lambda) &= 0, \qquad & (p\phi')(a;\lambda) &= 1.
\end{alignat*}
Then $\qD(\lambda)$, $\lambda\in\bb C\setminus\bb R$, is the unique number such that
$\theta(\cdot\,;\lambda)+\qD(\lambda)\phi(\cdot\,;\lambda)\in L^2_w(a,b)$
where $L^2_w(a,b)$ denotes the weighted $L^2$-space with inner product
$(f,g)=\int_a^b f\qu{g}w$.
Denote by $\AD$ and $\AN$ the operators $y\mapsto\frac{1}{w}\bigl(-(py')'+qy\bigr)$
in $L^2_w(a,b)$ associated with \eqref{Y181} and Dirichlet ($y(a)=0$) or
Neumann boundary condition ($(py')(a)=0$) at $a$ respectively.
The Titchmarsh--Weyl coefficient can be analytically continued
to $\bb C\setminus[\min\sigma(\AD),\infty)$.

Let us first consider the case when $q\equiv0$.  In this situation
we can apply our main result, \thref{Y98}, directly.
Note that $\qD$ is defined at least on $\bb C\setminus[0,\infty)$ in this case.

\begin{corollary}\thlab{Y180}
	Let $\kappa\in(0,\frac12-\frac{1}{2\sqrt{2}\,})$, set $\sigma\DE\frac{1}{(1-2\kappa)^2}-1$,
	and, for $r>0$, let $\hat x(r)\in(a,b)$ be the unique solution of the equation
	\begin{equation}\label{Y182}
		\int_a^{\hat x(r)}w(t)\DD t\cdot\int_a^{\hat x(r)}\frac{1}{p(t)}\DD t
		= \frac{\kappa^2}{r}\,.
	\end{equation}
	Then the Titchmarsh--Weyl coefficient $\qD$ for \eqref{Y181} with $q\equiv0$
	satisfies
	\begin{equation}\label{Y183}
		C_{1,\vartheta}B(r) \le |\qD(re^{i\vartheta})| \le C_{2,\vartheta}B(r),
		\qquad r>0,\,\vartheta\in(0,2\pi),
	\end{equation}
	where
	\begin{equation}\label{Y188}
		B(r) = \frac{r}{\kappa}\int_a^{\hat x(r)}w(t)\DD t
		= \frac{\kappa}{\int_a^{\hat x(r)}\frac{1}{p(t)}\DD t}
	\end{equation}
	and
	\[
		C_{2,\vartheta}
		= \frac{1+\sigma+\frac{1}{\kappa\sin(\vartheta/2)}}{1-\sigma},
		\qquad
		C_{1,\vartheta} = \frac{1}{C_{2,\vartheta}}\,.
	\]
\end{corollary}

\begin{proof}
	With the given $p,w$ we consider the Hamiltonian
	\[
		H = \begin{pmatrix} w & 0 \\[1ex] 0 & \frac{1}{p} \end{pmatrix}.
	\]
	Comparing \eqref{Y182} and \eqref{Y30} (with $\eta=2\kappa$) we see
	that $\hat x(r)=\mr t(r^{\frac12})$.
	It follows from \cite[(3.1)]{langer.woracek:lokinv} that $\qD(z^2)=zq_H(z)$
	for $z\in\bb C\setminus\bb R$.  Now \eqref{Y35} and \eqref{Y36} imply
	\begin{align*}
		& |\qD(re^{i\vartheta})|
		= r^{\frac12}\big|q_H\bigl(r^{\frac12}e^{i\frac{\vartheta}{2}}\bigr)\big|
		\le r^{\frac12}\frac{1+\sigma+\frac{1}{\kappa\sin(\vartheta/2)}}{1-\sigma}A(r^{\frac12})
		\\[1ex]
		&= C_{2,\vartheta}r^{\frac12}\frac{r^{\frac12}}{\kappa}m_1\bigl(\mr t(r^{\frac12})\bigr)
		= C_{2,\vartheta}\frac{r}{\kappa}\int_a^{\hat x(r)}w(t)\DD t
		= C_{2,\vartheta}\frac{\kappa}{\int_a^{\hat x(r)}\frac{1}{p(t)}\DD t}\,,
	\end{align*}
	which is the second inequality in \eqref{Y183}; the lower bound follows similarly.
\end{proof}

\noindent
Let us now consider the case where also the potential $q$ is present.
We use a transformation to reduce this situation to the previous case with no potential.
Asymptotic estimates have been proved in, e.g.\ \cite[Theorem~3]{hille:1963},
\cite[Theorems~1 and 2]{atkinson:1988} and \cite[Theorem~3.3]{bennewitz:1989}.
The following corollary is similar to the latter two references although
in those theorems also sign-changing $p$ is allowed.
On the other hand, in the following corollary the bounds and the range of validity
depend\,---\,apart from an a priori lower bound of the corresponding
Neumann operator\,---\,explicitly and uniformly on certain integrals over
the coefficients in a neighbourhood of the left endpoint $a$,
in contrast to the results in the literature we know of.
Note that in some of the papers the Neumann Titchmarsh--Weyl coefficient
$\qN=-\frac{1}{\qD}$ is considered.

\begin{corollary}\thlab{Y184}
	Let $p,q,w$ be as at the beginning of this subsection and let $\kappa$, $\sigma$,
	$C_{1,\vartheta}$, $C_{2,\vartheta}$, $\hat x(r)$ and $B$ be as in \thref{Y180}.
	Further, assume that the Neumann operator $\AN$ is bounded below,
	let $\lambda_0<\min\sigma(\AN)$, choose $x_0\in(a,b)$ such that
	\begin{equation}\label{Y186}
		\int_a^{x_0}\frac{1}{p(t)}\DD t \le \frac13, \qquad
		\int_a^{x_0}\big|q(t)-\lambda_0w(t)\big|\DD t \le \frac13
	\end{equation}
	and set
	\[
		r_0 \DE 9\kappa^2
		\biggl[\int_a^{x_0}w(t)\DD t\cdot\int_a^{x_0}\frac{1}{p(t)}\DD t\biggr]^{-1}.
	\]
	Then
	\begin{equation}\label{Y189}
		\frac{C_{1,\vartheta}}{36}B(9r)
		\le \big|\qD\bigl(\lambda_0+re^{i\vartheta}\bigr)\big|
		\le \frac{9C_{2,\vartheta}}{4}B(9r),
		\qquad r\ge r_0,\;\vartheta\in(0,2\pi).
	\end{equation}
\end{corollary}

\begin{proof}
	Let $v$ be the solution of the initial value problem
	\begin{equation}\label{Y185}
		-(pv')'+qv=\lambda_0wv, \qquad v(a)=1, \quad (pv')(a)=0.
	\end{equation}
	It follows in exactly the same way as in \cite[Lemma~2.3]{niessen.zettl:1992}
	that $v(x)>0$ for all $x\in[a,b)$.

	In order to obtain an explicit estimate for $v$, we rewrite the
	initial value problem \eqref{Y185} in a standard way: with $u=\binom{v}{pv'}$,
	\eqref{Y185} is equivalent to
	\[
		u(x) = u_0 + \int_a^x
		\begin{pmatrix} 0 & \frac{1}{p(t)} \\[1ex] q(t)-\lambda_0w(t) & 0 \end{pmatrix}
		u(t)\DD t
	\]
	where $u_0=\binom10$.
	Let $T$ be the operator in the space $C([a,x_0])^2$
	(with norm $\big\|\binom{f_1}{f_2}\big\|=\|f_1\|_\infty+\|f_2\|_\infty$)
	that maps $u$ onto the integral on the right-hand side.
	It follows from \eqref{Y186} that $\|T\|\le\frac13$.
	Hence
	\[
		\|u-u_0\| = \big\|(I-T)^{-1}u_0-u_0\big\| \le \sum_{n=1}^\infty \|T\|^n\|u_0\|
		\le \frac{1}{2}\,,
	\]
	and, in particular,
	\begin{equation}\label{Y187}
		\frac12 \le v(x) \le \frac32\,, \qquad x\in[a,x_0].
	\end{equation}

	We use the following transformation: set $P\DE v^2p$, $W\DE v^2w$
	and define the unitary mapping
	\[
		U: L^2_w(a,b) \to L^2_W(a,b), \;\; y \mapsto \frac{y}{v}\,.
	\]
	Let $\lambda\in\bb C\setminus\bb R$ and let $\psi$ be a non-trivial solution
	of $-(p\psi')'+q\psi-\lambda_0w\psi=\lambda w\psi$ which is in $L^2_w(a,b)$.
	Then $\qD(\lambda_0+\lambda)=\frac{(p\psi')(a)}{\psi(a)}$.
	Set $\wt\psi\DE U\psi$, which belongs to $L^2_W(a,b)$.
	It follows from \cite[Lemma~3.2]{niessen.zettl:1992}
	that $-(P\wt\psi')'=\lambda W\wt\psi$.
	Let $\wtqD$ be the Titchmarsh--Weyl coefficient corresponding to the equation
	$-(Py')'=\lambda Wy$; then $\wtqD(\lambda)=\frac{(P\wt\psi')(a)}{\wt\psi(a)}$.
	The two Titchmarsh--Weyl coefficients are related as follows:
	\begin{align*}
		\qD(\lambda_0+\lambda) &= \frac{(p\psi')(a)}{\psi(a)}
		= \frac{\bigl(p(v\wt\psi)'\bigr)(a)}{(v\wt\psi)(a)}
		= \frac{v(a)\lim_{x\to a}\bigl(p(x)\wt\psi'(x)\bigr)+(pv')(a)\wt\psi(a)}{v(a)\wt\psi(a)}
		\\[1ex]
		&= \frac{1}{v(a)^2}\cdot\frac{(P\wt\psi')(a)}{\wt\psi(a)}+\frac{(pv')(a)}{v(a)}
		= \wtqD(\lambda)
	\end{align*}
	since $v$ satisfies the initial conditions in \eqref{Y185}.
	
	We want to apply \thref{Y180} with $P$ and $W$ instead of $p$ and $w$ respectively.
	Let $\tilde x(r)$ be the unique solution of
	\[
		\int_a^{\tilde x(r)}W(t)\DD t\cdot \int_a^{\tilde x(r)}\frac{1}{P(t)}\DD t
		= \frac{\kappa^2}{r}
	\]
	for $r>0$.  It follows from \eqref{Y187} that, for $r$ such that $\tilde x(r)\le x_0$,
	we have
	\begin{align*}
		& \int_a^{\tilde x(r)}w(t)\DD t\cdot\int_a^{\tilde x(r)}\frac{1}{p(t)}\DD t
		= \int_a^{\tilde x(r)}\frac{W(t)}{v(t)^2}\DD t\cdot
		\int_a^{\tilde x(r)}\frac{v(t)^2}{P(t)}\DD t
		\\[1ex]
		&\ge \frac{\bigl(\frac12\bigr)^2}{\bigl(\frac32\bigr)^2}
		\int_a^{\tilde x(r)}W(t)\DD t\cdot\int_a^{\tilde x(r)}\frac{1}{P(t)}\DD t
		= \frac{1}{9}\cdot\frac{\kappa^2}{r}
		= \int_a^{\hat x(9r)}w(t)\DD t\cdot\int_a^{\hat x(9r)}\frac{1}{p(t)}\DD t
	\end{align*}
	and hence $\tilde x(r)\ge\hat x(9r)$.
	In a similar way one proves $\tilde x(r)\le\hat x(\frac{r}{9})$.
	In particular, $\tilde x(r)\le x_0$ is satisfied if $\hat x(\frac{r}{9})\le x_0$,
	which, in turn, is equivalent to $r\ge r_0$.

	Now \thref{Y180} applied to $-(Py')'=\lambda Wy$ yields
	\begin{align*}
		|\qD(\lambda_0+re^{i\vartheta})|
		&= |\wtqD(re^{i\vartheta})|
		\le C_{2,\vartheta}\,\kappa\biggl[\int_a^{\tilde x(r)}\frac{1}{P(t)}\DD t\biggr]^{-1}
		\le C_{2,\vartheta}\,\kappa\,\Bigl(\frac32\Bigr)^2
		\biggl[\int_a^{\tilde x(r)}\frac{1}{p(t)}\DD t\biggr]^{-1}
		\\[1ex]
		&\le \frac{9C_{2,\vartheta}}{4}\cdot\kappa
		\biggl[\int_a^{\hat x(9r)}\frac{1}{p(t)}\DD t\biggr]^{-1}
		= \frac{9C_{2,\vartheta}}{4}B(9r)
	\end{align*}
	and
	\begin{align*}
		|\qD(\lambda_0+re^{i\vartheta})|
		&= |\wtqD(re^{i\vartheta})|
		\ge C_{1,\vartheta}\frac{r}{\kappa}\int_a^{\tilde x(r)}W(t)\DD t
		\ge C_{1,\vartheta}\frac{r}{\kappa}\cdot\Bigl(\frac12\Bigr)^2
		\int_a^{\tilde x(r)}w(t)\DD t
		\\[1ex]
		&\ge \frac{C_{1,\vartheta}}{4}\cdot\frac{r}{\kappa}\int_a^{\hat x(9r)}w(t)\DD t
		= \frac{C_{1,\vartheta}}{36}\cdot\frac{9r}{\kappa}\int_a^{\hat x(9r)}w(t)\DD t
		= \frac{C_{1,\vartheta}}{36}B(9r)
	\end{align*}
	for $r\ge r_0$, which proves \eqref{Y189}.
\end{proof}

\begin{remark}\thlab{Y195}
	It follows from \thref{Y184} and the second form of $B(r)$ in \eqref{Y188} that
	$|\qD(\lambda_0+re^{i\vartheta})|$ is bounded above and below by constants
	times the function $B$, which is monotonic increasing and tends to $\infty$
	as $r\to\infty$.
\end{remark}

\subsection{The work of H.~Winkler}
\label{Y194}

Let us now discuss \cite{winkler:2000a,winkler:2000b}.
These papers were of utmost importance for us since the method to estimate Weyl discs
is taken from there.  H.~Winkler proves three theorems about membership of Kac classes;
in our notation, these are the classes $\mc M_{\ms g_\alpha}$
with $\ms g_\alpha(r)\DE r^\alpha$, $\alpha\in(0,2)$, where the classes $\mc M_{\ms g}$
are defined in \thref{Y88}.

We start with a slightly more general situation that covers Theorems~4.2 and 4.4
in \cite{winkler:2000a}.
Let $\rho_1,\rho_2>0$, $c_1,c_2>0$ and assume that the primitive $M$ of
a Hamiltonian $H$, defined on $[0,\infty)$, satisfies
\[
	m_i(t) = c_it^{\rho_i} + \Smallo(t^{\rho_i}), \qquad t\to0,
	\;\; i=1,2.
\]
Further, assume that one of the following conditions is satisfied:
\begin{equation}\label{Y61}
\begin{aligned}
	\text{(i)} \;\; & \rho_1\ne\rho_2;
	\\[1ex]
	\text{(ii)} \;\; & \rho_1=\rho_2, \quad
	m_3(t)=c_3t^{\rho_1}+\Smallo(t^{\rho_1}), \quad c_3^2<c_1c_2.
	\hspace*{20ex}
\end{aligned}
\end{equation}
It follows from \cite[Lemma~4.3]{winkler:2000a} in the case when $H$ is trace normed
(or from the much more general Theorem~6.1 in \cite{langer.pruckner.woracek:gapsatz-arXiv})
that if (i) is satisfied, then
\[
	\limsup_{t\to0}\frac{m_3(t)^2}{m_1(t)m_2(t)}
	\le \biggl(\frac{\sqrt{\rho_1\rho_2}}{\frac12(\rho_1+\rho_2)}\biggr)^2 < 1.
\]
In the case when (ii) is satisfied, we obtain
$\limsup_{t\to0}\frac{m_3(t)^2}{m_1(t)m_2(t)}=\frac{c_3^2}{c_1c_2}<1$.
It now follows from \thref{Y167} that
\begin{equation}\label{Y93}
	\mu_H \in \mc F_{\ms g_{\alpha_0}}\setminus\mc F_{\ms g_{\alpha_0}}^0
	\qquad \text{where} \;\; \alpha_0\DE\frac{2\rho_2}{\rho_1+\rho_2}
\end{equation}
and hence
\begin{equation}\label{Y174}
	\mu_H \in \mc M_{\ms g_\alpha} \quad\Leftrightarrow\quad
	\alpha > \alpha_0.
\end{equation}
The two special cases considered in \cite{winkler:2000a} are the following.
In that paper the Hamiltonian is always trace normed,
which implies $\min\{\rho_1,\rho_2\}=1$.
\begin{Steps}
\item
	Theorem~4.2 in \cite{winkler:2000a}.
	\\[1ex]
	In that theorem the situation (ii) with $\rho_1=\rho_2=1$ is considered,
	i.e.\ $m_i(t)=c_it+\Smallo(t)$, $i=1,2,3$, with $c_3^2<c_1c_2$.
	In this case we have $\alpha_0=1$.
	One should note that this situation is actually more specific:
	by \cite[Theorem~3.1]{eckhardt.kostenko.teschl:2018} one has
	\[
		\lim_{r\to\infty} q_H(ir) = \zeta \qquad\text{with some }\zeta\in\bb C^+;
	\]
	in particular, $\Im q_H(ir)\asymp 1$, and this also implies the result.
\item
	Theorem~4.4 in \cite{winkler:2000a}.
	\\[1ex]
	In that theorem the situation (i) is considered, i.e.\ either $\rho_1=1,\rho_2>1$
	or $\rho_1>1,\rho_2=1$,
	which leads to $\alpha_0=\frac{2\rho_2}{\rho_2+1}$
	and $\alpha_0=\frac{2}{\rho_1+1}$ respectively.
	We should note that in \cite{winkler:2000a} only the classes $\mc M_{\ms g}$
	are studied and hence only \eqref{Y174} is proved but not the statement
	in \eqref{Y93}.
\end{Steps}

\noindent
Finally, let us consider a situation that is slightly more complicated.
\begin{Steps}
\setcounter{StepsCount}{173}
\item
	Theorem~4.5 in \cite{winkler:2000a}.
	\\[1ex]
	Consider a Hamiltonian $H$ defined on $[0,\infty)$ such that $M$ satisfies
	\begin{equation}\label{Y73}
	\begin{aligned}
		m_i(t) &= c_it + d_it^{\gamma} + \Smallo(t^{\gamma}),
		\qquad i=1,2,
		\\
		m_3(t) &= c_3t + d_3t^{\delta} + \Smallo(t^{\delta}),
	\end{aligned}
	\end{equation}
	as $t\to0$, where
	\[
		c_1,c_2>0, \quad c_3^2=c_1c_2, \quad
		\gamma,\delta>1, \quad d_i\in\bb R.
	\]
	We cannot argue in the same way as in \ding{172} and \ding{173}
	since $\frac{\det M(t)}{(m_1m_2)(t)}\to0$ and hence $L(r)\ll A(r)$.
	However, we can apply the transformation from \S\ref{Y173}.
	In particular, \eqref{Y171} yields the primitive $\tildM$ of a
	new Hamiltonian $\wt H$ such that
	\begin{align*}
		\wt m_1(t) &= c_2m_1(t)+c_1m_2(t)-2c_3m_2(t)
		= (c_2d_1+c_1d_2)t^\gamma - 2c_2d_3t^\delta + \Smallo(t^\kappa),
		\\[1ex]
		\wt m_2(t) &= c_1m_1(t)+c_2m_2(t)+2c_3m_3(t)
		= (c_1^2+c_2^2+2c_3^2)t + \Smallo(t),
		\\[1ex]
		\wt m_3(t) &= c_3\bigl(m_1(t)-m_2(t)\bigr)+(c_2-c_1)m_3(t)
		\\[1ex]
		&= c_3(d_1-d_2)t^\gamma + (c_2-c_1)d_3t^\delta + \Smallo(t^\kappa),
	\end{align*}
	where $\kappa\DE\min\{\gamma,\delta\}$.
	Let us assume that $\wt m_1(t)\sim ct^\kappa$ with some $c>0$, i.e.\ that one of the
	following three conditions is satisfied:
	\begin{itemize}
	\item
		$\gamma<\delta$, \; $c_2d_1+c_1d_2 > 0$;
	\item
		$\delta<\gamma$, \; $d_3<0$;
	\item
		$\gamma=\delta$, \; $c_2d_1+c_1d_2-2c_2d_3>0$.
	\end{itemize}
	(If $\wt m_1(t) = \Smallo(t^\kappa)$, then one needs more information on
	the small o terms in the representations of $m_i$.)
	The Hamiltonian $\tildM$ satisfies (i) in \eqref{Y61} with $\rho_1=\kappa$,
	$\rho_2=1$.  Hence \eqref{Y93} holds with $H$ replaced by $\wt H$
	and $\alpha_0=\frac{2}{\kappa+1}$.
	By \thref{Y168} we have $\Im q_H(ir)\asymp\Im q_{\wt H}(ir)$
	and therefore also $\dblarrow\mu_H(r)\asymp\dblarrow\mu_{\wt H}(r)$;
	see \thref{Y75}.  This implies that \eqref{Y93} holds also for $H$ and that
	\[
		\mu_H \in \mc M_{\ms g_\alpha} \quad\Leftrightarrow\quad
		\alpha > \frac{2}{\kappa+1}.
	\]
	This covers most cases in \cite[Theorem~4.5]{winkler:2000a};
	some cases where $\wt m_1(t)=\Smallo(t^\kappa)$ and more information
	on $m_i$ is known are also treated there.
	
	Let us note that, as in \ding{172}, the situation in \eqref{Y73} is more specific.
	By the main theorem of the forthcoming paper \cite{langer.pruckner.woracek:asysupp-arXiv},
	$q_H(ir)$ has a power asymptotic for $r\to\infty$.
	However, contrasting \ding{172}, this does not lead to a proof of the assertion
	concerning Kac classes since the leading term of this asymptotic expansion is real.
\end{Steps}

%
%
\appendix
\makeatletter
\DeclareRobustCommand{\@seccntformat}[1]{%
  \def\temp@@a{#1}%
  \def\temp@@b{section}%
  \ifx\temp@@a\temp@@b
  Appendix\ \csname the#1\endcsname.\quad%
  \else
  \csname the#1\endcsname\quad%
  \fi
}
\makeatother
\renewcommand{\thelemma}{\Alph{section}.\arabic{lemma}}
%
%

%
%
%
\section{Regularly varying functions}
\label{Y50}

To quantify speed of growth, we use comparison functions which behave roughly
like a power in Karamata's sense of regular variation.
In this appendix we recall the definition and some facts about such functions.
A very good source for the theory of regular variation is \cite{bingham.goldie.teugels:1989};
and this is our standard reference.

\begin{definition}\thlab{Y113}
	A function $\ms g\DF(0,\infty)\to(0,\infty)$ is called \emph{regularly varying
	with index $\alpha\in\bb R$} if it is measurable and
	\begin{equation}\label{Y111}
		\forall \lambda>0\DP
		\lim_{r\to\infty}\frac{\ms g(\lambda r)}{\ms g(r)} = \lambda^\alpha.
	\end{equation}
\end{definition}

\noindent
Examples of regularly varying functions include functions $\ms g$ behaving,
for large $r$, like
\[
	r^\alpha\cdot\bigl(\log r\bigr)^{\beta_1}\cdot\bigl(\log\log r\bigr)^{\beta_2}
	\cdot\ldots\cdot
	\bigl(\underbrace{\log\cdots\log}_{\text{\footnotesize$m$\textsuperscript{th} iterate}}r\bigr)^{\beta_m},
\]
where $\alpha,\beta_1,\ldots,\beta_m\in\bb R$, which were studied already
in the context of entire functions.
Other examples are $\ms g(r) = r^\alpha e^{(\log r)^\beta}$ with $\beta\in(0,1)$,
or $\ms g(r) = r^\alpha e^{\frac{\log r}{\log\log r}}$;
see \cite[\S 1.3]{bingham.goldie.teugels:1989}.

The following theorem shows that regularly varying functions with index $\alpha$
are asymptotically strictly between powers with exponents strictly larger than $\alpha$
and powers with exponents strictly smaller than $\alpha$.
It follows, e.g.\ from the Potter bounds;
see \cite[Theorem~1.5.6\,(iii)]{bingham.goldie.teugels:1989}.

\begin{theorem}\thlab{Y114}
	Let $\ms g:(0,\infty)\to(0,\infty)$ be a regularly varying function with
	index $\alpha\in\bb R$.
	For every $\rho\in\bb R$, $\rho\ne\alpha$, there exist $r_0>0$ and $C>0$ such that
	\begin{alignat*}{2}
		\ms g(r) &\le Cr^\rho \qquad &&\text{if $\rho>\alpha$},
		\\[0.5ex]
		\ms g(r) &\ge Cr^\rho &&\text{if $\rho<\alpha$},
	\end{alignat*}
	for $r\ge r_0$.
\end{theorem}

\medskip

\noindent
A regularly varying function with a strictly positive index is asymptotically
equivalent to a monotonic increasing, regularly varying function with the same index,
as the following theorem shows;
see \cite[Theorems~1.5.3 and 1.3.1]{bingham.goldie.teugels:1989}.

\begin{theorem}\thlab{Y115}
	Let $\ms g:(0,\infty)\to(0,\infty)$ be a locally bounded, regularly varying
	function with index $\alpha>0$.  Then
	\[
		\bar{\ms g}(r) \DE \sup\{\ms g(t):0\le t\le r\} \sim \ms g(r)
	\]
	as $r\to\infty$, and $\bar{\ms g}$ is regularly varying with index $\alpha$.
\end{theorem}

\noindent
Another fundamental result, due to J.~Karamata, determines what happens when a
regularly varying function is integrated against a power.
We recall this theorem in a comprehensive formulation collecting what is proved in
\cite[Section~1.5.6]{bingham.goldie.teugels:1989}.

\begin{theorem}[Karamata]\thlab{Y48}
	Let $\ms g$ be regularly varying at $\infty$ with index $\alpha\in\bb R$
	and assume that $\ms g$ is locally bounded.
	\begin{Enumerate}
	\item
		Let $\delta\in\bb R$ and assume that $\delta+\alpha+1\ge 0$.
		Then the function $x\mapsto\int_1^x t^\delta \ms g(t)\DD t$
		is regularly varying with index $\delta+\alpha+1$, and
		\[
			\lim_{x\to\infty}
			\biggl(
			\raisebox{3pt}{$\displaystyle x^{\delta+1}\ms g(x)$}
			\bigg/
			\raisebox{-2pt}{$\displaystyle\int_1^x t^\delta \ms g(t)\DD t$}
			\biggr)
			= \delta+\alpha+1.
		\]
	\item
		Let $\delta\in\bb R$ and assume that $\int_1^\infty t^\delta \ms g(t)\DD t<\infty$.
		Then $\delta+\alpha+1\le 0$, the function
		$x\mapsto\int_x^\infty t^\delta \ms g(t)\DD t$
		is regularly varying with index $\delta+\alpha+1$, and
		\[
			\lim_{x\to\infty}\biggl(
			\raisebox{3pt}{$\displaystyle x^{\delta+1}\ms g(x)$}
			\bigg/
			\raisebox{-2pt}{$\displaystyle\int_x^\infty t^\delta \ms g(t)\DD t$}
			\biggr)
			= -(\delta+\alpha+1).
		\]
	\end{Enumerate}
\end{theorem}

\section{The generalised inverse of a non-decreasing function}
\label{Y108}

Let us recall the notion of a generalised inverse;
see, e.g.\ \cite[Definition~2.1]{embrechts.hofert:2013}.
We slightly adapt the definition and, in particular, allow that the given function
may attain the value $+\infty$; this is convenient when we apply our results
to Krein strings.  On the set $(-\infty,\infty]$ we use the obvious order structure.

\begin{definition}\thlab{Y109}
	Let $-\infty<a<b\le\infty$, let $f\DF [a,b)\to(-\infty,\infty]$
	be a non-decreasing function and set $\ranf{f}\DE\conv(\ran f)$,
	the convex hull of the range of $f$.
	The function $f^-$, defined by
	\begin{equation}\label{Y175}
		f^-(y) \DE \inf\bigl\{x\in[a,b)\DS f(x)\ge y\bigr\},
		\qquad y\in\ranf{f},
	\end{equation}
	is called the \emph{generalised inverse} of $f$.
\end{definition}

\noindent
Note that in this definition the function $f$ is neither assumed to be strictly increasing
nor to be continuous.
Further, note that, with $b'\DE\lim_{x\to b}f(x)$, we have $\ranf{f}=[f(a),b']$
if $b'\in\ran f$ and $\ranf{f}=[f(a),b')$ otherwise.

In the next two lemmata we state some facts about generalised inverses which
we use in the present paper.
They are folklore; some can be found in \cite{embrechts.hofert:2013},
some in \cite{kaltenbaeck.winkler.woracek:bimmel}, and most probably in
many other references.  For the sake of completeness and because the setting
is slightly different, we give proofs.

\begin{lemma}\thlab{Y116}
	Let $f$ be as in \thref{Y109} and set $b'\DE\lim_{x\to b}f(x)$.
	Then the following statements hold.
	\begin{Enumerate}
	\item
		$f^-(y)\in[a,b)$ \; for all $y\in\ranf{f}$.
	\item
		$f^-$ is non-decreasing and left-continuous.
	\item
		Let $x\in[a,b)$.  Then
		\[
			f^-\bigl(f(x)\bigr) = \inf\{\xi\in[a,x]:f(\xi)=f(x)\} \le x
		\]
		with equality if and only if $x$ is not right endpoint of an interval
		where $f$ is constant.
		In particular, $f^-(f(a))=a$.
	\item
		Let $y\in\ranf{f}$.  Then
		\begin{alignat*}{2}
			f\bigl(f^-(y)\bigr) &\le y \qquad &&\text{if $f$ is left-continuous at $f^-(y)$},
			\\[1ex]
			f\bigl(f^-(y)\bigr) &\ge y \qquad &&\text{if $f$ is right-continuous at $f^-(y)$}.
		\end{alignat*}
		In particular, if $f$ is continuous on $[a,b)$, then $f(f^-(y))=y$.
	\item
		Let $y_0\in\ranf{f}$ with $y_0<b'$ and assume that $f$ is strictly increasing
		on the interval $[f^-(y_0),f^-(y_0)+\varepsilon]$ for some $\varepsilon>0$.
		Then $f^-$ is continuous at $y_0$.
	\item
		Assume hat $f$ is continuous and let $x\in[a,b)$ and $y\in\ranf{f}$.  Then
		\[
			f^-(y) \le x \quad\Leftrightarrow\quad y \le f(x).
		\]
	\item
		Set $x_0\DE\sup\{\xi\in[a,b): f(\xi)<b'\}$.
		Then $\lim\limits_{y\nearrow b'}f^-(y)=x_0$.
	\item
		Let $g$ be another function as in \thref{Y109} defined on the same interval $[a,b)$.
		If $f(x)\le g(x)$ for all $x\in[a,b)$, then
		\[
			f^-(y) \ge g^-(y), \qquad y\in\ranf{f}\cap\ranf{g}.
		\]
	\item
		Let $-\infty<c<d\le\infty$, let $I'\subseteq(-\infty,\infty]$ be an interval,
		and let $\varphi\DF[c,d)\to[a,b)$ and $\psi:\ranf{f}\to I'$ be increasing
		bijections.  Then
		\[
			\bigl(\psi\circ f\circ\varphi\bigr)^-(v)
			= \bigl(\varphi^{-1}\circ f^-\circ\psi^{-1}\bigr)(v),
			\qquad v\in I'.
		\]
	\item
		Let $\tilde f$ be another function as in \thref{Y109} defined on $[a,\tilde b)$
		such that $\tilde b\ge b$ and $f(x)=\tilde f(x)$ for $x\in[a,b)$.
		Then $\tilde f^-(y)=f^-(y)$ for $y\in\ranf{f}$.
	\end{Enumerate}
\end{lemma}

\begin{proof}
	\phantom{}
	
	(i)
	For $y\in\ranf{f}$ the set in \eqref{Y175}, of which the infimum is taken,
	is non-empty and contained in $[a,b)$.

	(ii)
	To show that $f^-$ is non-decreasing, let $y_1\le y_2$.  Then
	\[
		\bigl\{x\in[a,b)\DS f(x)\ge y_1\bigr\}
		\supseteq \bigl\{x\in[a,b)\DS f(x)\ge y_2\bigr\},
	\]
	and hence $f^-(y_1)\le f^-(y_2)$.
	In order to prove the left-continuity of $f^-$, we first show the following implications
	\begin{equation}\label{Y135}
		\begin{alignedat}{3}
			x &< f^-(y) \quad&&\Rightarrow\quad & f(x) &< y,
			\\[0.5ex]
			x &> f^-(y) \quad&&\Rightarrow\quad & f(x) &\ge y.
		\end{alignedat}
	\end{equation}
	Indeed, if $x<f^-(y)$, then $x\notin\{\xi\in[a,b):f(\xi)\ge y\}$
	and hence $f(x)<y$;
	if $x>f^-(y)$, then there exists $\xi\in[a,x)$ such that $f(\xi)\ge y$
	and therefore $f(x)\ge f(\xi)\ge y$.

	Now let $y_0\in\ranf{f}$ with $y_0>f(a)$ and assume that
	\[
		x_- \DE \lim_{y\nearrow y_0}f^-(y) < f^-(y_0).
	\]
	Choose an arbitrary number $x'\in(x_-,f^-(x_0))$.  For every $y\in(f(a),y_0)$ we have
	$f^-(y)\le x_-<x'<f^-(y_0)$.
	Hence, two applications of \eqref{Y135} yield $y \le f(x') < y_0$,
	which is a contradiction to the arbitrariness of $y$ in $(f(a),y_0)$.

	(iii)
	For $x\in[a,b)$ we have
	\begin{align*}
		f^-\bigl(f(x)\bigr) &= \inf\{\xi\in[a,b): f(\xi)\ge f(x)\}
		= \inf\{\xi\in[a,x]: f(\xi)\ge f(x)\}
		\\
		&= \inf\{\xi\in[a,x]: f(\xi)=f(x)\}.
	\end{align*}
	The remaining assertions are now clear.

	(iv)
	The implications follow from \eqref{Y135}, e.g.\ if $f$ is left-continuous
	at $f^-(y)$, then $f(f^-(y))=\lim_{x\nearrow f^-(y)}f(x)\le y$.
	The last statement clearly follows from this when $f^-(y)\in(a,b)$;
	when $f^-(y)=a$, then $f(f^-(y))=f(a)\le y$ since $y\in\ranf{f}$,
	and the other inequality follows since $f$ is right-continuous at $f^-(y)$.

	(v)
	In light of (ii) it is sufficient to show right-continuity.  Suppose that
	\[
		f^-(y_0) < \lim_{y\searrow y_0}f^-(y)
	\]
	and set
	\[
		x_+ \DE \min\Bigl\{\lim_{y\searrow y_0}f^-(y),f^-(y_0)+\varepsilon\Bigr\}.
	\]
	Further, choose $x_1,x_2\in(f^-(y_0),x_+)$ with $x_1<x_2$.
	For every $y\in(y_0,b')$ we have $f^-(y_0)<x_1<x_2<x_+\le f^-(y)$.
	Hence, applying \eqref{Y135} twice and using the strict monotonicity of $f$
	on $[f^-(y_0),x_+]$ we obtain
	\[
		y_0 \le f(x_1) < f(x_2) < y,
	\]
	which is a contradiction to the arbitrariness of $y$ in $(y_0,b')$.

	(vi)
	Assume that $f^-(y)\le x$.  Since $f$ is continuous, it follows from (iv)
	that $y=f(f^-(y))\le f(x)$.
	Now assume that $f^-(y)>x$; then $y>f(x)$ by \eqref{Y135}.

	(vii)
	We distinguish two cases.
	When $x_0<b$, then $b'\in\ranf{f}$ and $f^-(b')=x_0$.
	Hence the left-continuity of $f^-$, shown in (ii), implies the assertion.

	Now assume that $x_0=b$.  It is clear from (i) that $f^-(y)<b$
	for every $y<b'$; therefore $\lim_{y\nearrow b'}f^-(y)\le x_0$.
	To show equality in the latter relation, let $x\in[a,x_0)$ be arbitrary.
	By the definition of $x_0$ we have $f(x)<b'$.
	Hence there exists $y\in(f(x),b')$, which, by \eqref{Y135},
	implies that $f^-(y)\ge x$.

	(viii)
	For $y\in\ranf{f}\cap\ranf{g}$ we have
	\[
		\bigl\{x\in[a,b)\DS f(x)\ge y\bigr\}
		\subseteq \bigl\{x\in[a,b)\DS g(x)\ge y\bigr\}
	\]
	and hence $f^-(y)\ge g^-(y)$.

	(ix)
	Let $v\in I'$.  Then
	\begin{align*}
		\bigl\{t\in[c,d)\DS (\psi\circ f\circ\varphi)(t) \ge v\bigr\}
		&= \bigl\{t\in[c,d)\DS f(\varphi(t)) \ge \psi^{-1}(v)\bigr\}
		\\[0.5ex]
		&= \varphi^{-1}\bigl(\bigl\{x\in[a,b)\DS f(x)\ge\psi^{-1}(v)\bigr\}\bigr),
	\end{align*}
	from which the desired relation follows.

	(x)
	For $y\in\ranf{f}$ we have
	\[
		\tilde f^-(y) = \inf\{x\in[a,\tilde b):\tilde f(x)\ge y\}
		= \inf\{x\in[a,b):\tilde f(x)\ge y\},
	\]
	which implies the assertion.
\end{proof}

\noindent
The following lemma is used in \S\ref{Y91}.
To avoid distinction of cases, we use the relation $\frac{\infty}{x}=\infty$
for $x\in(0,\infty)$.

\begin{lemma}\thlab{Y128}
	Let $b\in(0,\infty]$, and let $f\DF[0,b)\to[0,\infty]$ be
	a non-decreasing function such that $f(0)=0$, $\lim_{x\to b}f(x)=\infty$
	and $x\mapsto\frac{f(x)}{x^\rho}$ is non-decreasing on $(0,b)$ for some $\rho>0$.
	Then for all $c>0$ we have $f^-(cy)\asymp f^-(y)$, $y\in[0,\infty)$.
\end{lemma}

\begin{proof}
	First note that we have either $\ranf{f}=[0,\infty)$ or $\ranf{f}=[0,\infty]$.

	It is sufficient to consider the case when $c>1$.
	From the monotonicity of $f^-$ it follows that
	\begin{equation}\label{Y177}
		f^-(y) \le f^-(cy), \qquad y\in[0,\infty).
	\end{equation}

	Our aim is to show a reverse inequality with a multiplicative constant.
	If $b<\infty$, we extend $f$ as follows:
	\[
		\tilde f(x) \DE
		\begin{cases}
			f(x), & x\in[0,b), \\[0.5ex]
			+\infty, & x\in[b,\infty).
		\end{cases}
	\]
	In the case when $b=\infty$, we set $\tilde f\DE f$.
	It follows from \thref{Y116}\,(x) that $\tilde f^-(y)=f^-(y)$
	for $y\in\ranf{f}$.
	For $x\in(0,c^{-\frac1\rho}b)$ we have
	\[
		f(x) = x^\rho\frac{f(x)}{x^\rho}
		\le x^\rho\frac{f\bigl(c^{\frac1\rho}x\bigr)}{\bigl(c^{\frac1\rho}x\bigr)^\rho}
		= \frac{f\bigl(c^{\frac1\rho}x\bigr)}{c}\,.
	\]
	We can extend this inequality to
	\begin{equation}\label{Y176}
		\tilde f(x)\le \frac{1}{c}\tilde f\bigl(c^{\frac1\rho}x\bigr),
		\qquad x\in[0,\infty).
	\end{equation}
	Define the functions $\varphi:[0,\infty)\to[0,\infty)$, $\varphi(x)=c^{\frac1\rho}x$
	and $\psi:\ranf{\tilde f}\to\ranf{\tilde f}$, $\psi(y)=\frac{1}{c}y$.
	Then \eqref{Y176} can be written as $\tilde f(x) \le (\psi\circ\tilde f\circ\varphi)(x)$,
	$x\in[0,\infty)$.
	Hence \thref{Y116}\,(viii), (ix) imply that, for $y\in\ranf{f}$,
	\begin{align*}
		f^-(y) &= \tilde f^-(y) \ge \bigl(\psi\circ\tilde f\circ\varphi\bigr)^-(y)
		= \bigl(\varphi^{-1}\circ\tilde f^-\circ\psi^{-1}\bigr)(y)
		\\[1ex]
		&= c^{-\frac1\rho}\tilde f^-(cy)
		= c^{-\frac1\rho}f^-(cy).
	\end{align*}
	In particular, this inequality is true for $y\in[0,\infty)$,
	which, together with \eqref{Y177}, proves the assertion.
\end{proof}

\noindent
Note that \thref{Y128} says that under the given assumptions the function $f^-$
belongs to the class $\mathop{OR}$ defined in \cite[\S2.0.2]{bingham.goldie.teugels:1989};
see also \thref{Y163}.


{\footnotesize
\begin{flushleft}
	M.~Langer \\
	Department of Mathematics and Statistics \\
	University of Strathclyde \\
	26 Richmond Street \\
	Glasgow G1 1XH \\
	United Kingdom \\
	email: m.langer@strath.ac.uk \\[5mm]
\end{flushleft}
\begin{flushleft}
	R.~Pruckner \\
	Institute for Analysis and Scientific Computing \\
	Vienna University of Technology \\
	Wiedner Hauptstra{\ss}e 8--10/101 \\
	1040 Wien \\
	Austria \\
	email: raphael.pruckner@tuwien.ac.at \\[5mm]
\end{flushleft}
\begin{flushleft}
	H.~Woracek \\
	Institute for Analysis and Scientific Computing \\
	Vienna University of Technology \\
	Wiedner Hauptstra{\ss}e 8--10/101 \\
	1040 Wien \\
	Austria \\
	email: harald.woracek@tuwien.ac.at
\end{flushleft}
}


\ifthenelse{\Version=0}{
\newpage

\noindent
{\large\sc Labels:}\\[10mm]
	
Y:\hspace*{3mm}
\framebox{
\begin{tabular}{r@{\quad}r@{\quad}r@{\quad}r@{\quad}r@{\quad}r@{\quad}r@{\quad}r@{\quad}r@{\quad}r@{\quad}}
	** & 01 & 02 & 03 & 04 & 05 & 06 & 07 & 08 & 09 \rule{0pt}{20pt}\kern-5pt\\[2mm]
	10 & 11 & 12 & 13 & 14 & 15 & 16 & 17 & 18 & 19 \\[2mm]
	20 & 21 & 22 & 23 & 24 & 25 & 26 & 27 & 28 & 29 \\[2mm]
	30 & 31 & 32 & 33 & 34 & 35 & 36 & 37 & 38 & 39 \\[2mm]
	40 & 41 & 42 & 43 & 44 & 45 & 46 & 47 & 48 & 49 \\[2mm]
	50 & 51 & 52 & 53 & 54 & 55 & 56 & 57 & 58 & 59 \\[2mm]
	60 & 61 & 62 & 63 & 64 & 65 & 66 & 67 & 68 & 69 \\[2mm]
	70 & 71 & 72 & 73 & 74 & 75 & 76 & 77 & 78 & 79 \\[2mm]
	80 & 81 & 82 & 83 & 84 & 85 & 86 & 87 & 88 & 89 \\[2mm]
	90 & 91 & 92 & 93 & 94 & 95 & 96 & 97 & 98 & 99 \\[2mm]
	100 & 101 & 102 & 103 & 104 & 105 & 106 & 107 & 108 & 109 \\[2mm]
	110 & 111 & 112 & 113 & 114 & 115 & 116 & 117 & 118 & 119 \\[2mm]
	120 & 121 & 122 & 123 & 124 & 125 & 126 & 127 & 128 & 129 \\[2mm]
	130 & 131 & 132 & 133 & 134 & 135 & 136 & 137 & 138 & 139 \\[2mm]
	140 & 141 & 142 & 143 & 144 & 145 & 146 & 147 & 148 & 149 \\[2mm]
	150 & 151 & 152 & 153 & 154 & 155 & 156 & 157 & 158 & 159 \\[2mm]
	160 & 161 & 162 & 163 & 164 & 165 & 166 & 167 & 168 & 169 \\[2mm]
	170 & 171 & 172 & 173 & 174 & 175 & 176 & 177 & 178 & 179 \\[2mm]
	180 & 181 & 182 & 183 & 184 & 185 & 186 & 187 & 188 & 189 \\[2mm]
	190 & 191 & 192 & 193 & 194 & 195 & 196 & 197 & 198 & 199 \\[2mm]
	200 & 201 & 202 & 203 & 204 & . & . & . & . & . \\[2mm]
	. & . & . & . & . & . & . & . & . & . \\[2mm]
\end{tabular}
}
}
{}

\end{document}